\documentclass[journal,twoside,web]{ieeecolor}

\usepackage{silence}
\usepackage{xcolor}
\usepackage{generic}
\usepackage{cite}
\usepackage{amsmath,amssymb,amsfonts}
\usepackage{algorithmic}
\usepackage{graphicx}
\usepackage[caption=false,font=footnotesize]{subfig}
\usepackage{algorithm}
\usepackage{hyperref}
\hypersetup{hidelinks}
\usepackage{textcomp}

\usepackage{bm, mathtools, marvosym, accents}
\usepackage{epstopdf}
\usepackage{tikz, environ, float}
\usetikzlibrary{shapes, arrows, calc, patterns, decorations.pathreplacing, backgrounds, positioning, fit}
\let\labelindent\IEEElabelindent
\usepackage{enumitem}
\usepackage{stfloats}
\usepackage{placeins}
\usepackage{tabularx,array,booktabs}
\usepackage{setspace}
\usepackage{pdfcomment}
\newif\ifarxivversion
\arxivversiontrue       % 编译arXiv版本需要额外添加的内容
\newtheorem{theorem}{Theorem}
\newtheorem{lemma}{Lemma}
\newtheorem{assumption}{Assumption}
\newtheorem{remark}{Remark}
\newtheorem{proposition}{Proposition}
\newcommand{\ubar}[1]{\underaccent{\bar}{#1}}
\newcommand{\sint}{\textstyle\int}
\newtheorem{assumptionprime}{Assumption}

\DeclareMathOperator{\co}{co}

\DeclareMathOperator{\esssup}{ess~sup}

\DeclareMathOperator{\diag}{diag}

\newcommand{\argmin}{\operatornamewithlimits{argmin}}

\newcommand{\biographyvspace}{-0.8cm}

\definecolor{mclr}{rgb}{0,0,1}

\def\BibTeX{{\rm B\kern-.05em{\sc i\kern-.025em b}\kern-.08em
    T\kern-.1667em\lower.7ex\hbox{E}\kern-.125emX}}
\begin{document}

\title{\LARGE \bf Safety-Critical Output-Feedback Control of Mobile Robots with Finite-Range, Set-Valued Obstacle Measurements}

\author{Si Wu, Tengfei Liu, Jinliang Ding, Yuzhe Li, Tianyou Chai, and Zhong-Ping Jiang % <-this % stops a space
\thanks{This work was supported in part by National Key R \& D Program of China (2025YFB3313200), in part by NSFC under Grants 62325303, 62333004 and 62521001, in part by the Research Program of the Liaoning Liaohe Laboratory under Grant LLL23ZZ-05-01, in part by the Key Research and Development Program of Liaoning Province under Grant 2023JH26/10200011, and in part by NSF under Grant ECCS-2210320.}% <-this % stops a space
\thanks{S. Wu, T. Liu, J. Ding, Y. Li, and T. Chai are with the State Key Laboratory of Synthetical Automation for Process Industries, Northeastern University, Shenyang, 110004, China {\tt\small(e-mails:{wusi, tfliu, jlding, yuzheli, tychai}@mail.neu.edu.cn)} }
\thanks{Z. P. Jiang is with the Department of Electrical and Computer Engineering, New York University, 370 Jay Street, Brooklyn, NY 11201, USA {\tt\small (e-mail: zjiang@nyu.edu)} }
\thanks{Corresponding author: T. Liu.}
}

\maketitle

\begin{abstract}
This paper investigates safety-critical control for mobile robots using finite-range, set-valued obstacle measurements. Obstacles are represented by disjoint convex sets in the position-like output space of a linear time-invariant robot model. The objective is to maintain a prescribed distance from obstacles while tracking a time-varying velocity command as closely as possible. Finite sensing range and nonsmooth obstacles may cause feedback discontinuities and constraint qualification violations, making it difficult to guarantee robust safety as unmeasured internal states and observer errors interact with the safety-critical output. Our contribution is twofold. First, to ensure constraint qualification despite discontinuous set-valued obstacle measurements, we develop a new approach to constructing the safety constraints. Specifically, we define a constant constraint matrix whose rows are positive-basis vectors, and combine a new safe velocity-reference input with a Lipschitz-truncated regularization of the set-valued feedback along these vectors to encode constraints on safe velocity-reference inputs. Incorporating these constraints and the velocity command into a quadratic program yields a Lipschitz continuous safety-critical controller, enabling further robust safety synthesis. Second, to address potential safety violations arising from the interaction between the safety-critical-output and internal-state subsystems, we propose a convolution-based gain expression that explicitly characterizes this interaction, and then develop a refined small-gain tool to establish robust safety of the overall system under a safety margin condition independent of the initial internal states. These main contributions are complemented by space decomposition, velocity-regulation control, and reduced-order observer design to address the overall safety-critical control problem. Robust safety in the presence of moving obstacles, external disturbances, and measurement errors is also established. Numerical simulations and quadrotor experiments in cluttered environments demonstrate the effectiveness of the proposed method.
\end{abstract}


\section{Introduction}

\IEEEPARstart{S}{afe} navigation of mobile robots in cluttered environments under a higher-level decision-maker or operator is a practically important problem in control and robotics \cite{Siegwart-Nourbakhsh-Scaramuzza-2011-Book,Loizou-TRO-2017,Cheniouni-Berkane-Tayebi-TAC-2025}. When accurate obstacle maps are unavailable, onboard finite-range sensing that provides direction-distance measurements to obstacle points is often the primary means of acquiring obstacle information in real time \cite{Arslan-Koditschek-IJRR-2019,Chen-Herbert-Hu-Pu-Fisac-Bansal-Han-Tomlin-2021-TAC,Cheniouni-Berkane-Tayebi-TAC-2025,Wu-Liu-Zhang-Ding-Jiang-Chai-TAC-2026}. The control input, typically a velocity-reference signal, must then be determined from these measurements and high-level commands, such as time-varying velocity commands. This motivates an optimization-based controller that updates the velocity reference to satisfy safety constraints induced by measured obstacle points while minimizing its deviation from the velocity command \cite{Fiorini-Shiller-IJRR-1998,Ames-Xu-Grizzle-Tabuada-TAC-2017}.

A major challenge arises when navigation relies on finite-range direction-distance obstacle measurements (e.g., point clouds) and the robot's internal states are not directly measurable. In the considered setting, the safety requirement cannot be directly described by a finite collection of explicitly predefined barrier or potential functions, since the measured contours of the surrounding cluttered obstacles generally comprise infinitely many points \cite{Siegwart-Nourbakhsh-Scaramuzza-2011-Book,Arslan-Koditschek-IJRR-2019,Martins-Marinov-Youssef-NC-2022}, thereby yielding set-valued feedback in real time. Such feedback may be discontinuous due to nonsmooth obstacle geometries, mutual occlusion, and obstacles entering or leaving the sensing range. Moreover, incorporating time-varying velocity commands and obstacle-induced constraints generally precludes a closed-form expression for the optimization-based safety-critical controller \cite{Boyd-Vandenberghe-Book-2004}. 
Crucially, as the constraints on safety-oriented velocities change with the robot's position, the gradients of the active constraint functions with respect to velocity may become linearly dependent, violating the linear independence constraint qualification. Consequently, optimization-based controllers constructed from these constraints may lose Lipschitz continuity \cite{Mestres-Allibhoy-Cortes-EJC-2025}, potentially resulting in a loss of robust safety \cite{Wu-Liu-Egerstedt-Jiang-2023-TAC,Maghenem-Karaki-TAC-2024}. In the scenario considered in this paper, dynamic uncertainties associated with unmeasured internal states and state-estimation errors interact with the safety-critical output and the robust safety issue should be carefully addressed. The problem becomes more challenging when perturbations caused by moving obstacles, external disturbances, and measurement errors should be taken into account.

Recent advances in safety-critical control provide a promising foundation. Building on invariance-based safety characterization and ideas from control Lyapunov functions, control barrier functions (CBFs) extend barrier functions to controlled systems \cite{Ames-Xu-Grizzle-Tabuada-TAC-2017,Ames-Coogan-Egerstedt-ECC-2019,Wabersich-Taylor-Choi-Sreenath-Tomlin-Ames-Zeilinger-CSM-23,Krstic-TAC-2023}. This framework transforms state constraints into constraints on the control input and establishes a unified connection with the tangent-cone condition that ensures the invariance of dynamical systems. For control-affine systems, safety can be enforced through constraints affine in the control inputs \cite{Ames-Grizzle-Tabuada-CDC-2014}, which can be combined with additional constraints and exogenous commands in a quadratic program (QP) \cite{Ames-Xu-Grizzle-Tabuada-TAC-2017}. In the single-constraint case, the resulting controller admits a closed-form expression closely related to Sontag's universal formula for stabilization using control Lyapunov functions \cite{Jankovic-Auto-2018, Liu-Wu-Liu-Jiang-Book-2026}.

Regarding recent results related to safety-critical control of mobile robots based on finite-range set-valued measurements, we would like to mention references \cite{Arslan-Koditschek-IJRR-2019,Cheniouni-Berkane-Tayebi-TAC-2025,Paternain-Koditschek-Ribeiro-TAC-2018,Farivarnejad-Lafmejani-Berman-Auto-2024,Smaili-Berkane-Auto-2026}. In particular, references \cite{Arslan-Koditschek-IJRR-2019,Cheniouni-Berkane-Tayebi-TAC-2025,Paternain-Koditschek-Ribeiro-TAC-2018,Farivarnejad-Lafmejani-Berman-Auto-2024} consider disjoint convex obstacles, whereas reference \cite{Smaili-Berkane-Auto-2026} does not assume convexity on the obstacles. Note that all these results mainly consider the single integrator as the robot model, among which \cite{Arslan-Koditschek-IJRR-2019} further extends the study to unicycles. One may recognize that learning-based and probabilistic methods have been used to reconstruct obstacle geometries, generate safe gradient fields, and characterize environmental uncertainties for mobile robots with different dynamical models, as validated experimentally in \cite{Loizou-Rimon-RAL-2022,Lin-Peng-Bansal-TRO-2026,Long-Yi-Dai-Herbert-Cortes-Atanasov-IJRR-2026}.

On the other hand, for the scenarios where safety is characterized by barrier or potential functions, besides the aforementioned designs, considerable effort has also been devoted to dealing with unmeasured internal states. References \cite{Wang-Xu-ACC-2022,Agrawal-Panagou-LCSS-2023,Ogri-Qureshi-Bell-Kamalapurkar-Automatica-2026} consider control-affine systems whose desired safe sets are defined as the super-level sets of single CBFs, and propose safety-critical output-feedback controllers based on state observers of various forms. References \cite{Furieri-Guo-Martin-FerrariTrecate-TAC-2023, He-Tanaka-TAC-2025} consider linear systems subject to multiple safety constraints that are affine in system state and control input, and propose safety-critical output-feedback controllers using data-driven synthesis and observer-based design, respectively. 

In this paper, we address safety-critical control for mobile robots modeled as controllable and observable linear time-invariant systems with position-like outputs and unmeasured internal states. Safety constraints arise from irregularly shaped, disjoint convex obstacles in the output space. With finite-range, direction-distance obstacle measurements, the objective is to maintain a prescribed safety distance from obstacles while tracking a velocity command as closely as possible.

Our main contribution is twofold:
\begin{itemize}
\item {\it Qualification-enforcing constraint construction based on set-valued measurements.} To ensure constraint qualification despite discontinuous set-valued obstacle measurements, we develop a new approach to constructing the safety constraints. Specifically, we define a constant constraint matrix whose rows are positive-basis vectors \cite{Davis-AJM-1954}, and combine a new safe velocity-reference input with a Lipschitz-truncated regularization of the set-valued feedback along these vectors to encode constraints on safe velocity-reference inputs. Indeed, the carefully chosen constraint matrix ensures the required constraint qualification. Then, following standard practice, incorporating these constraints and the velocity command into a QP yields a Lipschitz continuous safety-critical controller, enabling further robust safety synthesis.
\item {\it Safety-oriented velocity regulation and convolution-based small-gain synthesis.} Using a space decomposition and an observer-based velocity-regulation controller, we reformulate the closed-loop system as an interconnection of two subsystems: a safety-critical output subsystem affected by velocity-regulation errors, and an internal-state subsystem driven by the velocity-reference input. We characterize their dynamic interaction through convolution-based gain representations that capture the fading-memory effects of input responses. This formulation separates initial-state and input-induced responses, allowing them to be handled independently through suitably chosen initial safety distances and safety margins, respectively. We further develop a refined small-gain tool to establish robust safety of the overall system under a safety margin condition independent of the initial internal states, provided that the initial distance is sufficiently large to handle the initial-state response. Our result improves upon the conventional intuition that transient safety requires initial-state-dependent safety margins.
\end{itemize}

With the above efforts, we present a systematic framework for safety-critical output-feedback control of mobile robots with finite-range, set-valued obstacle measurements. Moreover, we establish robust safety guarantees against external disturbances, measurement errors, and moving obstacles.

This paper substantially extends our conference paper \cite{Wu-Liu-Ding-Li-Chai-Jiang-CDC-2025} in both problem formulation and technical approach. Reference \cite{Wu-Liu-Ding-Li-Chai-Jiang-CDC-2025} studies safety-critical control of linear systems with a cascade structure for avoiding spherical obstacles described by barrier functions. In contrast, the present paper addresses the coexistence of discontinuous set-valued obstacle feedback and unmeasured internal dynamics, which is not covered by the results in \cite{Wu-Liu-Ding-Li-Chai-Jiang-CDC-2025}.

{\em Notations:} In this paper, $|\cdot|$ represents the Euclidean norm for real vectors and the induced $2$-norm for real matrices. For $a\in\mathbb{R}$, denote $[a]^+=\max\{a,0\}$. For a real matrix $A$, we use $[A]_{i,j}$, $[A]_{i,:}$, and $[A]_{:,j}$ to represent the element at row $i$ and column $j$, the $i$-th row vector, and the $j$-th column vector, respectively. For vectors $x,x'\in\mathbb{R}^n$, $x<x'$ means that the corresponding elements $[x]_i$ and $[x']_i$ satisfy $[x]_i<[x']_i$ for all $i=1,\ldots,n$. The notations $\le$, $>$, and $\ge$ are defined in the same way for vectors. The Moore-Penrose pseudoinverse of a full-row-rank real matrix $A$ is given by $A^\dagger=A^T(AA^T)^{-1}$. We use $[x_1;\ldots;x_m]$ to represent the column-stacked vector $[x_1^T,\ldots,x_m^T]^T$ of real column vectors $x_1,\ldots,x_m$. For a symmetric matrix $A\in\mathbb{R}^{n\times n}$, $\lambda_{\max}(A)$ and $\lambda_{\min}(A)$ denote the maximum and minimum eigenvalues, respectively. We use $\mathbb{H}$ to represent the set of Hurwitz matrices, i.e., real matrices whose eigenvalues all lie in the open left half-plane of the complex plane. Denote the convex hull of any set $\mathcal{S}$ by $\co\mathcal{S}$. Denote $\mathbb{R}^{n}\setminus\mathcal{S}$ by $\complement\mathcal{S}$ for any $\mathcal{S}\subseteq\mathbb{R}^{n}$. A matrix $A_L\in\mathbb{R}^{m\times n}$ with $m>n$ is said to be a normalized independent positive-spanning matrix with spanning coefficient $c_A\in(0,1)$ if
\begin{enumerate}[label=(\roman*),ref=(\roman*)]
\item \label{condition_AL_unit_rows} each row of $A_L$ is a unit vector;
\item \label{condition_AL_full_rank} any $n$ rows of $A_L$ are linearly independent;
\item \label{condition_AL_positive_combination} any unit vector $l_0\in\mathbb{R}^{n}$ is a positive combination of the elements in $\{[A_L]_{i,:}^T:[A_L]_{i,:}l_0\ge c_A,i=1,\ldots,m\}$, which contains at least $n$ elements.
\end{enumerate}

For a measurable and locally essentially bounded signal $u:\mathbb{R}_+\rightarrow\mathbb{R}^m$, denote $\|u\|_I=\esssup_{\tau\in I}|u(\tau)|$ for any interval $I\subseteq\mathbb{R}_+$, $\|u\|_t=\|u\|_{[0,t]}$, and $\|u\|_\infty=\|u\|_{[0,\infty)}$. For a locally Lipschitz signal $v:\mathbb{R}_+\to\mathbb{R}^m$, we denote by $\dot{v}$ its derivative at points where it exists; since $v$ is differentiable almost everywhere, $\dot{v}$ is defined almost everywhere (one may choose arbitrary values on the measure-zero set where $v$ is not differentiable when $\dot{v}$ is used in Lebesgue integrals, essential supremum norms, or inequalities holding almost everywhere). For locally integrable functions $f,g:\mathbb{R}_+\to\mathbb{R}$, their convolution is given by
\begin{align}
(f(\cdot)\ast g(\cdot))(t)=\int_{0}^{t}f(t-\tau)g(\tau)\mathrm{d}\tau.
\end{align}

% \newpage

\section{Problem Formulation} \label{section_problemformulation}

In this paper, we consider safety-critical control of a mobile robot navigating among multiple obstacles based on finite-range, set-valued obstacle measurements while responding to a time-varying velocity-command signal.

{\em Obstacles:} Consider $n_o$ obstacles denoted by sets
\begin{subequations}
\begin{align}
\mathcal{O}_1, \ldots, \mathcal{O}_{n_o}\subset\mathbb{R}^{n_y}.
\end{align}
The overall obstacle set $\mathcal{O}$ is defined as 
\begin{align}
\mathcal{O}=\bigcup_{i=1}^{n_o}\mathcal{O}_i.
\end{align} 
\end{subequations}

{\em Mobile robot model:} The mobile robot is modeled as a linear time-invariant system
\begin{subequations} \label{eq_plant_ABC}
\begin{align}
\dot{x} &= Ax + Bu, \\
y &= Cx,
\end{align}
\end{subequations}
where $x\in\mathbb{R}^{n_x}$, $u\in\mathbb{R}^{n_u}$, and $y\in\mathbb{R}^{n_y}$ are the state, control input, and output, respectively, and $A$, $B$, and $C$ are constant real matrices of compatible dimensions. Here, $y$ serves as the position of the mobile robot, while the full state $x$, except for the portion corresponding to $y$, is unavailable for feedback.

{\em Control objective:} It is desired that the distances between the robot position $y$ and the obstacle sets $\mathcal{O}_i$ for $i=1,\ldots,n_o$ remain no less than a prescribed safety distance $D>0$ at all times. That is, under suitable initial conditions, $x(t)$ is well defined for all $t\ge 0$, and
\begin{align}
|y(t)-o|\ge D,\quad \forall o\in\mathcal{O}_i,~i=1,\ldots,n_o
\label{eq_objective}
\end{align}
holds for all $t\geq 0$. Additionally, it is desired that the velocity $\dot{y}$ deviates as little as possible from a velocity command $v_c$. In practice, $v_c$ often represents primary control objectives. 

\begin{figure}[htbp]
\centering
\includegraphics[width=1\linewidth]{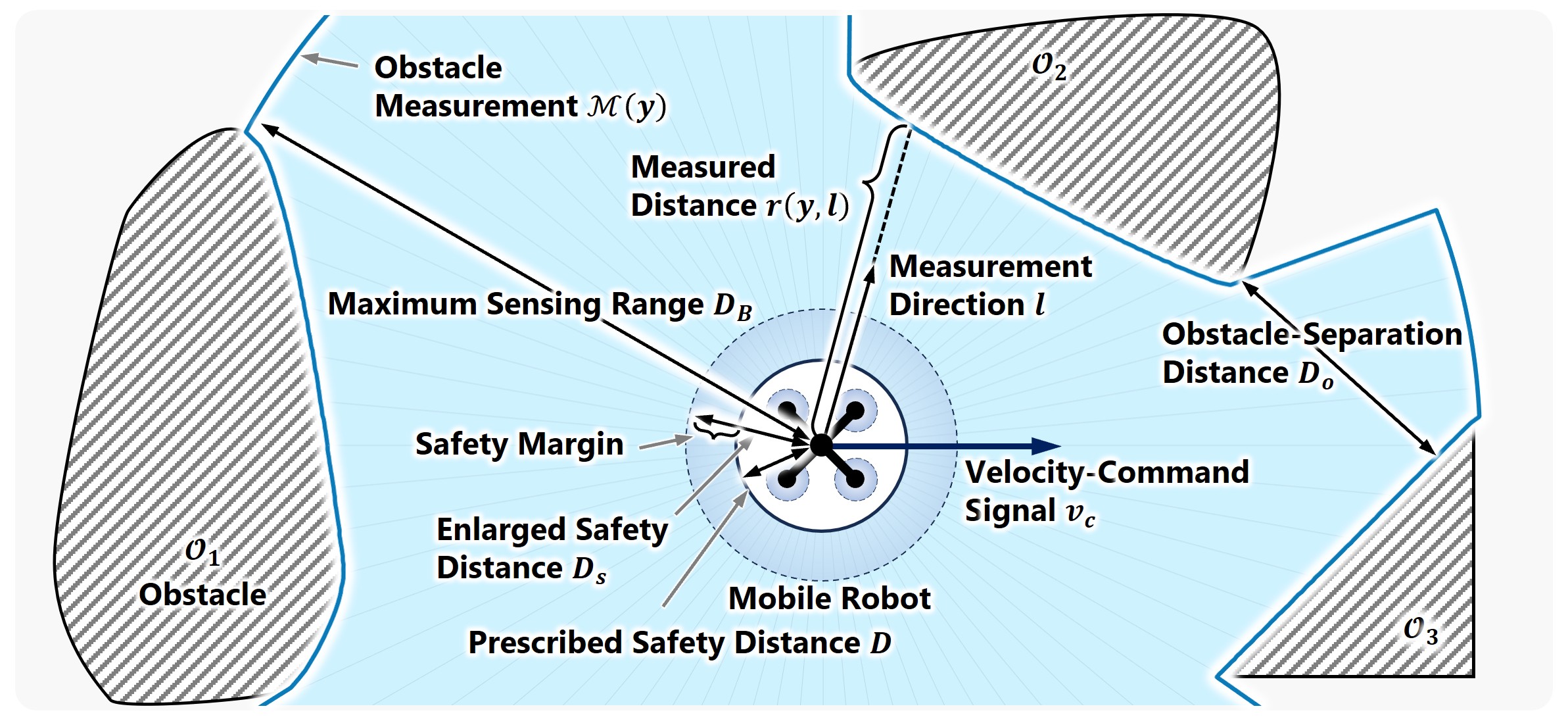}
\caption{A safety-critical control scenario.}
\label{fig_scenario}
\end{figure}

{\em Control system setup:} Although the control objective follows the main stream of the recent literature of safety-critical control, our study focuses on a control system setup in which only practically available relative measurements between $y$ and $\mathcal{O}$ can be used as obstacle information for feedback. In particular, the relative measurements between $y$ and $\mathcal{O}$ are represented by the set of vectors from $y$ to the nearest point of $\mathcal{O}$ along all directions:
\begin{subequations}
\begin{align}
\mathcal{M}(y)&=\{[l;r(y,l)]:|l|=1\}, \label{eq_setup_setvalued_measurement_map}
\end{align}
for $y\in\mathbb{R}^{n_y}$, where
\begin{align}
r(y,l)&=\min\bigl(\{s\ge0:y+sl\in\mathcal{O}\}\cup\{D_B\}\bigr) \label{eq_setup_setvalued_range_model}
\end{align}
\end{subequations} 
with $D_B>D$ being the radius of the maximum sensing range; see Figure \ref{fig_scenario}.

The map $\mathcal{M}:\mathbb{R}^{n_y}\leadsto \mathbb{R}^{n_y+1}$ in \eqref{eq_setup_setvalued_measurement_map} is set-valued, and $r(y,l)l$ represents the vector from $y$ to the nearest obstacle point along direction $l$ if it lies within the sensing range $D_B$, and equals $D_Bl$ otherwise. Clearly, our measurement model accounts for mutual occlusion among obstacles and uses only the obstacle contours captured by the finite-range onboard sensors as feedback. In practice, several types of sensors, such as LiDAR and stereo cameras, are dedicated to measuring relative positions \cite{Siegwart-Nourbakhsh-Scaramuzza-2011-Book,Corke-Book-2017}. The validity of the measurement model \eqref{eq_setup_setvalued_measurement_map} for safety-critical control has been verified in our previous works \cite{Wu-Liu-Niu-Jiang-RAL-2022,Wu-Liu-Zhang-Ding-Jiang-Chai-TAC-2026}. The set-valued feedback, together with the unmeasured internal state, poses challenges for safety-critical control.

The following assumptions are made on the obstacles, the mobile robot, and the velocity-command signal, respectively.

\begin{assumption} \label{assumption_obstacle}
The obstacles $\mathcal{O}_1,\ldots,\mathcal{O}_{n_o}$ are nonempty, closed, convex, and mutually disjoint. There exists an obstacle-separation distance $D_o>2D$ such that
\begin{align}
|o_i-o_j|\ge D_o,\qquad \forall o_i\in\mathcal{O}_i,~\forall o_j\in\mathcal{O}_j,
\label{assumption_obstacle_separation}
\end{align}
for any $i,j=1,\ldots,n_o$, $i\neq j$.
\end{assumption}

\begin{assumption} \label{assumption_plant}
For the robot model \eqref{eq_plant_ABC}, the pair $(A,C)$ is observable, the pair $(A,B)$ is controllable, $C$ has full row rank, and $AC^T=0$.
\end{assumption}

\begin{assumption}\label{assumption_command}
The velocity-command signal $v_c:\mathbb{R}_+\to\mathbb{R}^{n_y}$ is Lipschitz continuous, and there exist constants $\bar{v}_c,\bar{v}_c^d\ge 0$ such that
\begin{align}
\|v_c\|_{\infty}\le \bar{v}_c,\qquad \|\dot{v}_c\|_{\infty}\le \bar{v}_c^d. \label{eq_assumption_command}
\end{align}
\end{assumption}

\begin{remark}
It is natural to represent obstacles by closed convex sets; see, e.g., \cite{Cheniouni-Berkane-Tayebi-TAC-2025, Arslan-Koditschek-IJRR-2019}, as well as the experimental scenario in Section \ref{section_simulation_experiment} of this paper. Moreover, geometric transformations can be used for convexification when obstacles do not satisfy the convexity condition; see, e.g., \cite{Li-Tanner-TRO-2019,Loizou-TRO-2017}.
\end{remark}

\begin{remark}
Assumption \ref{assumption_plant} implies the translational symmetry of the model \eqref{eq_plant_ABC} in its output space: the vector field in \eqref{eq_plant_ABC} is invariant under translations $y \mapsto y + \mathrm{constant}$, and thus the output $y$ is a ``cyclic'' variable \cite{Marsden-Ratiu-Book-1999}. A variety of mobile robots can be modeled in the form of \eqref{eq_plant_ABC} satisfying Assumption \ref{assumption_plant}. Typical examples include near-identity diffeomorphism-transformed models of wheeled robots \cite{Wilson-Egerstedt-CSM-2020}, the controlled dynamics of quadrotors \cite{Quan-Book-17} (see also the identified model in Section \ref{section_simulation_experiment}), and linearized models of vertical takeoff and landing vehicles \cite{Fantoni-Lozano-EJC-2001}. Such a translational symmetry property enables the space decomposition in Section \ref{subsection_space_decomposition} and further facilitates safety-critical control based on relative measurements $\mathcal{M}(y)$, which can be easily obtained in real time by onboard sensors without requiring global localization \cite{Siegwart-Nourbakhsh-Scaramuzza-2011-Book,Corke-Book-2017}. Overall, our system setup is general enough to cover a wide range of safety-critical control systems.
\end{remark}

\begin{remark}\label{remark_challenges}
The main challenges lie in constructing a robust safety-critical controller from potentially discontinuous set-valued obstacle measurements $\mathcal{M}(y)$ to ensure safety in the presence of unmeasured internal dynamics. In particular, the measured distance $r(y,l)$ defined in \eqref{eq_setup_setvalued_range_model} may be discontinuous with respect to the position $y$ and sensing direction $l$; even if the distances to individual obstacles and their gradients can be obtained from $\mathcal{M}(y)$, the gradients of the resulting active constraint functions with respect to the velocity-reference input may become linearly dependent as $y$ varies, violating the linear independence constraint qualification and potentially resulting in a non-Lipschitz safety-critical controller (see Remark \ref{remark_direct_constraint_construction}) and potentially a loss of robust safety \cite{Wu-Liu-Egerstedt-Jiang-2023-TAC}. Moreover, the safety-critical output interacts with dynamic uncertainties associated with unmeasured internal states and state-estimation errors (see Figure \ref{fig_small_gain_interconnection}). This interaction calls for dedicated mathematical formulations and new tools to establish robust safety. The problem becomes even more challenging when perturbations due to moving obstacles, external disturbances, and measurement errors must also be considered (see Section \ref{section_robustness}).
\end{remark}

\section{Safety-Critical Controller Design} \label{section_control_design}

The challenges posed by finite-range set-valued feedback and unmeasured internal dynamics are addressed through a new co-design combining qualification-enforcing constraint construction with a safety-oriented, convolution-based extension of the nonlinear small-gain theorem. 

\subsection{Space Decomposition}
\label{subsection_space_decomposition}

Under Assumption \ref{assumption_plant}, the following linear coordinate transformation recasts the overall system as a cascade of safety-critical output and internal-state subsystems:
\begin{align}
[y; \eta]=[C; N^T]x, \label{eq_coordinate_transformation}
\end{align}
where $\eta\in\mathbb{R}^{n_x-n_y}$ is the internal state, and $N\in\mathbb{R}^{n_x\times(n_x-n_y)}$ is a constant matrix satisfying
\begin{align}
CN=0,\quad N^TN=I.\label{eq_coordinate_transformation_condition}
\end{align}

Through the coordinate transformation \eqref{eq_coordinate_transformation}, the robot model \eqref{eq_plant_ABC} can be rewritten as
\begin{subequations} \label{eq_yeta_system}
\begin{align}
y\text{-subsystem:}\quad \dot{y}&=v, \label{eq_yr_subsystem} \\
\eta\text{-subsystem:}\quad\dot{\eta}&=N^TAN\eta+N^TBu, \label{eq_eta_subsystem_state} \\
v&=CAN\eta+CBu,\label{eq_v}
\end{align}
\end{subequations}
where $v\in\mathbb{R}^{n_y}$ is the robot's velocity.

The following proposition establishes controllability and observability of the $\eta$-subsystem with input $u$ and output $v$.

\begin{proposition}\label{proposition_eta_controllability_observability}
Under Assumption \ref{assumption_plant}, with $N$ satisfying \eqref{eq_coordinate_transformation_condition}, the pair $(N^TAN,N^TB)$ is controllable and the pair $(N^TAN,CAN)$ is observable.
\end{proposition}

The proof of Proposition \ref{proposition_eta_controllability_observability} is given in Appendix \ref{appendix_proposition_eta_controllability_observability}.

\begin{remark}
The proposed space decomposition \eqref{eq_coordinate_transformation} resembles the coordinate transformations used in classical controllability and observability decompositions of linear systems \cite{Chen-Book-1999,Hespanha-Book-2018}, and the transformed system \eqref{eq_yeta_system} also separates the output-related and internal dynamics as in the normal forms of nonlinear systems \cite{Isidori-Book-1995}. The decomposition \eqref{eq_coordinate_transformation} enables the subsequent codesign of a safety-critical controller for the $y$-subsystem \eqref{eq_yr_subsystem} and an observer-based velocity-regulation controller for the $\eta$-subsystem \eqref{eq_eta_subsystem_state}.
\end{remark}

\subsection{Safety-Critical Control in the \texorpdfstring{$\mathit{y}$}{y}-Space}
\label{subsection_safety}

Taking advantage of the cascade structure of system \eqref{eq_yeta_system}, we consider the velocity $v$ as the virtual control input of the $y$-subsystem \eqref{eq_yr_subsystem}, and rewrite
\begin{align}
\dot{y}=v_*+\tilde{v}, \label{eq_def_y_subsystem_new}
\end{align}
where $v_*$ represents the velocity-reference input, and
\begin{align}
\tilde{v}=v-v_* \label{eq_def_vtilde}
\end{align}
is the velocity-regulation error. At the current stage, we consider $v_*$ as the control input and $\tilde{v}$ as an external disturbance, and design a safety-critical controller for the $y$-subsystem \eqref{eq_def_y_subsystem_new} based on finite-range, direction-distance obstacle measurements $\mathcal{M}(y)$.

We denote
\begin{subequations} \label{eq_safety_critical_controller}
\begin{align}
\mathcal{F}=\mathcal{M}(y), \label{eq_ideal_setvalued_feedback}
\end{align}
and propose a safety-critical controller in the following form:
\begin{align}
&v_* = \argmin_{\nu\in\mathbb{R}^{n_y}} |\nu-v_c|^2 \label{eq_safety_critical_controller_costfunction}\\
&\text{s.t.}~~~A_L(\nu-\rho_s(\mathcal{F}))\le b_L(\mathcal{F}), \label{eq_safety_critical_controller_constraints}
\end{align}
where $A_L\in\mathbb{R}^{n_L\times n_y}$ is a normalized independent positive-spanning matrix with spanning coefficient $c_A\in(0,1)$ (see Notations for the definition),
and
\begin{align}
\rho_s(\mathcal{F})&=-k_\alpha\ubar{l}(\mathcal{F})[D_s-\ubar{r}(\mathcal{F})]^+, \label{eq_feasibility_point}\\
[b_L(\mathcal{F})]_j&=[k_\alpha(r_L(\mathcal{F},[A_L]_{j,:}^T)-D_s)-[A_L]_{j,:}\rho_s(\mathcal{F})]^+, \notag \\
&\hspace{50pt} j=1,\ldots,n_L, \label{eq_def_bL_main}\\
\ubar{r}(\mathcal{F}) &= \min_{[l;r]\in\mathcal{F}} r, \\
\ubar{l}(\mathcal{F}) &= \argmin_{l'\in\co\{l:[l;\ubar{r}(\mathcal{F})]\in\mathcal{F}\}} |l'|, \\
r_L(\mathcal{F},l)&=\inf_{[l';r]\in\mathcal{F}}\{r+k_L[c_A-l^Tl']^+\} \label{eq_def_rL}
\end{align}
\end{subequations}
with constants $k_{\alpha},k_L>0$ and $D< D_s \le \min\{D_o/2,D_B\}$. Note that $D_s$ is an enlarged safety distance that provides an additional safety margin.

\begin{figure*}[!t]
\centering
\includegraphics[width=1\textwidth]{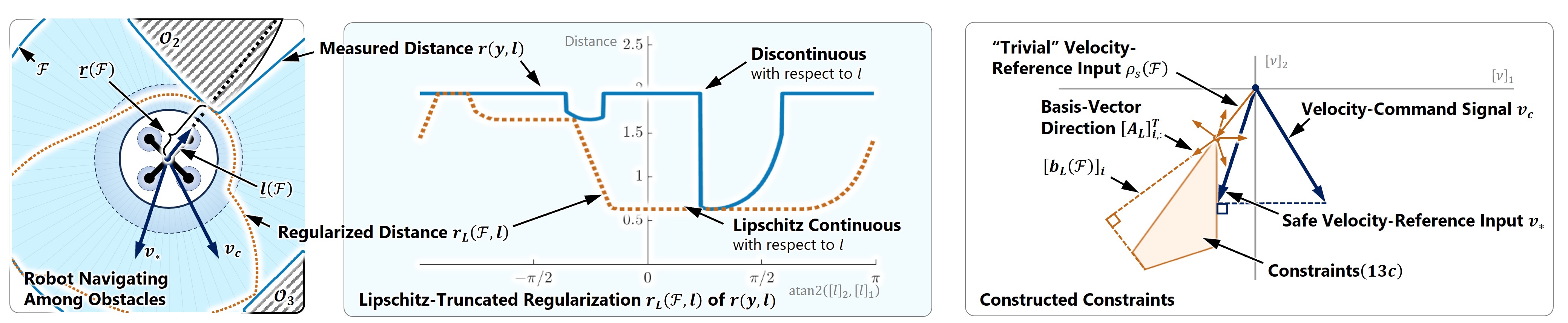}
\caption{Illustration of the Lipschitz-truncated regularization and the constructed constraints.}
\label{figure_safety_critical_controller}
\end{figure*}

Figure \ref{figure_safety_critical_controller} illustrates the construction of the constraints in \eqref{eq_safety_critical_controller_constraints} defining the set of safe velocity-reference inputs. In particular, for the set-valued feedback $\mathcal{F}$, $\rho_s(\mathcal{F})$ is constructed as a ``trivial'' velocity-reference input that ensures safety with respect to the nearest obstacle point within a radius of $D_s$, with $\ubar{r}(\mathcal{F})$ and $\ubar{l}(\mathcal{F})$ denoting the distance to and the direction toward the nearest obstacle point, respectively. The function $r_L(\mathcal{F},l)$ performs a truncated regularization of $\mathcal{F}$ along the positive-basis vectors $[A_L]_{j,:}^T$, for $j=1,\ldots,n_L$. Accordingly, $b_L(\mathcal{F})$ specifies the upper bounds on the projection of $v_*-\rho_s(\mathcal{F})$ onto each positive-basis vector, yielding the constraints \eqref{eq_safety_critical_controller_constraints}.

With the above construction, the regularized distance $r_L(\mathcal{F},l)$ varies continuously even when the measured distances $r(y,l)$ for $|l|=1$ change abruptly. Moreover, we can verify the linear independence of the ``binding constraints'' at $v_*$ \cite{Hager-SIAMControl-1979} since the constraint matrix $A_L$ is a normalized independent positive-spanning matrix, thereby ensuring constraint qualification. Overall, we can establish the feasibility and Lipschitz continuity of the QP-based controller \eqref{eq_safety_critical_controller}, as well as the robust safety of the controlled $y$-subsystem in the presence of the velocity-regulation error $\tilde{v}$; see Proposition \ref{proposition_safety}.

% Note that $r_L(\mathcal{F},l)$ applies a truncated regularization to the direction-distance obstacle measurements $\mathcal{F}=\mathcal{M}(y)$ along the basis-vector directions $[A_L]_{j,:}^T$, for $j=1,\ldots,n_L$, based on which $b_L$ is defined. This is the key to representing the set of safe velocity-reference inputs with finitely many constraints. Moreover, with the constraint matrix $A_L$ being a normalized independent positive-spanning matrix, we can verify the linear independence of the ``binding constraints'' at $v_*$ \cite{Hager-SIAMControl-1979}, which ensures constraint qualification. Then, the Lipschitz continuity of the QP-based controller \eqref{eq_safety_critical_controller} is ensured; see property (b) in Proposition \ref{proposition_safety} and its proof.

% Figure \ref{figure_safety_critical_controller} illustrates the construction of constraints defining the set of safe velocity-reference inputs. The measured distances $r(y,l)$ for $|l|=1$ can change abruptly, whereas their Lipschitz-truncated regularizations, $r_L(\mathcal{F},l)$, vary continuously. The figure also shows the set of safe velocity-reference inputs constructed using $A_L$ and $r_L(\mathcal{F},l)$.
% The following proposition establishes the feasibility and Lipschitz continuity of the QP-based controller, as well as the robust safety of the controlled $y$-subsystem in the presence of the velocity-regulation error $\tilde{v}$.

For notational convenience, define the distances from $y$ to each obstacle $\mathcal{O}_i$ and to the overall obstacle set $\mathcal{O}$ as
\begin{subequations} \label{eq_di_and_d}
\begin{align}
d_i(y)&=\inf_{o_i\in\mathcal{O}_i}|y-o_i|,
\qquad i=1,\ldots,n_o, \label{eq_d_i}\\
d(y)&=\min_{i=1,\ldots,n_o}d_i(y). \label{eq_d}
\end{align}
\end{subequations}

\begin{proposition}\label{proposition_safety}
Under Assumptions \ref{assumption_obstacle} and \ref{assumption_command}, consider the $y$-subsystem \eqref{eq_def_y_subsystem_new}, the safety-critical controller \eqref{eq_safety_critical_controller} and the obstacle measurement \eqref{eq_setup_setvalued_measurement_map} with $D_B>D$. Then, the following properties hold.
\begin{enumerate}[label=(\alph*)]
\item {\it Feasibility:} For any $(y,v_c)\in\mathbb{R}^{n_y}\times\mathbb{R}^{n_y}$, the QP \eqref{eq_safety_critical_controller} is feasible and has a unique solution.
\item {\it Boundedness and Lipschitz continuity:}
The map from $(\mathcal{M}(y),v_c)$ to $v_*$ generated by solving the QP \eqref{eq_safety_critical_controller}, denoted by
\begin{align}
v_*=\rho(\mathcal{M}(y),v_c),\label{proposition_rho}
\end{align}
satisfies
\begin{align}
|\rho(\mathcal{M}(y),v_c)|
&\le  k_{\alpha}[D_s-d(y)]^++|v_c|, \label{proposition_rho_bound}
\end{align}
for all $(y,v_c)\in\complement\mathcal{O}\times\mathbb{R}^{n_y}$; moreover, for any constant $\bar{v}_c>0$, there exists a constant $L_{\rho}>0$ such that
\begin{align}
&|\rho(\mathcal{M}(y_1),v_{c1})-\rho(\mathcal{M}(y_2),v_{c2})| \notag \\
&\qquad\le L_{\rho}(|y_1-y_2|+|v_{c1}-v_{c2}|), \label{proposition_rho_Lipschitz}
\end{align}
for all $(y_1,v_{c1})$ and $(y_2,v_{c2})$ in
\begin{align}
\{(y,v_c):d(y)\ge D, |v_c|\le \bar{v}_c\}. \label{proposition_yvcset}
\end{align}

\item {\it Robust safety with respect to the velocity-regulation error:} For each $i=1,\ldots,n_o$, if $d_i(y(0))\ge D_s$, then
\begin{align}
D_s-d_i(y(t))&\le e^{-k_\alpha t/c_A}(D_s-d_i(y(0)))\nonumber\\
&\quad+(e^{-k_\alpha(\cdot)}\ast|\tilde{v}(\cdot)|)(t) \label{eq_safety_dissipation}
\end{align}
holds on any time interval $[0,T)$ over which $d(y(t))\ge D$.
\end{enumerate}
\end{proposition}

Proposition \ref{proposition_safety} is proved in Appendix \ref{appendix_proposition_safety}. With our design, the Lipschitz continuity property \eqref{proposition_rho_Lipschitz} holds despite changes in the active constraints of the QP \eqref{eq_safety_critical_controller} due to the evolution of $(y,v_c)$, thereby facilitating the velocity-regulation design in Section \ref{subsection_regulation} and the robust safety synthesis in Section \ref{subsection_smallgain}.

\begin{remark} \label{remark_direct_constraint_construction}
Even when safety-critical distances and their gradients are available to define finitely many constraints, constructing a Lipschitz continuous controller remains nontrivial. Suppose that $d_i(y)$ and $\nabla d_i(y)$, $i=1,\ldots,n_o$, can be obtained from $\mathcal{M}(y)$. A QP-based controller could be defined with the cost function \eqref{eq_safety_critical_controller_costfunction} and constraints 
\begin{align} 
-\nabla d_i(y)^T\nu\le k_\alpha(d_i(y)-D_s), \quad i=1,\ldots,n_o. \label{eq_constraints_control} 
\end{align} 
However, the resulting map from $(y,v_c)$ to $v_*$ may not be Lipschitz: the active constraints vary with $y$ and may violate constraint qualification conditions, particularly when multiple constraints are simultaneously active; see \cite{Mestres-Allibhoy-Cortes-EJC-2025,Liu-Wu-Liu-Jiang-Book-2026,Wu-Liu-Egerstedt-Jiang-2023-TAC} for related discussions and \cite{Liu-Wu-Liu-Jiang-Book-2026} for numerical examples. 
\end{remark}

\subsection{Output-Feedback Velocity Regulation in the \texorpdfstring{$\mathit{\eta}$}{η}-Space}
\label{subsection_regulation}

Rewrite the $\eta$-subsystem composed of \eqref{eq_eta_subsystem_state} and \eqref{eq_v} as
\begin{subequations} \label{eq_eta_subsystem_new}
\begin{align}
\dot{\eta}&=A_{\eta}\eta+B_{\eta}u,\quad v=C_{\eta}\eta+D_{\eta}u\label{eq_eta_subsystem_state'}
\end{align}
by defining
\begin{align}
A_{\eta}=N^TAN,\, B_{\eta}=N^TB,\, C_{\eta}=CAN,\, D_{\eta}=CB. \label{eq_ABC_eta}
\end{align}
\end{subequations}

With $u$ as the control input and $v$ as the controlled output, we seek to design a controller that regulates the velocity $v$ to track the velocity-reference input $v_*$ generated by the safety-critical controller \eqref{eq_safety_critical_controller}, equivalently denoted in \eqref{proposition_rho}. 

The velocity-regulation controller may use only $y$ and $v_*$ according to the system setup in Section \ref{section_problemformulation}. Moreover, because $v_*=\rho(\mathcal{M}(y),v_c)$ depends on the robot's position $y$ and the external velocity command $v_c$, its time derivative, even if it exists, cannot be used in the controller design.

Our design builds on classical results in linear-system regulation \cite{Huang-Book-2004} and decoupling control \cite{Falb-Wolovich-TAC-1967} under specific structural conditions \cite{Ilchmann-Ryan-Automatica-1994}. For robust safety of our overall closed-loop system, we introduce a new design parameter $\gamma>0$ to quantify the effect of $\dot{v}_*$ on tracking error $\tilde{v}$.

\begin{figure}[htbp]
\centering
\includegraphics[width=1\linewidth]{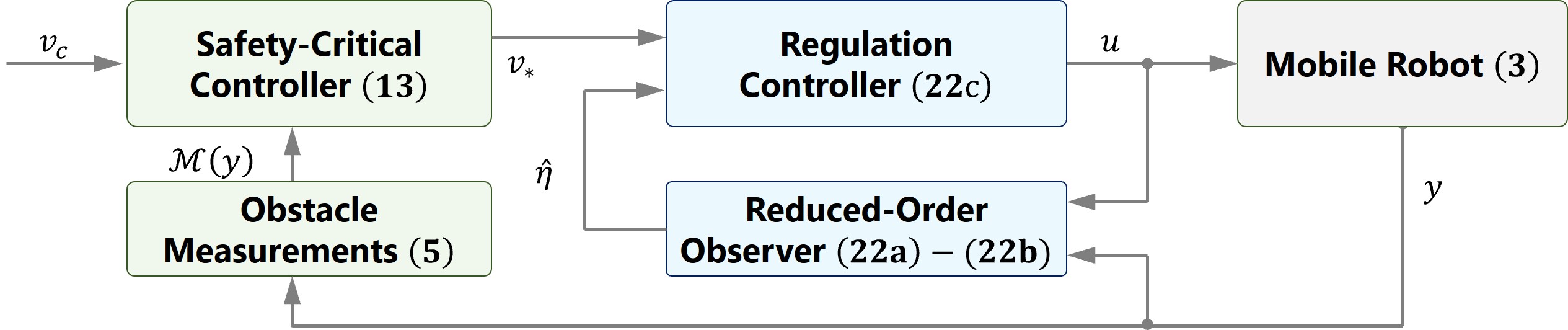}
\caption{Block diagram of the closed-loop system.}
\label{fig_control_diagram}
\end{figure}

By Proposition \ref{proposition_eta_controllability_observability}, the pair $(A_\eta,B_\eta)$ is controllable, and the pair $(A_\eta,C_\eta)$ is observable. We consider a standard reduced-order observer and a regulation controller:
\begin{subequations}\label{eq_regulation_controller}
\begin{align}
\dot{z}&=(A_{\eta}-LC_{\eta})(z+Ly)+(B_{\eta}-LD_{\eta})u, \label{eq_observer_state} \\
\hat{\eta}&=z+Ly,  \label{eq_observer} \\
u & = K\hat{\eta} + Tv_*, \label{eq_controller}
\end{align}
\end{subequations}
where $z\in\mathbb{R}^{n_x-n_y}$ is the observer state, $\hat{\eta}\in\mathbb{R}^{n_x-n_y}$ is an estimate of $\eta$, and $L\in\mathbb{R}^{(n_x-n_y)\times n_y}$, $K\in\mathbb{R}^{n_u\times(n_x-n_y)}$ and $T\in\mathbb{R}^{n_u\times n_y}$ are appropriately chosen constant matrices:
\begin{subequations}\label{eq_KT_selection_condition}
\begin{align}
\mathbb{H}&\ni A_{\eta}-LC_{\eta}, \label{eq_L_condition}\\
K&=-E_\eta^\dagger F_\eta +(I_{n_u}-E_\eta^\dagger E_\eta)K_0, \label{eq_KT_selection_condition_K}\\
T&=E_\eta^\dagger\diag(\lambda^{\sigma_1},\ldots,\lambda^{\sigma_{n_y}}), \label{eq_KT_selection_condition_T}
\end{align}
with $E_\eta$, $F_\eta$, $K_0$ and $\sigma_1,\ldots,\sigma_{n_y}$ satisfying
\begin{align}
[E_\eta]_{i,:}&=[C]_{i,:}A^{\sigma_i}B, \label{eq_def_Eeta}\\
[F_\eta]_{i,:}&=[C_\eta]_{i,:}\left(A_\eta+\lambda I\right)^{\sigma_i}, \label{eq_def_Feta}\\
\mathbb{H}&\ni A_\eta+B_\eta (-E_\eta^\dagger F_\eta +(I_{n_u}-E_\eta^\dagger E_\eta)K_0), \label{eq_KT_selection_condition_Hurwitz}\\
\sigma_i&=\min\{k\in\{0,\ldots,n_x\}:[C]_{i,:}A^kB\ne0\} \label{eq_def_ri}
\end{align}
\end{subequations}
for $i=1,\ldots,n_y$, and with $\lambda=\bar{\sigma}/\gamma$, $\bar{\sigma}=\max\{1, \sigma_1, \ldots, \sigma_{n_y}\}$ and $\gamma>0$. Here, $\gamma$, to be determined later, is used to characterize the tracking performance. One may first choose $E_\eta$, $F_\eta$ and $\sigma_1,\ldots,\sigma_{n_y}$, and then determine $K_0$ according to \eqref{eq_KT_selection_condition_Hurwitz}. Combining the designs in Sections \ref{subsection_safety} and \ref{subsection_regulation} yields the closed-loop system shown in Figure \ref{fig_control_diagram}.

\begin{proposition}\label{proposition_observer}
Under Assumption \ref{assumption_plant}, consider the $\eta$-subsystem given by \eqref{eq_eta_subsystem_state} and \eqref{eq_v}, the velocity-regulation controller \eqref{eq_regulation_controller} and the velocity-regulation error $\tilde{v}$ defined in \eqref{eq_def_vtilde}. Suppose that the velocity-reference input $v_*(t)$ is locally Lipschitz in $t$ on $\mathbb{R}_+$. Also suppose that the matrices $A,B,C$ in the robot model \eqref{eq_plant_ABC} satisfy:
\begin{itemize}[label=$\bullet$]
\item {\it Minimum-phase and right-invertibility, i.e.,}
\begin{align}
\mathrm{rank}\begin{bmatrix}sI-A&B\\ C& 0 \end{bmatrix} = n_x + n_y,
\quad \forall s\in\bar{\mathbb{C}}_+, \label{eq_rightinvertibility}
\end{align}
where $\bar{\mathbb{C}}_+=\{s\in\mathbb{C}:\operatorname{Re}(s)\ge0\}$;
\item {\it Full-row-rank decoupling matrix, i.e.,}
\begin{align}
\mathrm{rank}(E_\eta)=n_y. \label{eq_fullrowrank}
\end{align}
\end{itemize}
Then, given any constant $\gamma>0$, one can find $L$ satisfying \eqref{eq_L_condition} and $K_0$ satisfying \eqref{eq_KT_selection_condition_Hurwitz}. Moreover, there exist positive constants $c_1$ and $c_2$ such that, for any initial states $\eta(0)$, $y(0)$ and $z(0)$,
\begin{subequations} \label{eq_observer_tracking_convolution}
\begin{align}
|\tilde{v}(t)|
&\le c_1e^{-c_2t}\left(|\eta(0)|+|v_*(0)|+|\eta(0)-\hat{\eta}(0)|\right)\notag\\
&\quad+(h(\cdot)\ast|\dot{v}_*(\cdot)|)(t)  \label{eq_observer_tracking_convolution_property}
\end{align}
holds for all $t\ge 0$, where $\hat{\eta}$ is defined in \eqref{eq_observer}, and 
\begin{align}
h(t)&=e^{-{\bar{\sigma}}t/{\gamma}}\sum_{q=0}^{\bar{\sigma}-1}
\frac{\bar{\sigma}^qt^q}{q! \gamma^q}. \label{eq_observer_tracking_convolution_h}
\end{align}
\end{subequations}
\end{proposition}

The proof of Proposition \ref{proposition_observer} is given in Appendix \ref{appendix_proposition_observer}. 

\begin{remark}
The velocity-regulation controller \eqref{eq_regulation_controller} allows arbitrary attenuation of the influence of $\dot{v}_*$ on $\tilde{v}$, as quantified by $\gamma$ in \eqref{eq_observer_tracking_convolution}. For example, consider the $\eta$-subsystem \eqref{eq_eta_subsystem_state'} modeled as an $(n_x-n_y)$-th order integrator. In this case, $\sigma_1=\bar{\sigma}=n_x-n_y$, $E_\eta=1$, and
\begin{align}
[F_\eta]_{1,q}=\frac{\bar{\sigma}!}{(q-1)!(\bar{\sigma}-q+1)!}\lambda^{\bar{\sigma}-q+1},~ q=1,\ldots,\bar{\sigma}.
\end{align}
Given any $\gamma>0$, according to \eqref{eq_KT_selection_condition}, choose $K=-F_\eta$ and $T=\lambda^{\bar{\sigma}}$ with $\lambda=\bar{\sigma}/\gamma$. Then, $\det(sI-A_\eta-B_\eta K)=(s+\lambda)^{\bar{\sigma}}$, which ensures that $A_\eta+B_\eta K$ is Hurwitz. Moreover, $h$ in \eqref{eq_observer_tracking_convolution_h} satisfies $\sint_0^\infty h(t)\mathrm{d}t=\gamma$, indicating that the influence of $\dot{v}_*$ on $\tilde{v}$ can be reduced by choosing a smaller $\gamma$.
\end{remark}

\subsection{Convolution-Based Small-Gain Synthesis for Closed-Loop Safety}
\label{subsection_smallgain}

Properties \eqref{proposition_rho}, \eqref{eq_safety_dissipation} and \eqref{eq_observer_tracking_convolution} characterize the interaction between the controlled $y$-subsystem and the $\eta$-subsystem, which form the overall closed-loop system. We propose a refined small-gain synthesis to ensure robust safety of the overall system.

The designs in the preceding subsections rely on several conditions, recalled here for the subsequent synthesis:
\begin{enumerate}[label=(\alph*),ref=(\alph*)]
\item \label{condition_parameter_safety_controller} For the safety-critical controller \eqref{eq_safety_critical_controller}, $A_L\in\mathbb{R}^{n_L\times n_y}$ is a normalized independent positive-spanning matrix, $D_s\in(D,\min\{D_o/2,D_B\}]$, and $k_L,k_\alpha>0$.
\item \label{condition_parameter_regulation} For the velocity-regulation controller \eqref{eq_regulation_controller}, $L$, $K$ and $T$ satisfy \eqref{eq_KT_selection_condition} for any given $\gamma>0$.
\end{enumerate}

To ensure safety of the closed-loop system, we also choose the controller parameters to satisfy an additional condition:
\begin{enumerate}[label=(\alph*),ref=(\alph*),start=3]
\item \label{condition_parameter_safety_margin} With the constants $\bar{v}_c,\bar{v}_c^d\ge 0$ specified in Assumption \ref{assumption_command}, for any $D_s$ and $k_\alpha$ selected according to condition \ref{condition_parameter_safety_controller}, the constant $\gamma$ is chosen to satisfy
\begin{align}
0< \gamma <\frac{k_{\alpha}(D_s-D)}
{L_{\rho}\left(\bar{v}_c+\bar{v}_c^d+2k_{\alpha}(D_s-D)\right)},
\label{eq_command_bound_condition}
\end{align}
where $L_\rho$ is the Lipschitz constant in \eqref{proposition_rho_Lipschitz} for the map from $(y,v_c)$ to $v_*$ of the QP \eqref{eq_safety_critical_controller}. In the proof of the main result, condition \eqref{eq_command_bound_condition} ensures the small-gain condition.
\end{enumerate}

Our first main result is given by Theorem \ref{theorem_small_gain_safety}.

\begin{theorem}\label{theorem_small_gain_safety}
Consider the obstacles $\mathcal{O}_1,\ldots,\mathcal{O}_{n_o}$ satisfying Assumption \ref{assumption_obstacle}, the mobile robot \eqref{eq_plant_ABC} satisfying Assumption \ref{assumption_plant} and conditions \eqref{eq_rightinvertibility} and \eqref{eq_fullrowrank}, any velocity-command signal $v_c$ satisfying Assumption \ref{assumption_command}, the obstacle measurement \eqref{eq_setup_setvalued_measurement_map} with $D_B>D$, and the controller defined by \eqref{eq_safety_critical_controller}
and \eqref{eq_regulation_controller}. Then, there exist controller parameters $D_s$, $k_L$, $A_L$, $k_\alpha$, $L$, $K$, and $T$ satisfying conditions \ref{condition_parameter_safety_controller}--\ref{condition_parameter_safety_margin} with a suitably chosen $\gamma>0$. Moreover, there exist positive constants $c_0$ and $c_D$ such that, under initial conditions
\begin{subequations}\label{eq_initial_condition}
\begin{align}
d(Cx(0)) &\ge D_s + c_D,  \label{eq_initial_condition_y}\\
|N^Tx(0)|+|N^Tx(0)-z(0)-LCx(0)| &\le c_0, \label{eq_initial_condition_xz}
\end{align} 
\end{subequations}
the safety property \eqref{eq_objective} holds for all $t\ge 0$.
\end{theorem}

\begin{proof}
The controller parameters can be selected by first choosing $D_s$, $k_L$, $A_L$, and $k_\alpha$ according to condition \ref{condition_parameter_safety_controller}, then choosing $\gamma>0$ sufficiently small to satisfy condition \ref{condition_parameter_safety_margin}, and finally choosing $L$, $K$, and $T$ to satisfy condition \ref{condition_parameter_regulation}.

{\it 1) Existence and uniqueness of $(x(t),z(t))$ over a possibly infinite time interval.}
With $x=[C;N^T]^{-1}[y;\eta]$ in \eqref{eq_coordinate_transformation}, $z=\hat{\eta}-Ly$ in \eqref{eq_observer} and $v_*=\rho(\mathcal{M}(y),v_c)$ in \eqref{proposition_rho}, we represent the vector field of the closed-loop system composed of \eqref{eq_plant_ABC}, \eqref{eq_safety_critical_controller} and \eqref{eq_regulation_controller} as
\begin{align}
f(t,x,z)=\begin{bmatrix}
AN\eta+B(K\hat\eta+Tv_*)\\
(A_\eta-LC_\eta)\hat\eta+(B_\eta-LD_\eta)(K\hat\eta+Tv_*)
\end{bmatrix},
\label{eq_f_new}
\end{align}
for which $v_c$ is considered as a function of time $t$.

Under Assumption \ref{assumption_command}, the Lipschitz continuity property \eqref{proposition_rho_Lipschitz} of the safety-critical controller, together with the linearity of robot model \eqref{eq_plant_ABC} and velocity-regulation controller \eqref{eq_regulation_controller}, implies that $f(t,x,z)$ is continuous in $t$ and Lipschitz in $(x,z)$ on
\begin{align}
\mathcal{D} = \{(x,z):d(Cx)>D,z\in\mathbb{R}^{n_x-n_y}\}. \label{eq_domain}
\end{align}
Moreover, by \cite[Theorem 3.1]{Khalil-book-2002}, the closed-loop system admits a unique solution for any $x(0)$ and $z(0)$ satisfying \eqref{eq_initial_condition}, which places the initial states in the domain $\mathcal{D}$. Let $[0,T_m)$, with $0<T_m\le\infty$, denote the right maximal interval on which $(x(t),z(t))$ is defined and remains in $\mathcal{D}$. The continuity property of $f(t,x,z)$ established above ensures that the solution is continuously differentiable on $[0,T_m)$.

{\it 2) Safety over the right maximal interval $[0,T_m)$.}

{\it 2.1) Convolution-based gain interconnections}
The key for safety analysis of the closed-loop system lies in addressing the interaction between two variables: the distance deficit $[D_s-d(y)]^+$ and the velocity-regulation error $|\tilde{v}|$.

We first assess how $\tilde{v}$ affects $[D_s-d(y)]^+$. Under the initial condition \eqref{eq_initial_condition_y}, with property \eqref{eq_safety_dissipation} describing the influence of $\tilde{v}$ on $d_i(y)$, using the definition of $d(y)$ in \eqref{eq_d}, we have
\begin{align}
&[D_s-d(y(t))]^+=\max_{i=1,\ldots,n_o}[D_s-d_i(y(t))]^+\notag\\
&\le\max_{i=1,\ldots,n_o}[e^{-k_\alpha t/c_A} (D_s-d_i(y(0)))+(e^{-k_\alpha(\cdot)}\ast|\tilde{v}(\cdot)|)(t)]^+\notag\\
&\le(e^{-k_\alpha(\cdot)}\ast|\tilde{v}(\cdot)|)(t)
\label{eq_proof_small_gain_distance_convolution}
\end{align}
for all $t\in[0,T_m)$.

Now we analyze the influence of $[D_s-d(y)]^+$ on $\tilde{v}$. Note that property \eqref{eq_observer_tracking_convolution} shows the influence of $|\dot{v}_*|$ on $|\tilde{v}|$. We check the differentiability of $v_*$, generated by the safety-critical controller \eqref{proposition_rho}. Since $x(t)$ is continuously differentiable and $v_c(t)$ is Lipschitz under Assumption \ref{assumption_command}, we have that $y(t)=Cx(t)$ and $v_c(t)$ are locally Lipschitz on $[0,T_m)$. With \eqref{proposition_rho} satisfying \eqref{proposition_rho_Lipschitz}, by \cite[Proposition 2.3.1]{Cobzas-Miculescu-Nicolae-book-2019}, we have $v_*(t)$ is locally Lipschitz on $[0,T_m)$. Also recall the Lipschitz continuity of $v_c$ in Assumption \ref{assumption_command}. By \cite[Remark 3.4.2]{Cobzas-Miculescu-Nicolae-book-2019}, we have $v_*(t)$ and $v_c(t)$ are differentiable for almost every $t\in[0,T_m)$.

Then, taking the time derivative of $v_*$ defined in \eqref{proposition_rho} and using the Lipschitz continuity property \eqref{proposition_rho_Lipschitz}, we obtain
\begin{align}
|\dot{v}_*(t)|
&=\lim_{\tau\to t}\frac{|\rho(\mathcal{M}(y(\tau)),v_c(\tau))-\rho(\mathcal{M}(y(t)),v_c(t))|}{|\tau-t|}\notag\\
&\le \lim_{\tau\to t}\frac{L_\rho(|y(\tau)-y(t)|+|v_c(\tau)-v_c(t)|)}{|\tau-t|}\notag\\
&=L_\rho(|\dot{y}(t)|+|\dot{v}_c(t)|) \label{eq_proof_vstar_01}
\end{align}
for almost every $t\in[0,T_m)$. Substituting the right-hand side of \eqref{eq_def_y_subsystem_new} for $\dot{y}$ in \eqref{eq_proof_vstar_01}, applying the triangle inequality \cite[Theorem 3.3]{Apostol-book-1974}, and then replacing $|v_*(t)|$ by its upper bound in \eqref{proposition_rho_bound}, we have
\begin{align}
&|\dot{v}_*(t)|\le L_\rho(|v_*(t)|+|\tilde{v}(t)|+|\dot{v}_c(t)|) \notag \\
&~\le L_\rho(k_\alpha[D_s-d(y(t))]^+ +|\tilde{v}(t)|+|v_c(t)|+|\dot{v}_c(t)|) \label{eq_bounds}
\end{align}
for almost every $t\in[0,T_m)$. Since $h$ is nonnegative by \eqref{eq_observer_tracking_convolution_h}, substituting the upper bound on $|\dot{v}_*|$ in \eqref{eq_bounds} into the convolution term in \eqref{eq_observer_tracking_convolution_property} and expanding the resulting convolution, we obtain
\begin{align}
&|\tilde{v}(t)|\le c_1e^{-c_2t}\left(|\eta(0)|+|v_*(0)|+|\eta(0)-\hat{\eta}(0)|\right)\notag\\
&~~~+(k_\alpha L_\rho h(\cdot)\ast[D_s-d(y(\cdot))]^+)(t)+(L_\rho h(\cdot)\ast|\tilde{v}(\cdot)|)(t)\notag\\
&~~~+(L_\rho h(\cdot)\ast(|v_c(\cdot)|+|\dot{v}_c(\cdot)|))(t), \label{eq_proof_vtilde1}
\end{align}
for all $t\in[0,T_m)$. Property \eqref{eq_proof_vtilde1} provides a convolution-based upper bound on $|\tilde{v}|$ that captures the influence of both the distance deficit $[D_s-d(y)]^+$ and $|\tilde{v}|$ itself.

To simplify the subsequent analysis, we further estimate the terms involving $v_c$, $\dot{v}_c$, $v_*(0)$, $\eta(0)$, and $\hat{\eta}(0)$ in \eqref{eq_proof_vtilde1}. In particular, to estimate the convolution term involving $v_c$ and $\dot{v}_c$, we integrate $h$ in \eqref{eq_observer_tracking_convolution_h} and obtain
\begin{align}
\int_0^t h(\tau)\mathrm{d}\tau
&\le\int_0^\infty h(\tau)\mathrm{d}\tau
=\sum_{q=0}^{\bar{\sigma}-1}\frac{{\bar{\sigma}}^q}{{\gamma}^q q!}
\int_0^\infty e^{-{\bar{\sigma}\tau}/{\gamma}}\tau^q \mathrm{d}\tau\notag\\
&=\sum_{q=0}^{\bar{\sigma}-1}\frac{\left({\bar{\sigma}}/{\gamma}\right)^q}{q!}q!\left(\frac{\gamma}{\bar{\sigma}}\right)^{q+1}
=\sum_{q=0}^{\bar{\sigma}-1}\frac{\gamma}{\bar{\sigma}}
=\gamma
\label{eq_proof_h_bound}
\end{align}
for all $t\in\mathbb{R}_+$. This, together with the upper bounds of $|v_c|$ and $|\dot{v}_c|$ in \eqref{eq_assumption_command}, gives
\begin{align}
(h(\cdot)\ast(|v_c(\cdot)|+|\dot{v}_c(\cdot)|))(t) \le \gamma(\bar{v}_c+\bar{v}_c^d),~ \forall t\in\mathbb{R}_+. \label{eq_proof_estimate_bound1}
\end{align}
To estimate $|v_*(0)|$, evaluating the safety-critical controller \eqref{proposition_rho} at $t=0$ with its upper bound given by \eqref{proposition_rho_bound}, eliminating the term $[D_s-d(y(0))]^+$ by the initial condition \eqref{eq_initial_condition_y}, and replacing $|v_c(0)|$ by its upper bound $\bar{v}_c$ in \eqref{eq_assumption_command}, we have
\begin{align}
|v_*(0)| &=|\rho(\mathcal{M}(y(0)),v_c(0))|\le k_\alpha[D_s-d(y(0))]^+\!+|v_c(0)|\notag\\
&=|v_c(0)|\le\bar{v}_c.
\label{eq_proof_vstar0_bound}
\end{align}
We use the definitions of $\eta(0)$ and $\hat{\eta}(0)$ in \eqref{eq_coordinate_transformation} and \eqref{eq_observer} and the initial condition \eqref{eq_initial_condition} to estimate the terms involving $\eta(0)$ and $\hat{\eta}(0)$ in \eqref{eq_proof_vtilde1} and obtain $|\eta(0)|+|\eta(0)-\hat{\eta}(0)|\le c_0$. This, together with \eqref{eq_proof_estimate_bound1} and \eqref{eq_proof_vstar0_bound}, substituted into \eqref{eq_proof_vtilde1}, yields 
\begin{align}
|\tilde{v}(t)|&\le c_1e^{-c_2t}(\bar{v}_c+c_0)+(k_\alpha L_\rho h(\cdot)\ast[D_s-d(y(\cdot))]^+)(t)\notag\\
&~~~+(L_\rho h(\cdot)\ast|\tilde{v}(\cdot)|)(t) + \gamma L_\rho(\bar{v}_c+\bar{v}_c^d)
\label{eq_proof_small_gain_coupled_regulation}
\end{align}
for all $t\in[0,T_m)$.
With properties \eqref{eq_proof_small_gain_distance_convolution} and \eqref{eq_proof_small_gain_coupled_regulation}, we establish a convolution-based gain interconnection: the gains from $|\tilde{v}|$ to $[D_s-d(y)]^+$, from $[D_s-d(y)]^+$ to $|\tilde{v}|$, and from $|\tilde{v}|$ to itself are represented as $e^{-k_\alpha(\cdot)}\ast$, $k_\alpha L_\rho h(\cdot)\ast$, and $L_\rho h(\cdot)\ast$, respectively, as shown in Figure \ref{fig_small_gain_interconnection}.

\begin{figure}[htbp]
\centering
\includegraphics[width=1\linewidth]{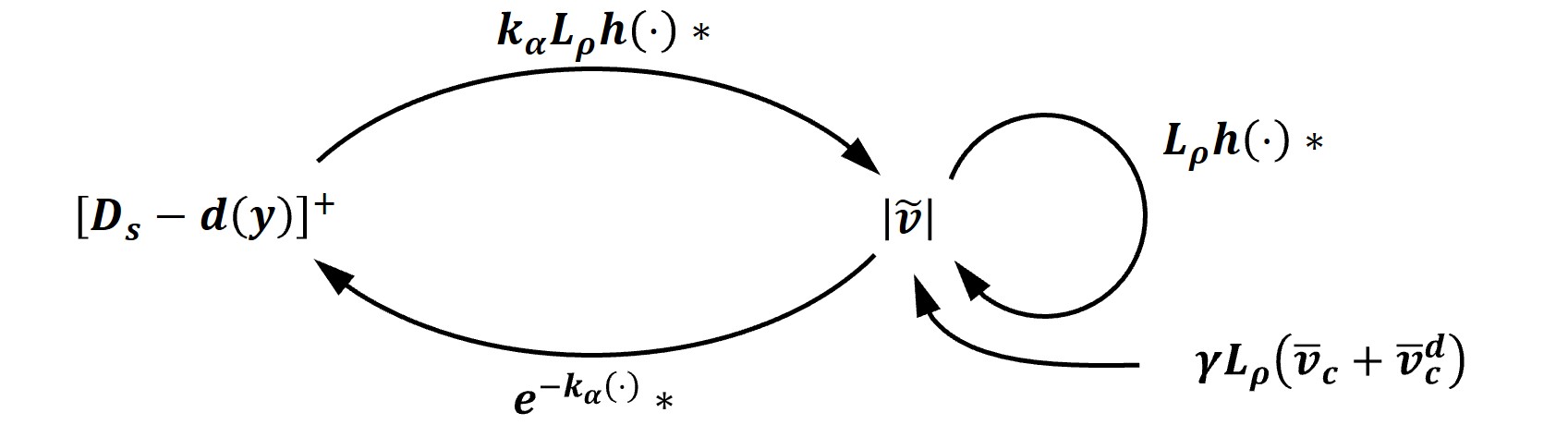}
\caption{The interconnection between $[D_s-d(y)]^+$ and $\tilde{v}$.}
\label{fig_small_gain_interconnection}
\end{figure}

{\it 2.2) Convolution-based small-gain synthesis.} 
We first derive a small-gain condition. Taking sup-norms of \eqref{eq_proof_small_gain_distance_convolution} and \eqref{eq_proof_small_gain_coupled_regulation} yields
\begin{subequations} \label{eq_remark_interconnection1and2}
\begin{align}
&[D_s-d(y(t))]^+\le \frac{1}{k_\alpha}\|\tilde{v}\|_{T_m},
\label{eq_remark_interconnection1}\\
&|\tilde{v}(t)|
\le c_1e^{-c_2t}(\bar{v}_c+c_0)+L_\rho\gamma\|\tilde{v}\|_{T_m}\notag\\
&\quad +k_\alpha L_\rho\gamma\|[D_s-d(y(\cdot))]^+\|_{T_m}
+\gamma L_\rho(\bar{v}_c+\bar{v}_c^d),
\label{eq_remark_interconnection2}
\end{align}
\end{subequations}
which gives the conventional interconnection gains, where the gains from $|\tilde{v}|$ to $[D_s-d(y)]^+$, from $[D_s-d(y)]^+$ to $|\tilde{v}|$, and from $|\tilde{v}|$ to itself are $1/k_{\alpha}$, $k_\alpha L_\rho\gamma$, and $L_\rho\gamma$, respectively. Thus, the boundedness and convergence of $[D_s-d(y)]^+$ and $|\tilde{v}|$ can be ensured by
the small-gain condition: the spectral radius of $[0, 1/k_\alpha; k_\alpha L_\rho\gamma, L_\rho\gamma]$ is less than $1$, that is, $(L_\rho \gamma + \sqrt{L_\rho^2\gamma^2 + 4L_\rho \gamma})/2<1$ \cite{Jiang-Teel-Praly-MCSS-1994}. Equivalently, we require
\begin{align}
0 < \gamma < \frac{1}{2L_{\rho}}, \label{eq_command_bound_condition_refine}
\end{align}
which, as stated at the beginning of the proof of Theorem \ref{theorem_small_gain_safety}, is guaranteed by condition \eqref{eq_command_bound_condition} through choosing $\gamma>0$ sufficiently small to satisfy condition \ref{condition_parameter_safety_margin}, since $k_{\alpha}(D_s-D)/(\bar{v}_c+\bar{v}_c^d+2k_{\alpha}(D_s-D))\le 1/2$.

We apply Lemma \ref{lemma_symmetric_convolution_small_gain} in Appendix \ref{appendix_lemmas} to estimate an upper bound of $|\tilde{v}|$ that is independent of $[D_s-d(y)]^+$. Define
\begin{align}
\Theta(s)&=
\int_0^\infty e^{-st} \begin{bmatrix*}[c]
0 &
e^{-k_\alpha t}\\
k_\alpha L_\rho h(t)&
L_\rho h(t)
\end{bmatrix*}\mathrm{d}t \label{eq_proof_small_gain_matrix_s}
\end{align}
for real $s>-\min\{k_\alpha,\bar{\sigma}/\gamma\}$. Here, $h$ defined in \eqref{eq_observer_tracking_convolution_h} ensures that $\Theta(s)$ is well defined. Clearly, under \eqref{eq_command_bound_condition_refine}, $\Theta(0)$ satisfies the condition \eqref{eq_lemma_smallgain_condition} in Lemma \ref{lemma_symmetric_convolution_small_gain}. 

Recall properties \eqref{eq_proof_small_gain_distance_convolution} and \eqref{eq_proof_small_gain_coupled_regulation} on the trajectories of $[D_s-d(y)]^+$ and $|\tilde{v}|$, respectively. By Lemma \ref{lemma_symmetric_convolution_small_gain}, there exists a constant $\mu$ satisfying $\max\{-k_\alpha/c_A,-c_2\}<\mu<0$ such that
\begin{align}
|\tilde{v}(t)|\le c_1'(\bar{v}_c+c_0)e^{\mu t}+c_3
\label{eq_proof_small_gain_convolution_bound}
\end{align}
for $t\in[0,T_m)$, with
\begin{align}
c_1'=c_1[(I_2-\Theta(\mu))^{-1}]_{2,2}, \quad c_3=\frac{\gamma L_\rho(\bar{v}_c+\bar{v}_c^d)}{1-2\gamma L_\rho}. \label{eq_proof_small_gain_convolution_bound_constant}
\end{align}

{\it 2.3) Establishing $d(y(t))\ge D+\epsilon$ with some $\epsilon>0$ on $[0,T_m)$.}
Substituting property \eqref{eq_proof_small_gain_convolution_bound} into property \eqref{eq_safety_dissipation} for $i=1,\ldots,n_o$, and using the initial condition \eqref{eq_initial_condition_y}, we have
\begin{align}
D_s-d_i(y(t))&\le-c_De^{-k_\alpha t/c_A }\notag\\
&+c_1'(\bar{v}_c+c_0)(e^{-k_\alpha(\cdot)}\ast e^{\mu(\cdot)})(t)
+\frac{c_3}{k_\alpha} \label{eq_proof_di}
\end{align}
for $t\in[0,T_m)$, where
\begin{align}
(e^{-k_\alpha(\cdot)}\ast e^{\mu(\cdot)})(t)=
\begin{cases}
\frac{e^{\mu t}-e^{-k_\alpha t}}{k_\alpha+\mu}, & k_\alpha\ne-\mu,\\
t e^{-k_\alpha t}, & k_\alpha=-\mu.
\end{cases} \label{eq_proof_ekalpha_ast_emu}
\end{align}

Under condition \eqref{eq_command_bound_condition}, choose $\epsilon>0$ such that
\begin{align}
2\epsilon<D_s-D-\frac{c_3}{k_\alpha}. \label{eq_proof_epsilon}
\end{align}
Given $\epsilon$, with \eqref{eq_proof_ekalpha_ast_emu}, choose $c_D>0$ such that
\begin{align}
&-c_De^{-{k_\alpha}t/{c_A}}
+c_1'(\bar{v}_c+c_0)(e^{-k_\alpha(\cdot)}\ast e^{\mu(\cdot)})(t)\le\epsilon
\label{eq_proof_epsilon_cD}
\end{align}
for all $t\ge 0$.

Substituting \eqref{eq_proof_epsilon_cD} and \eqref{eq_proof_epsilon} into \eqref{eq_proof_di}, and eliminating the convolution term and the $c_3/k_\alpha$ term, we obtain
\begin{align}
D_s-d_i(y(t))\le D_s-D-\epsilon, \quad \forall t\in [0,T_m),
\label{eq_proof_small_gain_strict_safety}
\end{align}
and thus,
\begin{align}
d(y(t))\ge D+\epsilon, \quad \forall t\in [0,T_m). \label{eq_proof_dy}
\end{align}

We finally prove forward completeness by contradiction. Suppose $T_m<\infty$. By \cite[Chapter I, Theorem 2.1]{Hale-Book-1980}, the solution $(x(t),z(t))$ approaches the boundary of $\mathcal{D}$ as $t\to T_m$. Together with \eqref{eq_proof_dy}, this implies $|[x(t);z(t)]|\to\infty$ as $t\to T_m$. Since the vector field $f$ is continuous in time $t$ and Lipschitz in $(x,z)$ on set $\mathcal{D}$, Gronwall-Bellman inequality \cite[Lemma A.1]{Khalil-book-2002} implies $|[x(t);z(t)]|$ remains bounded on $[0,T_m)$, a contradiction. Hence, $T_m=\infty$.
\end{proof}

\begin{remark}\label{remark_smallgain_comparison}
The proposed design allows the initial distance to obstacles to be arbitrarily close to the prescribed safety distance while maintaining safety. In particular, for any $D_s\in(D,\min\{D_o/2,D_B\}]$, consider the initial condition \eqref{eq_initial_condition} with $c_D=0$ and $c_0>0$. By \eqref{eq_proof_small_gain_convolution_bound_constant} and \eqref{eq_proof_di}, we have
\begin{align}
D_s-d_i(y(t))
&\le\frac{c_1'(\bar{v}_c+c_0)}{k_\alpha}
+\frac{\gamma L_\rho(\bar{v}_c+\bar{v}_c^d)}
{k_\alpha(1-2\gamma L_\rho)},
\label{eq_proof_di2}
\end{align}
where we used $(e^{-k_\alpha(\cdot)}\ast e^{\mu(\cdot)})(t)\le1/k_\alpha$. If $\gamma$, $\bar{v}_c$, and $c_0$ are sufficiently small, then the right-hand side of \eqref{eq_proof_di2} is smaller than $D_s-D$. Consequently, $d_i(y(t))>D$ for all $t\ge0$. 
\end{remark}

\section{Robustness Under External Disturbances, Measurement Errors, and Moving Obstacles}
\label{section_robustness}

This section evaluates the compatibility of the proposed design with perturbations caused by moving obstacles, external disturbances, and measurement errors. The primary challenge stems from obstacle measurement errors, which introduce set-valued disturbances into the constraints of the QP-based controller and render the velocity-reference input discontinuous. Consequently, the preceding analysis based on the Lipschitz continuity of the velocity-reference input requires nontrivial modifications, both in establishing the gain relationships \eqref{eq_safety_dissipation} and \eqref{eq_observer_tracking_convolution} for the $y$- and $\eta$-subsystems, respectively, and in the subsequent convolution-based small-gain synthesis of closed-loop safety.

The problem formulation is modified as follows.

{\em Moving obstacles:} There are $n_o$ moving obstacles, denoted by $\mathcal{O}_{d,i}(t)$ for $i=1,\ldots,n_o$, with obstacle $i$ modeled by
\begin{align}
\mathcal{O}_{d,i}(t)=\{o+\delta_{o,i}(t):o\in\mathcal{O}_i\}, \label{eq_Odi}
\end{align}
where $\delta_{o,i}:\mathbb{R}_+\to\mathbb{R}^{n_y}$ denotes the time-varying displacement of the obstacle $i$, and $\mathcal{O}_i\subset\mathbb{R}^{n_y}$ is the reference obstacle set. The set of all the moving obstacles is
\begin{align}
\mathcal{O}_d(t)=\bigcup_{i=1}^{n_o}\mathcal{O}_{d,i}(t). \label{eq_Od}
\end{align}
Corresponding to \eqref{eq_di_and_d}, define $d_i(y,t)=\inf_{o\in\mathcal{O}_{d,i}(t)}|y-o|$ and $d(y,t)=\min_{i=1,\ldots,n_o}d_i(y,t)$ for all $y\in\mathbb{R}^{n_y}$ and $t\in\mathbb{R}_+$.

{\em Mobile robot model:} We modify the robot model \eqref{eq_plant_ABC} as
\begin{align}
\dot{x} &= Ax+Bu+\delta_x,\quad y=Cx, \quad \bar{y}=y+\delta_y,  \label{eq_robust_plant}
\end{align}
where $\delta_x:\mathbb{R}_+\to\mathbb{R}^{n_x}$ represents external disturbances, $\delta_y:\mathbb{R}_+\to\mathbb{R}^{n_y}$ represents the output-measurement error, and $\bar{y}$ is the measured output available for feedback.

{\em Control objective:} The control objective is similar to that in Section \ref{section_problemformulation}, except that the stationary obstacle sets $\mathcal{O}_i$ in \eqref{eq_objective} are replaced by the moving obstacle sets $\mathcal{O}_{d,i}(t)$. That is, under suitable initial conditions, the closed-loop trajectories are defined for all $t\ge 0$, and
\begin{align}
|y(t)-o|\ge D,\quad \forall o\in\mathcal{O}_{d,i}(t),~i=1,\ldots,n_o
\label{eq_perturbed_objective}
\end{align}
holds for all $t\ge0$. 

{\em Control system setup:} For the moving obstacle set $\mathcal{O}_d(t)$, we consider disturbed obstacle measurement:
\begin{subequations}\label{eq_robust_range_measurement}
\begin{align}
\mathcal{M}_m(y,t)&=\{[l;r_m(y,l,t)]:|l|=1\},\, \forall t\in\mathbb{R}_+, \label{eq_disturbed_setvalued_feedback}
\end{align}
where
\begin{align}
r_m(y,l,t)&=\min\{[r_a(y,l,t)+\delta_m(l,t)]^+,D_B\}, \label{eq_robust_measured_range} \\
r_a(y,l,t)&=\min(\{s\ge0:y+sl\in\mathcal{O}_d(t)\}\cup\{D_B\}) \label{eq_robust_range}
\end{align}
\end{subequations}
for $y,l\in\mathbb{R}^{n_y}$, $|l|=1$, and $t\ge0$, with $\delta_m:\mathbb{R}^{n_y}\times\mathbb{R}_+\to\mathbb{R}$ representing the obstacle-measurement error. 

Let
\begin{align}
\mathcal{M}_a(y,t)&=\{[l;r_a(y,l,t)]:|l|=1\},\, \forall t\in\mathbb{R}_+. \label{eq_setvalued_feedback_moving}
\end{align}
If $\delta_{o,i}(t)\equiv 0$ and $\delta_m(l,t)\equiv 0$, then $\mathcal{M}_m(y,t)$, $r_m(y,l,t)$ and $r_a(y,l,t)$ defined above reduce to $\mathcal{M}(y)$, $r(y,l)$ and $r(y,l)$ defined in Section \ref{section_problemformulation}, respectively.

Assumption \ref{assumption_obstacle_new} extends Assumption \ref{assumption_obstacle} to moving obstacles.

\begin{assumptionprime} \label{assumption_obstacle_new}
The obstacle sets $\mathcal{O}_1,\ldots,\mathcal{O}_{n_o}$ are nonempty, closed, and convex; the displacements $\delta_{o,i}:\mathbb{R}_+\to\mathbb{R}^{n_y}$ for $i=1,\ldots,n_o$ are Lipschitz continuous; there exist an obstacle-separation distance $D_o>2D$ and a constant $\bar{\delta}_o^d\ge 0$ such that
\begin{align}
\|\dot{\delta}_{o,i}\|_\infty &\le \bar{\delta}_o^d, \label{assumption_obstacle_velocity}\\
|o_i-o_j| &\ge D_o,~
\forall o_i\in\mathcal{O}_{d,i}(t),
o_j\in\mathcal{O}_{d,j}(t), t\ge 0, \label{assumption_obstacle_separation_new}
\end{align}
for all $i,j=1,\ldots,n_o$ with $i\neq j$.
\end{assumptionprime}

Assumptions \ref{assumption_plant} and \ref{assumption_command} are still made on the matrices $A,B,C$ of model \eqref{eq_robust_plant} and the velocity-command signal $v_c$, respectively. Assumption \ref{assumption_disturbance} is made on the boundedness of the disturbances and measurement errors. 

\begin{assumption} \label{assumption_disturbance}
The functions $\delta_x:\mathbb{R}_+\to\mathbb{R}^{n_x}$ and $\delta_y:\mathbb{R}_+\to\mathbb{R}^{n_y}$ are measurable, and $\delta_m:\mathbb{R}^{n_y}\times\mathbb{R}_+\to\mathbb{R}$ is Borel measurable. The function $\delta_m(\cdot,t)$ is lower semicontinuous on $\{l\in\mathbb{R}^{n_y}:|l|=1\}$ for every $t\in\mathbb{R}_+$. Moreover, there exist positive constants $\bar{\delta}_x$, $\bar{\delta}_y$, and $\bar{\delta}_m$ such that
\begin{align}
|\delta_x(t)| \le \bar{\delta}_x,\quad
|\delta_y(t)| \le \bar{\delta}_y,\quad
|\delta_m(l,t)|\le\bar{\delta}_m \label{eq_assumption_delta}
\end{align}
for all $l\in\mathbb{R}^{n_y}$ satisfying $|l|=1$ and $t\in\mathbb{R}_+$.
\end{assumption}

\begin{figure}[htbp]
\centering
\includegraphics[width=1\linewidth]{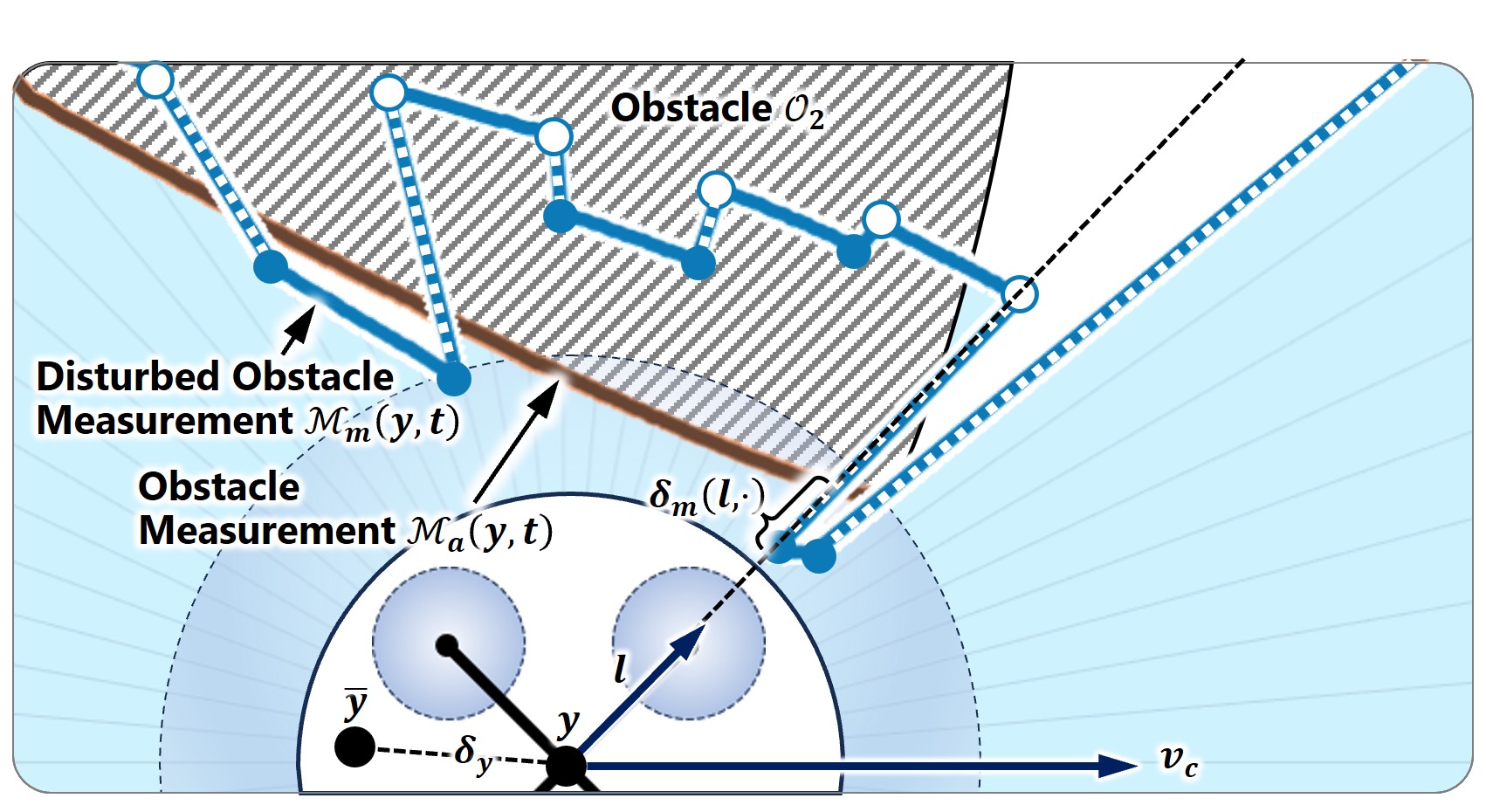}
\caption{An illustration of measurement errors.}
\label{figure_robust_illustration}
\end{figure}

Figure \ref{figure_robust_illustration} illustrates the measurement errors.

For the disturbed robot model \eqref{eq_robust_plant}, we still consider the controller composed of \eqref{eq_safety_critical_controller} and \eqref{eq_regulation_controller}, but with $\mathcal{F}$ replaced by 
\begin{align}
\mathcal{F}=\mathcal{M}_m(y,t), \label{eq_noisy_setvalued_feedback}
\end{align}
and $y$ replaced by $\bar{y}$. Parameter conditions \ref{condition_parameter_safety_controller}--\ref{condition_parameter_safety_margin} listed in Section \ref{subsection_smallgain} remain required, with $L_\rho>0$ being a Lipschitz constant for the function $(y,v_c)\mapsto\rho(\mathcal{M}_a(y,t),v_c)$ on $\{(y,v_c):d(y,t)\ge D,\ |v_c|\le\bar{v}_c\}$ for all fixed $t\ge0$.

Proposition \ref{proposition_noisy_controller_bound} enables us to express the perturbed closed-loop dynamics as the sum of a nominal vector field and a bounded set-valued disturbance. Based on this representation, Theorem \ref{theorem_robust_safety} extends Theorem \ref{theorem_small_gain_safety} to the robust safety setting.

\begin{proposition}\label{proposition_noisy_controller_bound}
Under Assumptions \ref{assumption_obstacle_new}, \ref{assumption_command} and \ref{assumption_disturbance}, for any $\bar{v}_c>0$, there exists a continuous nondecreasing function $\kappa_m:\mathbb{R}_+\to\mathbb{R}_+$ satisfying $\kappa_m(0)=0$ such that
\begin{align}
&|\rho(\mathcal{M}_m(y,t),v_c)-\rho(\mathcal{M}_a(y,t),v_c)|
\le\kappa_m(\bar\delta_m) \label{eq_noisy_controller_bound}
\end{align}
holds for any $(y,t)$ satisfying $d(y,t)>D$.
\end{proposition}

The proof of Proposition \ref{proposition_noisy_controller_bound} is given in Appendix \ref{appendix_proposition_noisy_controller_bound}.

\begin{theorem}\label{theorem_robust_safety}
Consider the moving obstacles \eqref{eq_Odi} satisfying Assumption \ref{assumption_obstacle_new}, the mobile robot \eqref{eq_robust_plant} satisfying Assumption \ref{assumption_plant} and conditions \eqref{eq_rightinvertibility} and \eqref{eq_fullrowrank}, any velocity-command signal $v_c$ satisfying Assumption \ref{assumption_command} for arbitrary constants $\bar{v}_c,\bar{v}_c^d\ge 0$, and the controller composed of \eqref{eq_safety_critical_controller} and \eqref{eq_regulation_controller}, with $\mathcal{M}$ and $y$ replaced by $\mathcal{M}_m$ and $\bar{y}$, respectively. Then, there exist controller parameters $D_s$, $k_L$, $A_L$, $k_\alpha$, $L$, $K$, and $T$ satisfying the conditions specified before Proposition \ref{proposition_noisy_controller_bound} with a suitably chosen $\gamma>0$. Moreover, there exist sufficiently large $c_D>0$ and sufficiently small $c_0$, $\bar{\delta}_x$, $\bar{\delta}_y$, $\bar{\delta}_m$, $\bar{\delta}_o^d>0$ such that, for any $\delta_{o,1},\ldots,\delta_{o,n_o}$ satisfying \eqref{assumption_obstacle_velocity} and any $\delta_x$, $\delta_y$, and $\delta_m$ satisfying Assumption \ref{assumption_disturbance}, under initial conditions
\begin{subequations}\label{eq_initial_condition_robust}
\begin{align}
&|y(0)-o|\ge D_s+c_D,\qquad \forall o\in\mathcal{O}_{d}(0), \label{eq_initial_conditionP_y}\\
&|N^Tx(0)|+|N^Tx(0)-z(0)-LCx(0)|\le c_0, \label{eq_initial_conditionP_x}
\end{align}
\end{subequations}
the safety property \eqref{eq_perturbed_objective} holds for all $t\ge 0$.
\end{theorem}

\begin{proof}
% Due to the space limitation, we only present a brief outline of the proof. See \cite{Wu-Liu-Ding-Li-Chai-Jiang-arXiv-2026} for the complete proof. 
Let $f_e(t,x,z)$ denote the vector field of the perturbed closed-loop system composed of the robot model \eqref{eq_robust_plant} and the controllers \eqref{eq_safety_critical_controller} and \eqref{eq_regulation_controller}, with $\mathcal{M}$ and $y$ replaced by $\mathcal{M}_m$ and $\bar{y}$, respectively. Let $f_n(t,x,z)$ denote the vector field of the corresponding nominal closed-loop system obtained by setting $\delta_x(t)\equiv0$, $\delta_y(t)\equiv0$, and $\delta_m(l,t)\equiv0$, while retaining the displacement of the obstacles.

It can be verified that $f_e-f_n$ is a sum of terms linear in $\delta_x$, $\delta_y$ and $\rho(\mathcal{M}_m,v_c)-\rho(\mathcal{M}_a,v_c)$. Therefore, under Assumptions \ref{assumption_obstacle_new}, \ref{assumption_command}, and \ref{assumption_disturbance}, the disturbance bounds \eqref{eq_assumption_delta} and property \eqref{eq_noisy_controller_bound} ensure the existence of $\bar{\delta}>0$ such that
\begin{align}
|f_e(t,x,z)-f_n(t,x,z)|\le\bar{\delta} \label{eq_proof_fefn_delta}
\end{align}
for all $(t,x,z)\in\mathcal{D}_r$, where 
\begin{align}
\mathcal{D}_r=\{(t,x,z):t\ge0,\ d(Cx,t)>D\}. \label{eq_proof_Dr}
\end{align}
Under the stated assumptions, one can verify that $f_e(t,x,z)$ is measurable and locally bounded on $\mathcal{D}_r$. Therefore, for any $(x(0),z(0))$ satisfying the initial condition \eqref{eq_initial_conditionP_y}, there exists a local absolutely continuous solution $(x(t),z(t))$ of the perturbed closed-loop system in the sense of Filippov \cite[Theorem 8 in Section 2.7]{Filippov-Book-1988}, and every such solution satisfies
\begin{align}
[\dot{x};\dot{z}]=f_n(t,x,z)+w,\qquad |w(t)|\le\bar{\delta}
\label{eq_proof_robust_differential_inclusion}
\end{align}
for almost every $t$ on any time interval over which $d(Cx,t)>D$, for some measurable function $w=[w_x;w_z]$.

Following the proof of Theorem \ref{theorem_small_gain_safety}, we characterize the interaction between $[D_s-d(y,t)]^+$ and $|C_\eta\eta+D_\eta K(z+Ly)+(D_\eta T-I)\rho(\mathcal{M}_a(y,t),v_c)|$ with convolution-based gains. With $\bar{\delta}_x$, $\bar{\delta}_y$, $\bar{\delta}_m$, and $\bar{\delta}_o^d$ describing the perturbations sufficiently small, a similar small-gain synthesis, with a sufficiently large initial safety distance, therefore yields $d(y(t),t)\ge D+\epsilon$ for some $\epsilon>0$ on the right maximal interval of each solution. The Gronwall-Bellman inequality \cite[Lemma A.1]{Khalil-book-2002} and the continuation property \cite[Theorem 9 in Section 2.7]{Filippov-Book-1988} then establish forward completeness and property \eqref{eq_perturbed_objective}.
\ifarxivversion

Choose $D_s$, $k_L$, $A_L$, and $k_\alpha$ according to condition \ref{condition_parameter_safety_controller}, then choose $\gamma>0$ to satisfy condition \ref{condition_parameter_safety_margin} with the time-uniform constant $L_\rho$ specified before Proposition \ref{proposition_noisy_controller_bound}, and finally choose $L$, $K$, and $T$ according to condition \ref{condition_parameter_regulation}.

{\it 1) Existence of a solution over a possibly infinite time interval.} 

{\it 1.1) Lipschitz continuity of $f_n(t,x,z)$ and local boundedness of $f_e(t,x,z)$.} We first establish the Lipschitz continuity of $f_n(t,x,z)$ on $\mathcal{D}_r$. By the linearity of the robot model \eqref{eq_plant_ABC} and velocity-regulation controller \eqref{eq_regulation_controller}, it suffices to show that $(t,x,z)\mapsto\rho(\mathcal{M}_a(Cx,t),v_c(t))$ is Lipschitz on $\mathcal{D}_r$. The proof of Proposition \ref{proposition_safety} establishes the Lipschitz dependence of the QP solution defining $\rho$ on $\rho_s$, $b_L$, and $v_c$. Since $v_c$ is Lipschitz by Assumption \ref{assumption_command}, it remains to establish the Lipschitz continuity of $\rho_s$ and $b_L$. The Lipschitz continuity of $\rho_s(\mathcal{M}_a(Cx,t))$ follows directly from the argument for $\rho_s(\mathcal{M}(y))$ in the proof of Proposition \ref{proposition_safety} by replacing $(d_i(y),\nabla d_i(y))$ by $(d_i(y,t),\nabla_y d_i(y,t))$, which are Lipschitz by \eqref{eq_Odi} and Assumption \ref{assumption_obstacle_new}. Consequently, recalling \eqref{eq_def_bL_main}, the Lipschitz continuity of $b_L$ follows directly if $r_L$ is Lipschitz. Rewrite the definition \eqref{eq_def_rL} as
\begin{align}
r_L(\mathcal{M}_a(y,t),l)&=\min\{D_B,\min_{i=1,\ldots,n_o}\inf_{o\in\mathcal{O}_i}
(|o+\delta_{o,i}(t)-y|\notag\\
&\quad+k_L[c_A-l^T\frac{o+\delta_{o,i}(t)-y}{|o+\delta_{o,i}(t)-y|}]^+)\}, 
\label{eq_proof_robust_rL_representation}
\end{align}
whose Lipschitz continuity on $\mathcal{D}_r$ follows from Assumption \ref{assumption_obstacle_new} and the Lipschitz properties of algebraic operations, infima, and minima given in \cite[Section 2.3]{Cobzas-Miculescu-Nicolae-book-2019}.
These properties establish the Lipschitz continuity of the reference map and hence of $f_n$ on $\mathcal{D}_r$.

We then prove the local boundedness of $f_e(t,x,z)$. For every nonempty compact set $\mathcal{K}\subset\mathcal{D}_r$, we have
\begin{align}
|f_e(t,x,z)|
&\le |f_n(t,x,z)|+|f_e(t,x,z)-f_n(t,x,z)|\notag\\
&\le\max_{(s,\xi,\zeta)\in\mathcal{K}}|f_n(s,\xi,\zeta)|+\bar{\delta}<\infty,
\label{eq_proof_robust_local_bound}
\end{align}
for all $(t,x,z)\in\mathcal{K}$, where we used \eqref{eq_proof_fefn_delta} for the second inequality, and used the Lipschitz continuity of $f_n$ and the compactness of $\mathcal{K}$ for the last inequality. Thus, $f_e(t,x,z)$ is locally bounded on $\mathcal{D}_r$.

{\it 1.2) Measurability of $f_e(t,x,z)$.}
By the robot model \eqref{eq_robust_plant} and velocity-regulation controller \eqref{eq_regulation_controller}, together with the measurability of $\delta_x$ and $\delta_y$ in Assumption \ref{assumption_disturbance}, it suffices to show that $(t,x,z)\mapsto\rho(\mathcal{M}_m(Cx,t),v_c(t))$ is measurable. The proof of Proposition \ref{proposition_safety} establishes the continuous dependence of the QP solution defining $\rho$ on $\rho_s$, $b_L$, and $v_c$. Since $v_c$ is continuous, it remains to establish the measurability of $\rho_s(\mathcal{M}_m(Cx,t))$ and $b_L(\mathcal{M}_m(Cx,t))$.

By the definition of $\rho_s$ in \eqref{eq_feasibility_point}, it suffices to prove the measurability of $\ubar{r}(\mathcal{M}_m(Cx,t))$ and $\ubar{l}(\mathcal{M}_m(Cx,t))$. For the former, write
\begin{align}
\ubar{r}(\mathcal{M}_m(Cx,t))=\min_{l\in\mathcal{S}}r_m(Cx,l,t),~
\mathcal{S}=\{l:|l|=1\}.
\end{align}
The closedness of the obstacle sets and the continuity of their displacements imply that $r_a(y,l,t)$ is lower semicontinuous in $(y,l,t)$. Together with Assumption \ref{assumption_disturbance} and \eqref{eq_robust_measured_range}, this implies that $r_m(Cx,l,t)$ is Borel measurable in $(t,x,z,l)$ and lower semicontinuous in $l$. Since $\mathcal{S}$ is compact, the minimum is attained, and \cite[Corollary 14.34 and Theorem 14.37]{Rockafellar-Wets-1998-book} imply that the function $\ubar{r}(\mathcal{M}_m(Cx,t))$ and the set-valued map
\begin{align}
\mathcal{Q}(t,x,z)=\argmin_{l\in\mathcal{S}}r_m(Cx,l,t)
\label{eq_proof_robust_minimizing_directions}
\end{align}
are Lebesgue measurable in $(t,x,z)$. To prove the measurability of $\ubar{l}(\mathcal{M}_m(Cx,t))$, recall that
\begin{align}
\ubar{l}(\mathcal{M}_m(Cx,t))=\argmin_{q\in\co\mathcal{Q}(t,x,z)}|q|^2.
\label{eq_proof_robust_measurable_direction}
\end{align}
The lower semicontinuity of $r_m(Cx,l,t)$ in $l$ and the compactness of $\mathcal{S}$ ensure that $\mathcal{Q}(t,x,z)$ is nonempty and compact. Its convex hull is therefore nonempty, compact, and convex. Moreover, the map $\co\mathcal{Q}$ is measurable by \cite[Exercise 14.12(a)]{Rockafellar-Wets-1998-book}. The projection theorem \cite[Proposition B.11]{Bertsekas-Book-1997} ensures that the minimization problem in \eqref{eq_proof_robust_measurable_direction} has a unique solution, while \cite[Example 14.32 and Theorem 14.37]{Rockafellar-Wets-1998-book} establish its measurability in $(t,x,z)$. This, together with the measurability of $\ubar{r}(\mathcal{M}_m(Cx,t))$ proved above, establishes the measurability of $\rho_s(\mathcal{M}_m(Cx,t))$.

We next prove the measurability of $b_L(\mathcal{M}_m(Cx,t))$. For each $j=1,\ldots,n_L$, write
\begin{align}
&r_L(\mathcal{M}_m(Cx,t),[A_L]_{j,:}^T)\notag\\
&\quad=\min_{l'\in\mathcal{S}}\{r_m(Cx,l',t)+k_L[c_A-[A_L]_{j,:}l']^+\}.
\label{eq_proof_robust_rL_measurability}
\end{align}
The objective function in \eqref{eq_proof_robust_rL_measurability} is Borel measurable in $(t,x,z,l')$ and lower semicontinuous in $l'$, since $r_m$ has these properties, as proved above, and the term $k_L[c_A-[A_L]_{j,:}l']^+$ is continuous in $l'$. The compactness of $\mathcal{S}$ ensures that the minimum is attained. Thus, \cite[Corollary 14.34 and Theorem 14.37]{Rockafellar-Wets-1998-book} establish the measurability of $r_L(\mathcal{M}_m(Cx,t),[A_L]_{j,:}^T)$ in $(t,x,z)$ for every $j=1,\ldots,n_L$. Together with the measurability of $\rho_s(\mathcal{M}_m(Cx,t))$ proved above, this establishes the measurability of $b_L(\mathcal{M}_m(Cx,t))$ by \eqref{eq_def_bL_main}.

{\it 1.3) Existence of $(x(t),z(t))$ over a possibly infinite time interval.} By \cite[Theorems 8 and 9 in Section 2.7]{Filippov-Book-1988}, the local boundedness and measurability of $f_e$ established above ensure that the perturbed closed-loop system composed of \eqref{eq_robust_plant}, \eqref{eq_safety_critical_controller}, and \eqref{eq_regulation_controller} admits a maximal absolutely continuous Filippov solution for any $x(0)$ and $z(0)$ satisfying \eqref{eq_initial_conditionP_y}, which places the initial point $(0,x(0),z(0))$ in $\mathcal{D}_r$ in \eqref{eq_proof_Dr}. Fix an arbitrary such solution, and let $[0,T_m)$, with $0<T_m\le\infty$, denote the right maximal interval on which $(x(t),z(t))$ is defined and $(t,x(t),z(t))$ remains in $\mathcal{D}_r$.

{\it 2) Safety over the right maximal interval $[0,T_m)$.}

We employ a convolution-based small-gain synthesis to establish $d(y(t),t)\ge D+\epsilon$ with some $\epsilon>0$ on $[0,T_m)$. As in the proof of Theorem \ref{theorem_small_gain_safety}, the key is to characterize the interaction between the distance deficit $[D_s-d(y(t),t)]^+$ and the magnitude $|\tilde{v}_r(t)|$ of the auxiliary velocity error defined below. For this purpose, define
% \begin{align}
% v_r&=\rho(\mathcal{M}_a(y,t),v_c),\notag\\
% \tilde{v}_r&=v-v_r,\qquad \tilde{\eta}_r=\eta-z-Ly.
% \label{eq_proof_robust_auxiliary_signals}
% \end{align}
\begin{subequations}\label{eq_proof_robust_auxiliary_signals}
\begin{align}
v_r&=\rho(\mathcal{M}_a(y,t),v_c),\qquad
\tilde{\eta}_r=\eta-z-Ly,\\
\tilde{v}_r&=C_\eta\eta+D_\eta(K(z+Ly)+Tv_r)-v_r.
\end{align}    
\end{subequations}
Here, $\tilde{v}_r$ is referred to as the auxiliary velocity error, while $\eta=N^Tx$ retains its definition in \eqref{eq_coordinate_transformation}.

Using \eqref{eq_coordinate_transformation}, \eqref{eq_ABC_eta}, and the expression \eqref{eq_f_new} for the nominal vector field in \eqref{eq_proof_robust_differential_inclusion}, we obtain
\begin{align}
\dot{y}=v_r+\tilde{v}_r+Cw_x,
\label{eq_proof_disturbed_y_subsystem}
\end{align}
and
\begin{subequations}\label{eq_proof_robust_internal_dynamics}
\begin{align}
\dot{\eta}&=(A_\eta+B_\eta K)\eta-B_\eta K\tilde{\eta}_r
+B_\eta Tv_r+N^Tw_x,
\label{eq_proof_robust_eta_dynamics}\\
\dot{\tilde{\eta}}_r&=(A_\eta-LC_\eta)\tilde{\eta}_r
+(N^T-LC)w_x-w_z.
\label{eq_proof_robust_observation_error}
\end{align}
\end{subequations}
These relations hold for almost every $t\in[0,T_m)$.

{\it 2.1) Influence of $|\tilde{v}_r|$ on $[D_s-d(y,t)]^+$.}
We first evaluate the influence of the auxiliary velocity error $\tilde{v}_r$ on the distance deficit $[D_s-d(y,t)]^+$. By the translation model \eqref{eq_Odi}, $d_i(y,t)=\inf_{o\in\mathcal{O}_i}|y-\delta_{o,i}(t)-o|$. 
Since $y(t)$ is absolutely continuous and $\delta_{o,i}(t)$ is Lipschitz, we have
\begin{align}
\frac{\mathrm{d}}{\mathrm{d}t}d_i(y(t),t)
=\nabla_y d_i(y(t),t)^T
(\dot{y}(t)-\dot{\delta}_{o,i}(t))
\label{eq_proof_robust_distance_derivative}
\end{align}
for almost every $t\in[0,T_m)$. Following the proof of property (c) in Proposition \ref{proposition_safety}, and using \eqref{eq_proof_robust_distance_derivative}, \eqref{eq_proof_disturbed_y_subsystem}, \eqref{eq_proof_robust_differential_inclusion}, and \eqref{assumption_obstacle_velocity}, we obtain
\begin{align}
\frac{\mathrm{d}}{\mathrm{d}t}(D_s-d_i(y(t),t))
&\le-\alpha_p(D_s-d_i(y(t),t))\notag\\
&\quad+|\tilde{v}_r(t)|+|C|\bar{\delta}+\bar{\delta}_o^d
\label{eq_proof_robust_distance_differential}
\end{align}
for almost every $t\in[0,T_m)$, where $\alpha_p$ is defined in \eqref{eq_proof_pi}. Integrating \eqref{eq_proof_robust_distance_differential} from $0$ to $t$ as in the proof of \eqref{eq_safety_dissipation}, and using the initial condition \eqref{eq_initial_conditionP_y}, we obtain
\begin{align}
D_s&-d_i(y(t),t)
\le e^{-k_\alpha t/c_A}(D_s-d_i(y(0),0))\notag\\
&\quad+(e^{-k_\alpha(\cdot)}
\ast(|\tilde{v}_r(\cdot)|+|C|\bar{\delta}
+\bar{\delta}_o^d))(t)\notag\\
&\le e^{-k_\alpha t/c_A}(D_s-d_i(y(0),0))\notag\\
&\quad+(e^{-k_\alpha(\cdot)}\ast|\tilde{v}_r(\cdot)|)(t)
+\frac{|C|\bar{\delta}+\bar{\delta}_o^d}{k_\alpha}
\label{eq_proof_robust_distance_bound}
\end{align}
for all $t\in[0,T_m)$ and $i=1,\ldots,n_o$. Since the initial-distance term is nonpositive, taking positive parts and the maximum over $i$ yields
\begin{align}
[D_s&-d(y(t),t)]^+
=\max_{i=1,\ldots,n_o}[D_s-d_i(y(t),t)]^+\notag\\
&\le(e^{-k_\alpha(\cdot)}\ast|\tilde{v}_r(\cdot)|)(t)
+\frac{|C|\bar{\delta}+\bar{\delta}_o^d}{k_\alpha}
\label{eq_proof_robust_coupled_distance}
\end{align}
for all $t\in[0,T_m)$.

{\it 2.2) Influences of $[D_s-d(y,t)]^+$ on $|\tilde{v}_r|$ and of $|\tilde{v}_r|$ on itself.} 

{\it 2.2.1) An upper bound on $|\dot{v}_r|$.}
We first check the differentiability of $v_r$. In particular, since $(x(t),z(t))$ is absolutely continuous, the local boundedness of $f_n$ and \eqref{eq_proof_robust_differential_inclusion} imply that $(x(t),z(t))$ is locally Lipschitz on $[0,T_m)$; see \cite[Example 8.1.8]{Cobzas-Miculescu-Nicolae-book-2019}. Then, under Assumption \ref{assumption_command}, $y(t)=Cx(t)$ and $v_c(t)$ are locally Lipschitz on $[0,T_m)$. From the expression of $v_r$ in \eqref{eq_proof_robust_auxiliary_signals} and the Lipschitz continuity established in Step 1.1, by \cite[Proposition 2.3.1]{Cobzas-Miculescu-Nicolae-book-2019}, $v_r(t)$ is locally Lipschitz on $[0,T_m)$. By \cite[Remark 3.4.2]{Cobzas-Miculescu-Nicolae-book-2019}, $v_r(t)$ and $v_c(t)$ are differentiable for almost every $t\in[0,T_m)$.

Using the Lipschitz estimates established in Step 1.1 and \eqref{assumption_obstacle_velocity}, with $L_\rho$ specified before Proposition \ref{proposition_noisy_controller_bound}, there exists a constant $L_o>0$, independent of the initial states and $\bar{\delta}_o^d$, such that
\begin{align}
&|\rho(\mathcal{M}_a(y_1,t_1),v_{c1})
-\rho(\mathcal{M}_a(y_2,t_2),v_{c2})|\notag\\
&\quad\le L_\rho(|y_1-y_2|+|v_{c1}-v_{c2}|)
+L_o\bar{\delta}_o^d|t_1-t_2|
\label{eq_proof_robust_rho_Lipschitz}
\end{align}
for all $(t_j,x_j,z_j)\in\mathcal{D}_r$, with $y_j=Cx_j$ and $|v_{cj}|\le\bar{v}_c$, $j=1,2$. Then, taking the time derivative of $v_r$ defined in \eqref{eq_proof_robust_auxiliary_signals} and using \eqref{eq_proof_robust_rho_Lipschitz} as in \eqref{eq_proof_vstar_01}, we obtain
\begin{align}
|\dot{v}_r(t)|
\le L_\rho(|\dot{y}(t)|+|\dot{v}_c(t)|)
+L_o\bar{\delta}_o^d
\label{eq_proof_robust_reference_derivative}
\end{align}
for almost every $t\in[0,T_m)$. Substituting \eqref{eq_proof_disturbed_y_subsystem} into \eqref{eq_proof_robust_reference_derivative}, and then using \eqref{proposition_rho_bound}, \eqref{eq_assumption_command}, and \eqref{eq_proof_robust_differential_inclusion}, yields
\begin{align}
|\dot{v}_r(t)|
&\le L_\rho(|v_r(t)|+|\tilde{v}_r(t)| +|C|\bar{\delta}+|\dot{v}_c(t)|)+L_o\bar{\delta}_o^d\notag\\
&\le L_\rho(k_\alpha[D_s-d(y(t),t)]^++|\tilde{v}_r(t)|\notag\\
&\quad+\bar{v}_c+\bar{v}_c^d+|C|\bar{\delta})
+L_o\bar{\delta}_o^d
\label{eq_proof_robust_dvref}
\end{align}
for almost every $t\in[0,T_m)$.

{\it 2.2.2) Convolution-based gains from $[D_s-d(y,t)]^+$ to $|\tilde{v}_r|$ and from $|\tilde{v}_r|$ to itself.} Following the proof of Proposition \ref{proposition_observer}, we evaluate the influence of $|\dot{v}_r|$ and $|w|$ on $|\tilde{v}_r|$. When $w=0$, \eqref{eq_observer_tracking_convolution_property} applies with $v_*$ and $\tilde{v}$ replaced by $v_r$ and $\tilde{v}_r$, respectively. In \eqref{eq_proof_robust_internal_dynamics}, $w$ enters linearly, and the state matrix is block triangular with Hurwitz diagonal blocks $A_\eta+B_\eta K$ and $A_\eta-LC_\eta$ by \eqref{eq_KT_selection_condition_Hurwitz} and \eqref{eq_L_condition}. Thus, by the solution formula \cite[Theorem 5.2]{Hespanha-Book-2018}, there exist constants $a_r,b_r>0$ such that
\begin{align}
|\tilde{v}_r(t)|
&\le c_1e^{-c_2t}
(|\eta(0)|+|v_r(0)|+|\tilde{\eta}_r(0)|)\notag\\
&\quad+(h(\cdot)\ast|\dot{v}_r(\cdot)|)(t)
+a_r(e^{-b_r(\cdot)}\ast|w(\cdot)|)(t)\notag\\
&\le c_1e^{-c_2t}
(|\eta(0)|+|v_r(0)|+|\tilde{\eta}_r(0)|)\notag\\
&\quad+(h(\cdot)\ast|\dot{v}_r(\cdot)|)(t)
+\frac{a_r}{b_r}\bar{\delta},
\label{eq_proof_robust_tracking_bound}
\end{align}
for all $t\in[0,T_m)$, where we used \eqref{eq_proof_robust_differential_inclusion} for the last inequality.

We estimate the initial terms in \eqref{eq_proof_robust_tracking_bound}. The initial conditions \eqref{eq_initial_conditionP_y} and \eqref{eq_initial_conditionP_x}, together with \eqref{proposition_rho_bound}, \eqref{eq_assumption_command}, and the definitions in \eqref{eq_proof_robust_auxiliary_signals}, give
\begin{subequations}\label{eq_proof_robust_initial_bounds}
\begin{align}
&|v_r(0)|
\le k_\alpha[D_s-d(y(0),0)]^++|v_c(0)|
\le\bar{v}_c,
\label{eq_proof_robust_initial_reference_bound}\\
&|\eta(0)|+|\tilde{\eta}_r(0)|
=|N^Tx(0)|+|N^Tx(0)-z(0)-LCx(0)|\notag\\
&\quad\le c_0.
\label{eq_proof_robust_initial_state_bound}
\end{align}
\end{subequations}
Since $h$ is nonnegative by \eqref{eq_observer_tracking_convolution_h}, substituting \eqref{eq_proof_robust_dvref} and \eqref{eq_proof_robust_initial_bounds} into \eqref{eq_proof_robust_tracking_bound}, and expanding the convolution, gives
\begin{align}
|\tilde{v}_r(t)|
&\le c_1e^{-c_2t}(\bar{v}_c+c_0)
+(L_\rho h(\cdot)\ast|\tilde{v}_r(\cdot)|)(t)\notag\\
&\quad+(k_\alpha L_\rho h(\cdot)
\ast[D_s-d(y(\cdot),\cdot)]^+)(t)\notag\\
&\quad+(L_\rho(\bar{v}_c+\bar{v}_c^d+|C|\bar{\delta})
+L_o\bar{\delta}_o^d)
\int_0^t h(\tau)\mathrm{d}\tau\notag\\
&\quad+\frac{a_r}{b_r}\bar{\delta}.
\label{eq_proof_robust_tracking_convolution}
\end{align}
By \eqref{eq_proof_h_bound}, the integral is no greater than $\gamma$. Hence,
\begin{align}
|\tilde{v}_r(t)|
&\le c_1e^{-c_2t}(\bar{v}_c+c_0)
+(L_\rho h(\cdot)\ast|\tilde{v}_r(\cdot)|)(t)\notag\\
&\quad+(k_\alpha L_\rho h(\cdot)
\ast[D_s-d(y(\cdot),\cdot)]^+)(t)\notag\\
&\quad+\gamma L_\rho(\bar{v}_c+\bar{v}_c^d)\notag\\
&\quad+\left(\frac{a_r}{b_r}+\gamma L_\rho|C|\right)\bar{\delta}
+\gamma L_o\bar{\delta}_o^d
\label{eq_proof_robust_coupled_regulation}
\end{align}
for all $t\in[0,T_m)$.

Properties \eqref{eq_proof_robust_coupled_distance} and \eqref{eq_proof_robust_coupled_regulation} characterize the convolution-based gain interconnection between $[D_s-d(y(t),t)]^+$ and $|\tilde{v}_r(t)|$. The gains from $|\tilde{v}_r|$ to $[D_s-d(y,t)]^+$, from $[D_s-d(y,t)]^+$ to $|\tilde{v}_r|$, and from $|\tilde{v}_r|$ to itself are respectively
$e^{-k_\alpha(\cdot)}\ast$,
$k_\alpha L_\rho h(\cdot)\ast$, and
$L_\rho h(\cdot)\ast$.
These are the same gains as in \eqref{eq_proof_small_gain_distance_convolution} and \eqref{eq_proof_small_gain_coupled_regulation}, with additional constant terms caused by the disturbances and obstacle motion.

{\it 2.3) Convolution-based small-gain synthesis.}
The interconnection characterized by \eqref{eq_proof_robust_coupled_distance} and \eqref{eq_proof_robust_coupled_regulation} has the same convolution-based gains as that in the proof of Theorem \ref{theorem_small_gain_safety}. Thus, we use the same matrix $\Theta(s)$ in \eqref{eq_proof_small_gain_matrix_s}. In particular,
\begin{align}
\Theta(0)=
\begin{bmatrix}
0 & 1/k_\alpha\\
k_\alpha L_\rho\gamma & L_\rho\gamma
\end{bmatrix}.
\end{align}
Under \eqref{eq_command_bound_condition_refine}, which is guaranteed by \eqref{eq_command_bound_condition}, the spectral radius of $\Theta(0)$ is less than $1$. Moreover, $\Theta(s)$ is defined on a left-hand neighborhood of $0$ and is continuous from the left at $s=0$, as verified in the proof of Theorem \ref{theorem_small_gain_safety}.

We apply Lemma \ref{lemma_symmetric_convolution_small_gain} to \eqref{eq_proof_robust_coupled_distance} and \eqref{eq_proof_robust_coupled_regulation}, with
$y_1(t)=[D_s-d(y(t),t)]^+$ and
$y_2(t)=|\tilde{v}_r(t)|$.
Both signals $y_1(t)$ and $y_2(t)$ are nonnegative and continuous on $[0,T_m)$, and hence locally bounded and measurable. The convolution-based gains in \eqref{eq_proof_robust_coupled_distance} and \eqref{eq_proof_robust_coupled_regulation} are the same as those in the proof of Theorem \ref{theorem_small_gain_safety}. Therefore, the same constants $\mu$ and $c_1'$ in \eqref{eq_proof_small_gain_convolution_bound} can be used to obtain
\begin{align}
|\tilde{v}_r(t)|\le c_1'(\bar{v}_c+c_0)e^{\mu t}+c_3'
\label{eq_proof_robust_convolution_bound}
\end{align}
for all $t\in[0,T_m)$, where
\begin{align}
c_3'
&=\frac{1}{1-2\gamma L_\rho}
(\gamma L_\rho(\bar{v}_c+\bar{v}_c^d)\notag\\
&\quad+\left(\frac{a_r}{b_r}+2\gamma L_\rho|C|\right)\bar{\delta}
+\gamma(L_o+L_\rho)\bar{\delta}_o^d).
\label{eq_proof_robust_c3}
\end{align}
To verify \eqref{eq_proof_robust_c3}, note that
\begin{align}
[I_2-\Theta(0)]^{-1}
&=\frac{1}{1-2\gamma L_\rho}
\begin{bmatrix}
1-\gamma L_\rho & 1/k_\alpha\\
k_\alpha\gamma L_\rho & 1
\end{bmatrix}.
\label{eq_proof_robust_inverse_gain_matrix}
\end{align}
The constant terms in \eqref{eq_proof_robust_coupled_distance} and \eqref{eq_proof_robust_coupled_regulation} are respectively
$(|C|\bar{\delta}+\bar{\delta}_o^d)/k_\alpha$ and
$\gamma L_\rho(\bar{v}_c+\bar{v}_c^d)
+(a_r/b_r+\gamma L_\rho|C|)\bar{\delta}
+\gamma L_o\bar{\delta}_o^d$.
By \eqref{eq_lemma_symmetric_convolution_bounds}, multiplying these terms by the second row of \eqref{eq_proof_robust_inverse_gain_matrix} gives \eqref{eq_proof_robust_c3}.

{\it 2.4) Establishing $d(y(t),t)\ge D+\epsilon$ on $[0,T_m)$.}
Substituting \eqref{eq_proof_robust_convolution_bound} into \eqref{eq_proof_robust_distance_bound}, and using the initial condition \eqref{eq_initial_conditionP_y}, we obtain
\begin{align}
D_s&-d_i(y(t),t)
\le-c_De^{-k_\alpha t/c_A}\notag\\
&\quad+c_1'(\bar{v}_c+c_0)
(e^{-k_\alpha(\cdot)}\ast e^{\mu(\cdot)})(t)+c_4
\label{eq_proof_robust_distance_transient}
\end{align}
for all $t\in[0,T_m)$ and $i=1,\ldots,n_o$, where
\begin{align}
c_4=\frac{c_3'+|C|\bar{\delta}+\bar{\delta}_o^d}{k_\alpha}.
\label{eq_proof_robust_c4}
\end{align}
As in the proof of Theorem \ref{theorem_small_gain_safety}, we handle the constant term $c_4$ and the initial-state response separately.

By \eqref{eq_proof_fefn_delta}, \eqref{eq_noisy_controller_bound}, \eqref{eq_proof_robust_c3}, and \eqref{eq_proof_robust_c4}, $c_4$ can be made arbitrarily close to $c_3/k_\alpha$ by choosing $\bar{\delta}_x$, $\bar{\delta}_y$, $\bar{\delta}_m$, and $\bar{\delta}_o^d$ sufficiently small, where $c_3$ is defined in \eqref{eq_proof_small_gain_convolution_bound_constant}. Since condition \eqref{eq_command_bound_condition} ensures that $c_3/k_\alpha<D_s-D$, we can choose $\epsilon>0$ satisfying \eqref{eq_proof_epsilon}, and then choose $\bar{\delta}_x$, $\bar{\delta}_y$, $\bar{\delta}_m$, and $\bar{\delta}_o^d$ sufficiently small that
\begin{align}
2\epsilon+c_4<D_s-D.
\label{eq_proof_robust_epsilon}
\end{align}
These choices do not depend on the initial states.

We next choose $c_D$ to account for the initial-state response in \eqref{eq_proof_robust_distance_transient}. Given any $c_0>0$, by \eqref{eq_proof_ekalpha_ast_emu}, choose $c_D>0$ sufficiently large that
\begin{align}
&-c_De^{-k_\alpha t/c_A}
+c_1'(\bar{v}_c+c_0)
(e^{-k_\alpha(\cdot)}\ast e^{\mu(\cdot)})(t)
\le\epsilon
\label{eq_proof_robust_transient_compensation}
\end{align}
for all $t\ge0$.

Substituting \eqref{eq_proof_robust_transient_compensation} and \eqref{eq_proof_robust_epsilon} into \eqref{eq_proof_robust_distance_transient}, we obtain
\begin{align}
D_s-d_i(y(t),t)\le\epsilon+c_4<D_s-D-\epsilon
\label{eq_proof_robust_distance_margin}
\end{align}
for all $t\in[0,T_m)$ and $i=1,\ldots,n_o$. Taking the minimum distance over the obstacles yields
\begin{align}
d(Cx(t),t)\ge D+\epsilon,\qquad t\in[0,T_m).
\label{eq_proof_robust_strict_safety}
\end{align}

{\it 3) Forward completeness of the perturbed closed-loop system.}
We finally prove that $T_m=\infty$ by contradiction. Suppose that $T_m<\infty$. By \cite[Theorem 9 in Section 2.7]{Filippov-Book-1988}, together with \eqref{eq_proof_robust_strict_safety}, the solution $(x(t),z(t))$ cannot remain bounded on $[0,T_m)$. Since $f_n(t,x,z)$ is Lipschitz on $\mathcal{D}_r$ by Step 1.1 and $T_m<\infty$, there exist constants $a,b>0$ such that
$|f_n(t,x,z)|\le a|[x;z]|+b$
for all $(t,x,z)\in\mathcal{D}_r$ with $t<T_m$. Together with \eqref{eq_proof_robust_differential_inclusion}, this gives
\begin{align}
|[\dot{x}(t);\dot{z}(t)]|\le a|[x(t);z(t)]|+b+\bar{\delta}
\label{eq_proof_robust_linear_growth}
\end{align}
for almost every $t\in[0,T_m)$. Applying the Gronwall-Bellman inequality \cite[Lemma A.1]{Khalil-book-2002}, we obtain the boundedness of $(x(t),z(t))$ on $[0,T_m)$, contradicting the preceding continuation property. Thus, $T_m=\infty$. Since the preceding arguments apply to every maximal Filippov solution satisfying \eqref{eq_initial_condition_robust}, all such solutions are defined for all $t\ge0$ and satisfy \eqref{eq_perturbed_objective}.
\fi
\end{proof}

\section{Numerical Simulation and Experiment}
\label{section_simulation_experiment}

Numerical simulations and hardware experiments are used to validate the proposed methods. The code is available at \url{https://github.com/EXPSIN/set-valued-and-output-feedback}.

\subsection{Numerical Simulation}

We consider a quadrotor navigating among three obstacles above a moving freighter's deck, focusing on safety in the horizontal plane $\mathbb{R}^2$. The moving obstacles, one disc-shaped and two triangular, are modeled as in \eqref{eq_Odi} with
\begin{subequations}
\begin{align}
\mathcal{O}_1 &=\{o\in\mathbb{R}^2:|o-[5.9;2.5]|\le0.4\},\\
\mathcal{O}_2 &=\co\{[2.2;4.0],[3.0;4.0],[3.0;3.2]\},\\
\mathcal{O}_3 &=\co\{[1.1;1.0],[1.1;0.2],[0.3;0.2]\},\\
\delta_{o,i}(t) &=[t-0.3\sin(\pi t/10+2\pi i/3);0],~ i=1,2,3.
\end{align}
\end{subequations}
The moving obstacles satisfy Assumption \ref{assumption_obstacle_new} with $D_o=2.0$ and $\bar{\delta}_o^d=1+3\pi/100$.

Suppose that the quadrotor can be modeled by \eqref{eq_plant_ABC} with
\begin{align}
A=[0,I_2;0,0], \quad B=[0;I_2], \quad C=[I_2,0], \label{eq_simulation_plant}
\end{align}
which satisfies Assumption \ref{assumption_plant}. In the simulation, we consider two setups of disturbances:
\begin{subequations} \label{eq_disturbances_setup2}
\begin{align}
\text{Setup 1:}~&\delta_x\equiv 0,~~~\delta_y\equiv 0,~~~\delta_m\equiv 0, \\
\text{Setup 2:}~&\|\delta_x\|_\infty\le1,~~~ \|\delta_y\|_\infty\le1,~~~ |\delta_m(l,t)|\le1
\end{align}
\end{subequations}
for every unit vector $l\in\mathbb{R}^{n_y}$ and $t\in\mathbb{R}_+$. We design a controller to achieve the control objective \eqref{eq_perturbed_objective} with $D=0.7$ m. Also, it is desired that the velocity $\dot{y}$ deviates as little as possible from a velocity-command signal $v_c=[2;0]$.

For coordinate transformation \eqref{eq_coordinate_transformation}, choose $N=[0,0;0,0;-1,0;0,1]$, which satisfies \eqref{eq_coordinate_transformation_condition}. The definitions in \eqref{eq_ABC_eta} give $A_\eta=0_2$, $B_\eta=C_\eta=\diag\{-1,1\}$ and $D_\eta=0_2$. Thus, $\sigma_1=\sigma_2=1$ and $E_\eta=CAB=I_2$, verifying \eqref{eq_fullrowrank}. Condition \eqref{eq_rightinvertibility} can be directly verified.

For controller \eqref{eq_safety_critical_controller} with measurement \eqref{eq_disturbed_setvalued_feedback}, choose
\begin{subequations}\label{eq_simulation_parameters}
\begin{align}
&D_s=1.0,\,D_B=3.0,\,k_\alpha=1.0,\,k_L=2.0,\,n_L=15,\\
&[A_L]_{q+1,:} =\left[\cos\frac{2\pi q}{n_L},~\sin\frac{2\pi q}{n_L}\right], c_A=\cos\frac{2\pi}{n_L}
\label{eq_AL_uniform_directions}
\end{align}
\end{subequations}
for $q=0,\ldots,n_L-1$. The parameters fulfill the requirement that $D<D_s\le\min\{D_o/2,D_B\}$, and $A_L$ is a normalized independent positive-spanning matrix with spanning coefficient $c_A$. Using the method in \cite{Wood-Zhang-JGO-1996} gives a numerical estimate $L_\rho=3.2$ of the Lipschitz constant.

For output-feedback velocity-regulation controller \eqref{eq_regulation_controller}, we choose $L=[-1,-0.5;-0.5,1]$. The eigenvalues of the resulting $A_\eta-LC_\eta$ are $-1\pm0.5\mathrm{i}$, verifying \eqref{eq_L_condition}. For $\bar{v}_c=2$ and $\bar{v}_c^d=0$, choose $\gamma=1/32$, which satisfies \eqref{eq_command_bound_condition} with $L_\rho=3.2$. The definitions in \eqref{eq_KT_selection_condition} give $K=\diag(32,-32)$ and $T=32I_2$. Then, $A_\eta+B_\eta K=-32I_2$, verifying \eqref{eq_KT_selection_condition_Hurwitz}.

The initial conditions are $x(0)=[0;2;0;0]$ and $z(0)=[0;-2]$, and the simulation step size is $0.05\mathrm{s}$.

\begin{figure*}[hbtp]
\centering
\includegraphics[width=1\textwidth]{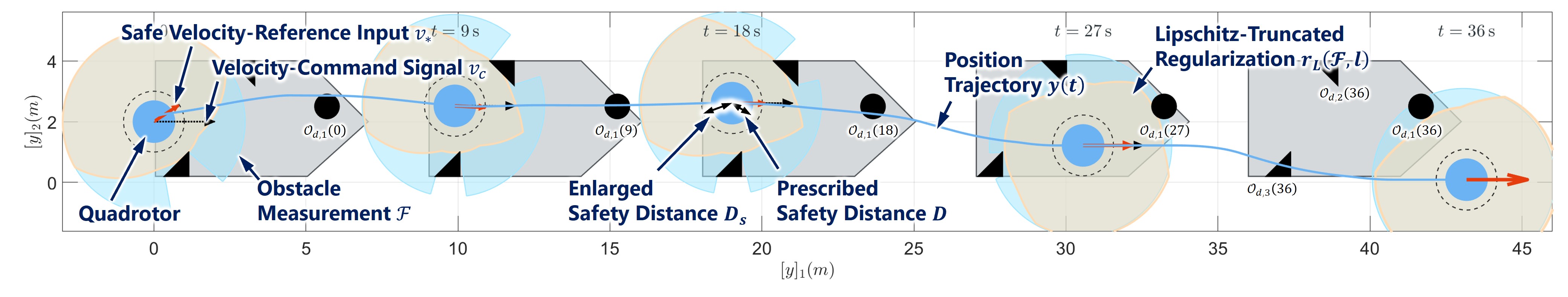}
\caption{Quadrotor navigation among moving obstacles. The blue curve shows the position trajectory $y(t)$. The cyan regions represent the areas scanned by the sensor. The orange regions represent the regularized obstacle measurements, with boundaries given by $\{y+r_L(\mathcal{F},l)l:|l|=1\}$.}
\label{figure_simulation_scene}
\end{figure*}

\begin{figure*}[htbp]
\centering
\includegraphics[width=1\textwidth]{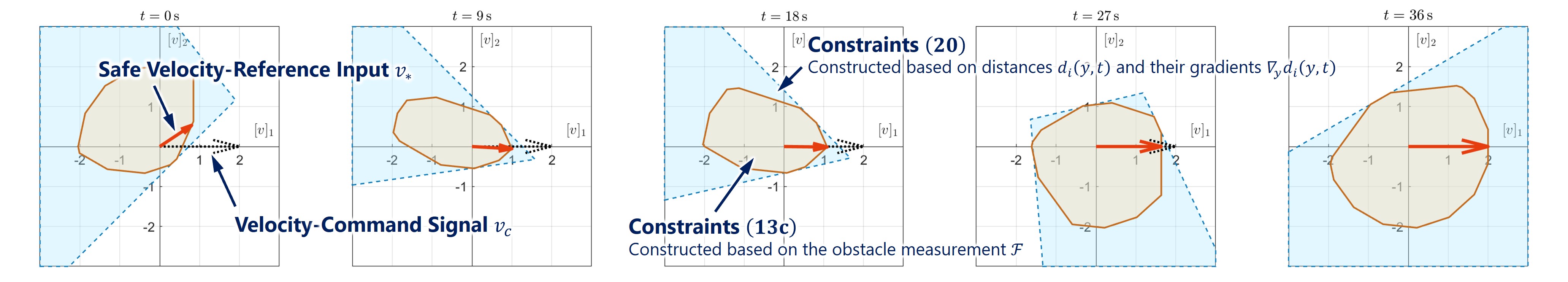}
\caption{Comparison of sets of velocity-reference inputs at $t\in\{0,9,18,27,36\}\,\mathrm{s}$. The cyan regions with dashed boundaries and the orange regions with solid boundaries are defined by the standard constraints and the proposed constraints \eqref{eq_safety_critical_controller_constraints}, respectively.}
\label{figure_simulation_feasible_set}
\end{figure*}

\begin{figure}[htbp]
\centering
\includegraphics[width=0.9\linewidth]{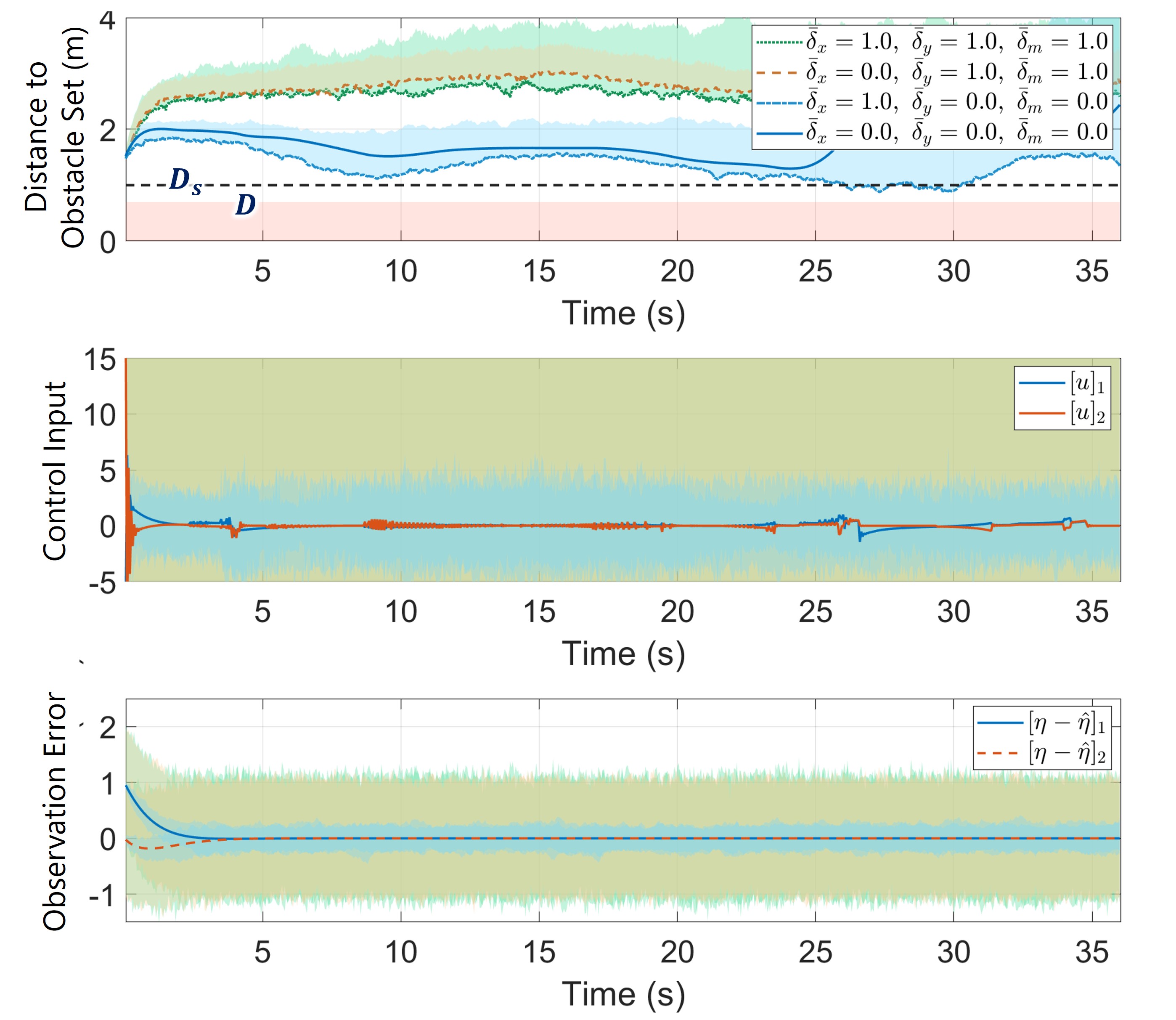}
\caption{Minimum distance between quadrotor and obstacles, control input, and observation error.}
\label{figure_simulation_robust_responses}
\end{figure}

For Setup 1, Figure \ref{figure_simulation_scene} shows that the output trajectory $y(t)$ satisfies condition \eqref{eq_perturbed_objective} defining the control objective, despite the presence of moving obstacles. Figure \ref{figure_simulation_feasible_set} visualizes the sets of velocity-reference inputs defined by the proposed constraints \eqref{eq_safety_critical_controller_constraints} and the standard constraints \eqref{eq_constraints_control}, showing that the shape of our constructed set changes ``more gradually'' across the displayed time instants. Figure \ref{figure_simulation_robust_responses} shows the minimum distance $d(y,t)$, control input $u$, and observation error $\eta-\hat{\eta}$ for Setups 1 and 2. In Setup 2, the minimum distance may drop below the enlarged safety distance $D_s$, but it remains above the prescribed safety distance $D$, thereby ensuring the safety property in \eqref{eq_perturbed_objective}. The control input and observation error remain bounded in both setups.

% \FloatBarrier
\subsection{Experiment} \label{subsection_experiment}

An indoor quadrotor experiment was conducted to study safety in a horizontal plane with several obstacles. The platform is a DJI Matrice 4T. Measured obstacle distances are obtained via the DJI Mobile SDK, while velocity control is provided through the DJI Payload SDK (PSDK). The onboard estimator and built-in velocity controller realize the inner-loop mapping from $v_*$ to $v=\dot{y}$; thus, the reduced-order observer \eqref{eq_observer_state}--\eqref{eq_observer} and velocity-regulation controller \eqref{eq_controller} are not implemented. Using MATLAB System Identification Toolbox, we identify a dynamical model from $v_*$ to $v$:
\begin{align}
\dot{\eta}'&=A'\eta'+B'v_*,&
v=C'\eta', \label{eq_experiment_identified_model}
\end{align}
where $\eta'\in\mathbb{R}^3$ is the state of the identified model, and
{\small
\begin{subequations}
\begin{align}
&A'=\left[\begin{array}{@{}r@{~~}r@{~~}r@{}}
-1.569 & 0.624 & 0.548\\
-0.586 & -0.977 & -0.221\\
0.124 & 0.957 & 0.304
\end{array}\right],\\
&B'=\left[\begin{array}{@{}r@{~~}r@{}}
1.533 & -1.863\\
-2.113 & -0.544\\
2.117 & 0.020
\end{array}\right],~~
{C'}^T=\left[\begin{array}{@{}r@{~~}r@{}}
-0.091 & -0.828\\
-0.404 & -0.104\\
0.073 & -0.129
\end{array}\right].
\end{align}
\end{subequations}
}

We design a controller to achieve the control objective \eqref{eq_perturbed_objective} with $D=0.6$ m. Additionally, it is desired that the velocity $\dot{y}$ deviates as little as possible from the velocity-command signal $v_c$, which guides the quadrotor along a square path with a side length of $3$ m. For the safety-critical controller \eqref{eq_safety_critical_controller}, choose 
\begin{align}
D_s=0.9,~ D_B=60,~ k_\alpha=0.75,~ k_L=4.5,~ n_L=37
\end{align}
with $A_L$ and $c_A$ defined in \eqref{eq_AL_uniform_directions}.

\begin{figure}[htbp]
\centering
\includegraphics[width=0.9\linewidth]{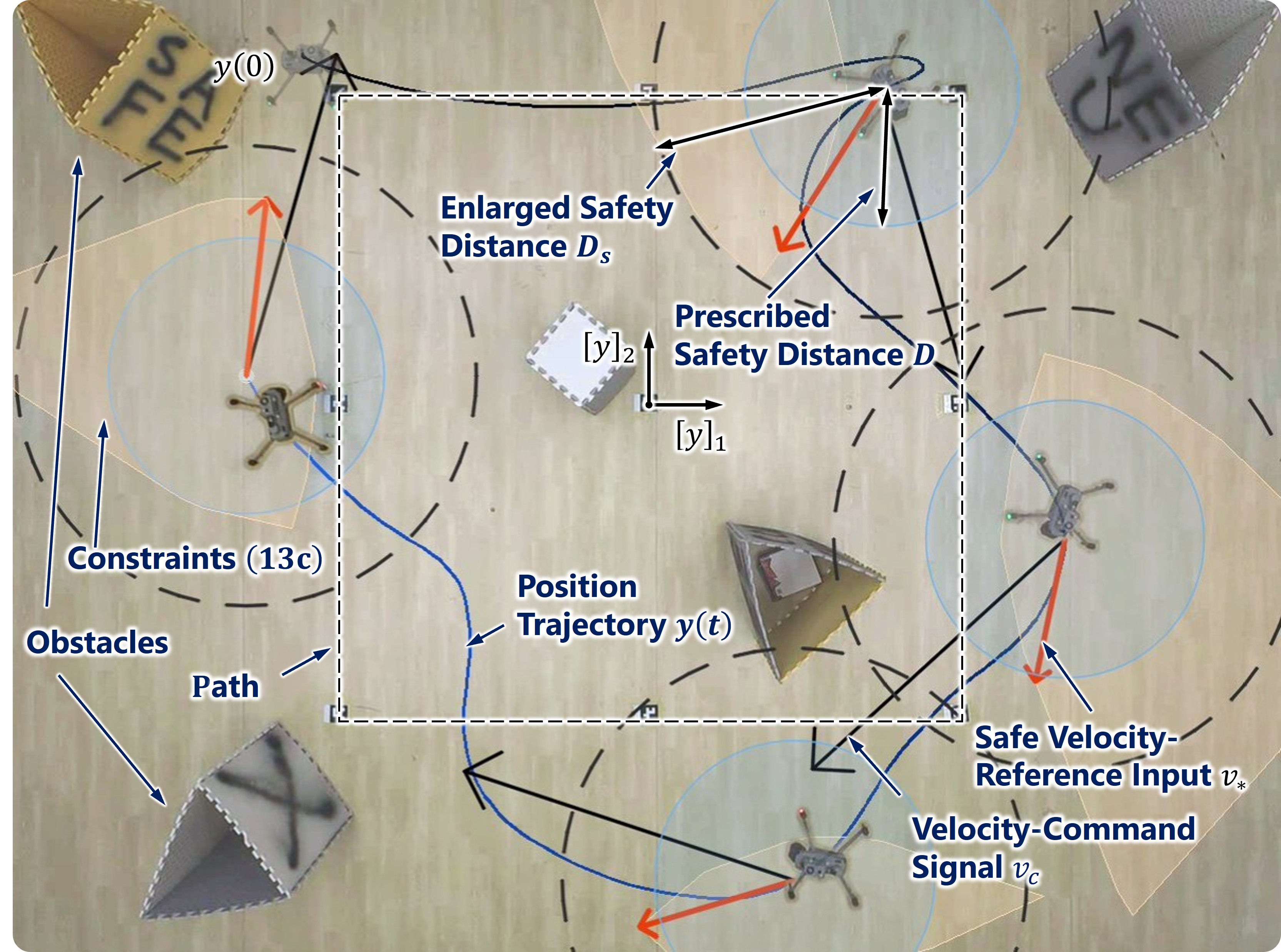}
\caption{Indoor quadrotor navigation among obstacles.}
\label{figure_experiment_position}
\end{figure}

\begin{figure}[htbp]
\centering
\includegraphics[width=0.9\linewidth]{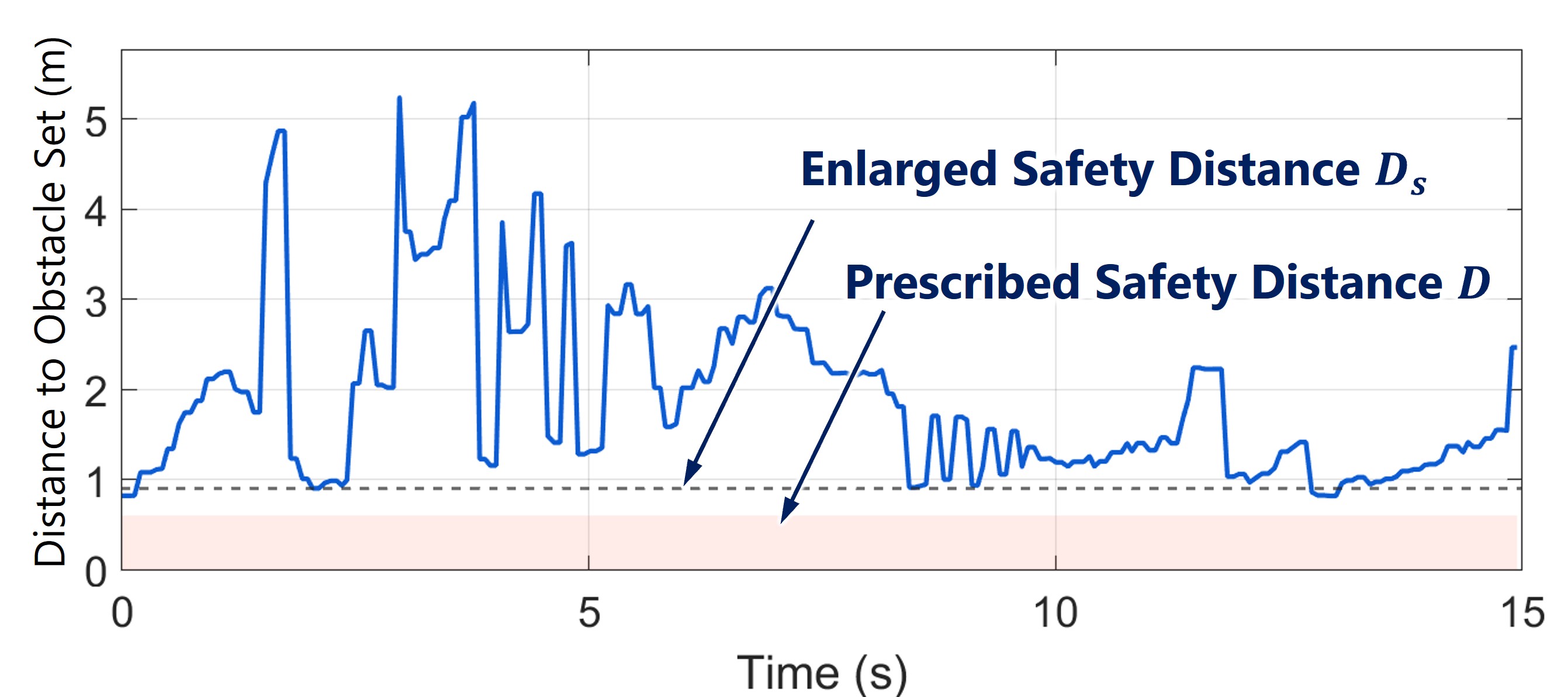}
\caption{Minimum distance between quadrotor and obstacles.}
\label{figure_experiment_distance}
\end{figure}

Figure \ref{figure_experiment_position} shows the position trajectory $y(t)$ when the quadrotor follows the prescribed path while maintaining the safety distance $D$ from the obstacles. The circles around the quadrotor indicate $D$ and the enlarged safety distance $D_s$. As shown in Figure \ref{figure_experiment_distance}, the minimum distance to all obstacles remains larger than $D$ throughout the experiment. These results verify that the control objective \eqref{eq_perturbed_objective} is achieved.

% \FloatBarrier
% \flushbottom
\section{Conclusions}\label{section_conclusion}

This paper has addressed the problem of safety-critical output-feedback control of mobile robots with finite-range, set-valued obstacle measurements. The fundamental challenges arise from the potential violation of constraint qualification conditions for the safety constraints constructed directly from set-valued obstacle measurements, as well as the interaction between the safety-critical output and unmeasured internal dynamics. To address the challenges, we have developed a new qualification-enforcing constraint-construction technique that ensures the feasibility and Lipschitz continuity of the resulting QP-based safety-critical controller. We have developed a new safety-oriented, convolution-based small-gain synthesis that preserves the fading-memory effects of subsystem inputs. Our main result establishes the safety of the closed-loop system under suitable initial conditions, while providing robustness guarantees against moving obstacles, external disturbances, and measurement errors. This result would provide a foundation for further investigating safety-critical control of more general dynamical systems with set-valued output feedback.

\appendices

\section{Proof of the Propositions}

\subsection{Proof of Proposition \ref{proposition_eta_controllability_observability}}\label{appendix_proposition_eta_controllability_observability}

The controllability and observability can be proved by contradiction. Suppose that $(N^TAN,N^TB)$ is not controllable. Then, by the Popov-Belevitch-Hautus (PBH) test for controllability \cite[Theorem 12.3]{Hespanha-Book-2018}, there exist a nonzero vector $\xi\in\mathbb{C}^{n_x-n_y}$ and a scalar $\lambda\in\mathbb{C}$ such that
\begin{align}
\xi^T[N^TAN-\lambda I, N^TB]=0. \label{PBH_equality}
\end{align}
Let $p=N\xi$. Since $N^TN=I$, we have $p\neq0$. Moreover,
\begin{align}
p^T[A-\lambda I, B]
&=\xi^TN^T[(A-\lambda I)\left(C^\dagger C+NN^T\right),B]\notag\\
&=\xi^TN^T[(A-\lambda I)NN^T, B]=0, \label{PBH_equality_contradict}
\end{align}
where we used $C^\dagger C+NN^T=I$ (due to the full-row-rank property of $C$ and \eqref{eq_coordinate_transformation_condition}) for the first equality, used $AC^T=0$ in Assumption \ref{assumption_plant} and $CN=0$ in \eqref{eq_coordinate_transformation_condition} for the second equality, and used $N^TN=I$ and \eqref{PBH_equality} for the last equality. Note that \eqref{PBH_equality_contradict} contradicts the controllability of $(A,B)$. Therefore, $(N^TAN,N^TB)$ is controllable. The observability of $(N^TAN,CAN)$ can be verified similarly by the PBH test for observability \cite[Theorem 15.9]{Hespanha-Book-2018}.

\subsection{Proof of Proposition \ref{proposition_safety}}\label{appendix_proposition_safety}

{\it (a) Feasibility.}
From \eqref{eq_feasibility_point}, $\rho_s(\mathcal{F})=\rho_s(\mathcal{M}(y))$ is defined for all $y\in\mathbb{R}^{n_y}$. Moreover, \eqref{eq_def_bL_main} gives $b_L(\mathcal{M}(y))\ge0$ for all $y\in\mathbb{R}^{n_y}$. Hence, $\nu=\rho_s(\mathcal{M}(y))$ is a feasible point of the QP \eqref{eq_safety_critical_controller} for all $y\in\mathbb{R}^{n_y}$. Uniqueness follows from the projection theorem \cite[Proposition B.11]{Bertsekas-Book-1997}.

{\it (b) Boundedness and Lipschitz continuity.}
For any $i\neq j$, the definition \eqref{eq_di_and_d} implies
\begin{align}
d_i(y)+d_j(y)
&=|y-o_i^*(y)|+|y-o_j^*(y)|\notag\\
&\ge|o_i^*(y)-o_j^*(y)|\ge D_o\ge2D_s, \label{eq_proof_feasibility_01}
\end{align}
where $o_i^*(y)$ denotes the closest point in $\mathcal{O}_i$ to $y$, condition \eqref{assumption_obstacle_separation} is used for the second inequality, and condition $D_s\le D_o/2$ given below \eqref{eq_def_rL} is used for the last inequality. For $y\in\complement\mathcal{O}$, property \eqref{eq_proof_feasibility_01} ensures that at most one of $d_i(y)$, with $i=1,\ldots,n_o$, is less than $D_s$. If such an obstacle exists, with its index denoted by $\ubar{i}$, the definition of $\rho_s$ in \eqref{eq_feasibility_point} gives $\rho_s(\mathcal{M}(y))=k_\alpha\nabla d_{\ubar{i}}(y)[D_s-d_{\ubar{i}}(y)]^+$. Otherwise, $\rho_s(\mathcal{M}(y))=0$. Hence, we have
\begin{align}
\rho_s(\mathcal{M}(y)) &=k_\alpha\sum_{i=1}^{n_o}\nabla d_i(y)[D_s-d_i(y)]^+
\label{eq_proof_rhos_relative}
\end{align}
for $y\in\complement\mathcal{O}$. Based on the discussion above \eqref{eq_proof_rhos_relative}, using the definition of $d(y)$ in \eqref{eq_di_and_d}, we have 
\begin{align}
|\rho_s(\mathcal{M}(y))| &\le k_\alpha[D_s-d(y)]^+ \label{eq_rhos_bound}
\end{align}
for $y\in\complement\mathcal{O}$. For any $y\in\complement\mathcal{O}$ and $v_c\in\mathbb{R}^{n_y}$, the projection theorem \cite[Proposition B.11]{Bertsekas-Book-1997} gives 
\begin{subequations} 
\begin{align}
|\rho(\mathcal{M}(y),v_c)-\rho(\mathcal{M}(y),0)|&\le|v_c|, \label{eq_rhos_bound_proj1}\\
|\rho(\mathcal{M}(y),0)|&\le|\rho_s(\mathcal{M}(y))|. \label{eq_rhos_bound_proj2}
\end{align}
\end{subequations}
Applying the triangle inequality \cite[Theorem 3.3]{Apostol-book-1974} to \eqref{eq_rhos_bound_proj1}, and using \eqref{eq_rhos_bound} and \eqref{eq_rhos_bound_proj2}, we obtain \eqref{proposition_rho_bound}.

We next prove the Lipschitz continuity property \eqref{proposition_rho_Lipschitz}. Let
\begin{subequations} \label{eq_safety_critical_controller_new}
\begin{align}
\bar{v} &= \argmin_{\nu'\in\mathbb{R}^{n_y}}
|\nu'-(v_c-\rho_s(\mathcal{M}(y)))|^2,\\
&\quad \text{s.t.}~A_L \nu'\le b_L(\mathcal{M}(y)). 
\end{align}
\end{subequations}
Clearly, $\nu'=0$ is a feasible point of the QP \eqref{eq_safety_critical_controller_new}, as $b_L(\cdot)\ge0$. The solution to the QP \eqref{eq_safety_critical_controller} takes the  form $\rho(\mathcal{M}(y),v_c)=\rho_s(\mathcal{M}(y))+\bar{v}$, where $\bar{v}$ is the solution of \eqref{eq_safety_critical_controller_new}. For the QP \eqref{eq_safety_critical_controller_new}, let $\breve{A}_L \bar{v}\le \breve{b}_L(y)$ represent the constraints that directly affect the optimal solution. Condition \ref{condition_AL_full_rank} defining a normalized independent positive-spanning matrix in Notations implies the existence of $c_{A_L}>0$ such that $|\breve{A}_L^T\xi|\ge c_{A_L}|\xi|$ for all $\xi$. Then, one can verify that all the conditions required by \cite[Theorem 3.1]{Hager-SIAMControl-1979} are satisfied and thus $\bar v$ is Lipschitz with respect to $v_c-\rho_s(\mathcal{M}(y))$ and $b_L(\mathcal{M}(y))$.

It remains to show that $\rho_s(\mathcal{M}(y))$ and $b_L(\mathcal{M}(y))$ are Lipschitz in $y$ on $\{y\in\mathbb{R}^{n_y}:d(y)\ge D\}$. On this domain, $d_i(y)$, $\nabla d_i(y)$, and $[D_s-d_i(y)]^+$ are Lipschitz, and $[D_s-d_i(y)]^+\le D_s-D$. Property \eqref{eq_proof_rhos_relative} then implies that $\rho_s(\mathcal{M}(y))$ is Lipschitz in $y$, while property \eqref{eq_rhos_bound} gives $|\rho_s(\mathcal{M}(y))|\le k_\alpha(D_s-D)$. Moreover, for each fixed unit vector $l\in\mathbb{R}^{n_y}$, $r_L(\mathcal{M}(y),l)$ is Lipschitz in $y$ \cite{Wu-Liu-Zhang-Ding-Jiang-Chai-TAC-2026}. Hence, the definition of $b_L$ in \eqref{eq_def_bL_main} and the algebraic properties of Lipschitz functions \cite{Cobzas-Miculescu-Nicolae-book-2019} imply that $b_L(\mathcal{M}(y))$ is Lipschitz in $y$. Combining these properties with the Lipschitz continuity established above for the solution of the QP \eqref{eq_safety_critical_controller_new} yields property \eqref{proposition_rho_Lipschitz}.

{\it (c) Robust safety with respect to the velocity-regulation error.}
We first establish some properties before proving \eqref{eq_safety_dissipation}. For any $i\in\{1,\ldots,n_o\}$ and any $y\in\mathbb{R}^{n_y}$ satisfying $d(y)\ge D$, condition \ref{condition_AL_positive_combination} for a normalized independent positive-spanning matrix in Notations ensures the existence of $\lambda_j\ge0$ such that
\begin{align}
-\nabla d_i(y)=\sum_{j\in\mathcal{J}_i}\lambda_j[A_L]_{j,:}^T,\qquad
\sum_{j\in\mathcal{J}_i}\lambda_j\le c_A^{-1}, \label{eq_proof_positive_combination}
\end{align}
where $\mathcal{J}_i=\{j:-[A_L]_{j,:}\nabla d_i(y)\ge c_A\}$. Using the definition of $r_L$ in \eqref{eq_def_rL}, we obtain
\begin{align}
&r_L(\mathcal{M}(y),[A_L]_{j,:}^T)\le r(y,-\nabla d_i(y))+k_L[c_A+[A_L]_{j,:}\nabla d_i(y)]^+\notag\\
&=r(y,-\nabla d_i(y))\le d_i(y),\quad j\in\mathcal{J}_i,
\label{eq_bounds_rL_y}
\end{align}
where the first inequality follows by choosing $l'=-\nabla d_i(y)$ in \eqref{eq_def_rL} and using the definition of $\mathcal{J}_i$ to make the term $[c_A+[A_L]_{j,:}\nabla d_i(y)]^+$ zero, while the second follows from \eqref{eq_setup_setvalued_range_model}. When $d_i(y)\le D_s$, direct calculation yields
\begin{align}
&k_\alpha(r_L(\mathcal{M}(y),[A_L]_{j,:}^T)-D_s)
-[A_L]_{j,:}\rho_s(\mathcal{M}(y))\notag\\
&\le k_\alpha(1+[A_L]_{j,:}\nabla d_i(y))(d_i(y)-D_s)
\le0,~~j\in\mathcal{J}_i, \label{eq_proof_bL_zero}
\end{align}
where we used \eqref{eq_proof_feasibility_01}, \eqref{eq_proof_rhos_relative}, and \eqref{eq_bounds_rL_y} for the first inequality, and used condition \ref{condition_AL_unit_rows} for a normalized independent positive-spanning matrix in Notations for the last inequality. This, together with the definition of $b_L$ in \eqref{eq_def_bL_main}, ensures $[b_L(\mathcal{M}(y))]_j=0$ for all $j\in\mathcal{J}_i$.

Let $v_*$ be any vector satisfying the constraints \eqref{eq_safety_critical_controller_constraints}. If $d_i(y)\le D_s$, we have
\begin{align}
\nabla d_i(y)^Tv_*&\ge-\sum_{j\in\mathcal{J}_i}\lambda_j
([A_L]_{j,:}\rho_s(\mathcal{M}(y))+[b_L(\mathcal{M}(y))]_j)\notag\\
&=\nabla d_i(y)^T\rho_s(\mathcal{M}(y))
=-k_\alpha(d_i(y)-D_s),\!\! \label{eq_proof_safety_dissipation_near}
\end{align}
by using \eqref{eq_proof_positive_combination} and \eqref{eq_safety_critical_controller_constraints} for the first inequality, using $[b_L(\mathcal{M}(y))]_j=0$ and \eqref{eq_proof_positive_combination} for the first equality, and using \eqref{eq_proof_feasibility_01} and \eqref{eq_proof_rhos_relative} for the last equality. If $d_i(y)>D_s$, we have
\begin{align}
&\nabla d_i(y)^Tv_*\ge-\sum_{j\in\mathcal{J}_i}\lambda_j
([A_L]_{j,:}\rho_s(\mathcal{M}(y))+[b_L(\mathcal{M}(y))]_j)\notag\\
&\quad \ge-k_\alpha(d_i(y)-D_s)\sum_{j\in\mathcal{J}_i}\lambda_j\ge-\frac{k_\alpha}{c_A}(d_i(y)-D_s),
\label{eq_proof_safety_dissipation_far}
\end{align}
where the first inequality follows from \eqref{eq_proof_positive_combination} and \eqref{eq_safety_critical_controller_constraints}, and the last from \eqref{eq_proof_positive_combination}. Recall the definition of $b_L$ in \eqref{eq_def_bL_main}. The second inequality in \eqref{eq_proof_safety_dissipation_far} follows from
\begin{align}
&[A_L]_{j,:}\rho_s(\mathcal{M}(y))+[b_L(\mathcal{M}(y))]_j\notag\\
&\le\max\{|\rho_s(\mathcal{M}(y))|,k_\alpha(d_i(y)-D_s)\}\notag\\
&\le k_\alpha(d_i(y)-D_s),\qquad j\in\mathcal{J}_i,
\end{align}
where the first inequality follows from \eqref{eq_bounds_rL_y} and $|[A_L]_{j,:}|=1$, and the last from \eqref{eq_proof_feasibility_01} and \eqref{eq_rhos_bound}.

To establish robust safety with respect to the velocity-regulation error $\tilde{v}$, we combine inequalities \eqref{eq_proof_safety_dissipation_near} and \eqref{eq_proof_safety_dissipation_far} and use the fact $|\nabla d_i(y)|=1$ for $y\in\complement\mathcal{O}$. We obtain
\begin{subequations}\label{eq_proof_vtilde}
\begin{align}
\nabla d_i(y)^T(v_*+\tilde{v}) \ge -k_\alpha(d_i(y)-D_s) - |\tilde{v}|, \\
\nabla d_i(y)^T(v_*+\tilde{v}) \ge -\frac{k_\alpha}{c_A}(d_i(y)-D_s) - |\tilde{v}|
\end{align}
\end{subequations}
for $d_i(y)\le D_s$ and $d_i(y)>D_s$, respectively. Let $p_i=D_s-d_i(y)$ for $i=1,\ldots,n_o$. Property \eqref{eq_proof_vtilde}, together with the $y$-subsystem \eqref{eq_yr_subsystem}, gives
\begin{align}
\dot{p}_i(t) &\le-\alpha_p(p_i(t))+|\tilde{v}(t)|,~\alpha_p(s)=\begin{cases}
k_\alpha s,&s\ge0,\\
\frac{k_\alpha}{c_A}s,&s<0,
\end{cases} \label{eq_proof_pi}
\end{align}
for all $t$ on any time interval over which $p_j(t)\le D_s-D$ for all $j=1,\ldots,n_o$. Then, integrating \eqref{eq_proof_pi} from $0$ to $t$ and substituting $p_i(t)=D_s-d_i(y(t))$ into the resulting estimate, we obtain \eqref{eq_safety_dissipation} whenever $d_i(y(0))\ge D_s$. 

\subsection{Proof of Proposition \ref{proposition_observer}}\label{appendix_proposition_observer}

\unless\ifarxivversion
{\it 1) Minimum phase and right invertibility.}
Under Assumption \ref{assumption_plant} and condition \eqref{eq_rightinvertibility}, the coordinate transformation \eqref{eq_coordinate_transformation} ensures that $\mathrm{rank}[sI-A_\eta,B_\eta;C_\eta,-D_\eta]=n_x$ for all $s\in\bar{\mathbb{C}}_+$.

{\it 2) Existence of $L$, $K$, and $T$ satisfying condition \eqref{eq_KT_selection_condition}.}
By Proposition \ref{proposition_eta_controllability_observability} and \cite[Theorem 16.9]{Hespanha-Book-2018}, there exists $L$ satisfying \eqref{eq_L_condition}. Using the rank condition established in Step 1 and condition \eqref{eq_fullrowrank}, the PBH test and \cite[Theorem 14.5]{Hespanha-Book-2018} establish the existence of $K$ and $T$ satisfying \eqref{eq_KT_selection_condition}. A detailed proof of Steps 1) and 2) is provided in \cite{Wu-Liu-Ding-Li-Chai-Jiang-arXiv-2026}.
\fi

% ~~~~放到arxiv版本的部分
\ifarxivversion
{\it 1) Minimum phase and right invertibility.}
We first show that $(A_\eta,B_\eta,C_\eta,D_\eta)$ defined in \eqref{eq_ABC_eta} satisfies a minimum-phase and right-invertibility condition analogous to \eqref{eq_rightinvertibility}, namely,
\begin{align}
\mathrm{rank}\begin{bmatrix}sI-A_\eta&B_\eta\\C_\eta&-D_\eta\end{bmatrix}&=n_x,\quad \forall s\in\bar{\mathbb{C}}_+. \label{eq_proof_observer_rank_equivalence_arxiv}
\end{align}

Under Assumption \ref{assumption_plant}, applying the coordinate transformation \eqref{eq_coordinate_transformation} with a nonsingular $[C;N^T]$, and performing elementary row operations, we have
\begin{align}
&\mathrm{rank}\begin{bmatrix}sI-A&B\\C&0\end{bmatrix}\notag\\
&=\mathrm{rank}\left(\begin{bmatrix}C&0\\N^T&0\\0&I_{n_y}\end{bmatrix}\begin{bmatrix}sI-A&B\\C&0\end{bmatrix}\begin{bmatrix}C^\dagger&N&0\\0&0&I_{n_u}\end{bmatrix}\right)\notag\\
&=\mathrm{rank}\begin{bmatrix}sI&-CAN&CB\\0&sI-N^TAN&N^TB\\I&0&0\end{bmatrix}\notag\\
&=\mathrm{rank}\begin{bmatrix}0&-CAN&CB\\0&sI-N^TAN&N^TB\\I&0&0\end{bmatrix}\notag\\
&=n_y+\mathrm{rank}\begin{bmatrix}sI-N^TAN&N^TB\\CAN&-CB\end{bmatrix}\notag\\
&=n_y+\mathrm{rank}\begin{bmatrix}sI-A_\eta&B_\eta\\C_\eta&-D_\eta\end{bmatrix}, \label{eq_proof_observer_rank_equivalence_procedure_arxiv}
\end{align}
for all $s\in\bar{\mathbb{C}}_+$, where the third equality follows by subtracting $s$ times the last block row from the first block row, and the last equality uses \eqref{eq_ABC_eta}. Therefore, under condition \eqref{eq_rightinvertibility}, property \eqref{eq_proof_observer_rank_equivalence_procedure_arxiv} gives \eqref{eq_proof_observer_rank_equivalence_arxiv}, which establishes the minimum-phase property of the $\eta$-subsystem \eqref{eq_eta_subsystem_new} and guarantees the right invertibility of the matrix $[sI-A_\eta,B_\eta;C_\eta,-D_\eta]$ for all $s\in\bar{\mathbb{C}}_+$.

{\it 2) Existence of $L$, $K$, and $T$ satisfying condition \eqref{eq_KT_selection_condition}.} 
The existence of $L$ satisfying \eqref{eq_L_condition} is guaranteed by \cite[Theorem 16.9]{Hespanha-Book-2018}, as Proposition \ref{proposition_eta_controllability_observability} ensures the observability of $(A_\eta,C_\eta)$ under Assumption \ref{assumption_plant}. Moreover, since $E_\eta$ has full row rank, $K$ and $T$ in \eqref{eq_KT_selection_condition_K} and \eqref{eq_KT_selection_condition_T} are defined for any $K_0$ satisfying \eqref{eq_KT_selection_condition_Hurwitz}. Let $P_\eta=I_{n_u}-E_\eta^\dagger E_\eta$ and $A_0=A_\eta-B_\eta E_\eta^\dagger F_\eta$. Condition \eqref{eq_KT_selection_condition_Hurwitz} can be satisfied if $A_0+B_\eta P_\eta K_0\in\mathbb{H}$. By \cite[Theorem 14.5]{Hespanha-Book-2018}, it therefore suffices to prove the stabilizability of $(A_0,B_\eta P_\eta)$.

Before proving the stabilizability of $(A_0,B_\eta P_\eta)$, we establish the following facts. Under Assumption \ref{assumption_plant}, with $N$ satisfying condition \eqref{eq_coordinate_transformation_condition}, we have $A=ANN^T$. Using this identity and the definitions of $A_\eta$, $B_\eta$, and $C_\eta$ in \eqref{eq_ABC_eta}, we obtain
\begin{subequations}
\begin{align}
C_\eta A_\eta^kB_\eta
&=CAN(N^TAN)^kN^TB\notag\\
&=C(ANN^T)^{k+1}B\notag\\
&=CA^{k+1}B \label{eq_proof_observer_equivalence_arxiv}
\end{align}
for any nonnegative integer $k$. Recall the definition of $\sigma_i$ in \eqref{eq_def_ri}. Using the property \eqref{eq_proof_observer_equivalence_arxiv}, we obtain
\begin{align}
[C_\eta]_{i,:}A_\eta^rB_\eta&=0, \label{eq_proof_CArB_arxiv}
\end{align}
for all $i=1,\ldots,n_y$ and all nonnegative integers $r<\sigma_i-1$. The definitions of $\sigma_i$ and $D_\eta$ in \eqref{eq_def_ri} and \eqref{eq_ABC_eta} give $[D_\eta]_{i,:}=0$ for $\sigma_i\ge1$, while $(s+\lambda)^{\sigma_i}=1$ for $\sigma_i=0$. Hence,
\begin{align}
(s+\lambda)^{\sigma_i}[D_\eta]_{i,:}&=[D_\eta]_{i,:}, \label{eq_proof_weighted_Deta_arxiv}
\end{align}
for all $i=1,\ldots,n_y$ and all $s\in\bar{\mathbb{C}}_+$.
\end{subequations}

We establish the stabilizability of $(A_0,B_\eta P_\eta)$ by contradiction. In particular, if $(A_0,B_\eta P_\eta)$ is not stabilizable, then the PBH test \cite[Theorem 14.3]{Hespanha-Book-2018} gives some $s\in\bar{\mathbb{C}}_+$ and a row vector $p\neq 0$ such that
\begin{align}
p[sI-A_0,B_\eta P_\eta]=0.
\label{eq_proof_K0_PBH}
\end{align}
We use $p$ to construct a row vector $\ell$, which is used to show $p=0$ under the conditions of this proposition, contradicting $p\ne0$:
\begin{align}
\ell&=p-\sum_{i=1}^{n_y}S_i(s), \label{eq_def_ell}
\end{align}
where 
\begin{align}
S_i(s)&=\begin{cases}
0,&\sigma_i=0,\\
\displaystyle[q]_i[C_\eta]_{i,:}\sum_{j=0}^{\sigma_i-1}(A_\eta+\lambda I)^{\sigma_i-1-j}(s+\lambda)^j,&\sigma_i\ge1
\end{cases} \label{eq_def_Si}
\end{align}
for $i=1,\ldots,n_y$ with $q=pB_\eta E_\eta^\dagger$.

We establish the following two properties:
\begin{enumerate}[label=(\alph*)]

\item $\ell B_\eta=pB_\eta P_\eta+qD_\eta$. For $i=1,\ldots,n_y$ satisfying $\sigma_i\ge1$, direct calculation yields
\begin{align}
S_i(s)B_\eta
&=[q]_i[C_\eta]_{i,:}(A_\eta+\lambda I)^{\sigma_i-1}B_\eta\notag\\
&=[q]_i[C_\eta]_{i,:}A_\eta^{\sigma_i-1}B_\eta\notag\\
&=[q]_i[C]_{i,:}A^{\sigma_i}B\notag\\
&=[q]_i[E_\eta]_{i,:}\notag\\
&=[q]_i\bigl([E_\eta]_{i,:}-[D_\eta]_{i,:}\bigr), \label{eq_proof_ellBeta_01}
\end{align}
where we have used the definition of $S_i(s)$ in \eqref{eq_def_Si} and the property \eqref{eq_proof_CArB_arxiv} for the first equality, the binomial formula \cite[Section 3.2.5]{Horn-Johnson-book-2012} and the property \eqref{eq_proof_CArB_arxiv} for the second equality, property \eqref{eq_proof_observer_equivalence_arxiv} with $k=\sigma_i-1$ for the third equality, the definition of $E_\eta$ in \eqref{eq_def_Eeta} for the fourth equality, and the fact that $[D_\eta]_{i,:}=0$, which follows from $\sigma_i\ge1$ and the definitions in \eqref{eq_def_ri} and \eqref{eq_ABC_eta}, for the last equality.

In addition, for $i=1,\ldots,n_y$ satisfying $\sigma_i=0$, the definitions of $S_i(s)$, $E_\eta$, and $D_\eta$ in \eqref{eq_def_Si}, \eqref{eq_def_Eeta}, and \eqref{eq_ABC_eta}, respectively, give $S_i(s)=0$ and $[E_\eta]_{i,:}=[D_\eta]_{i,:}$. Thus,
\begin{align}
S_i(s)B_\eta&=0=[q]_i\bigl([E_\eta]_{i,:}-[D_\eta]_{i,:}\bigr). \label{eq_proof_ellBeta_02}
\end{align}

Recall the definition of $\ell$ in \eqref{eq_def_ell}. Combining the properties \eqref{eq_proof_ellBeta_01} and \eqref{eq_proof_ellBeta_02} and using $q=pB_\eta E_\eta^\dagger$ and $P_\eta=I_{n_u}-E_\eta^\dagger E_\eta$, we have
\begin{align}
\ell B_\eta
&=pB_\eta-\sum_{i=1}^{n_y}S_i(s)B_\eta=pB_\eta-q(E_\eta-D_\eta)\notag\\
&=pB_\eta(I_{n_u}-E_\eta^\dagger E_\eta)+qD_\eta\notag\\
&=pB_\eta P_\eta+qD_\eta. \label{eq_proof_proposition_K0_2}
\end{align}

\item $\ell(sI-A_\eta)+\Sigma_{i=1}^{n_y}[q]_i(s+\lambda)^{\sigma_i}[C_\eta]_{i,:}=p(sI-A_0)$. Specifically, \eqref{eq_def_Feta} gives
\begin{align}
[F_\eta]_{i,:}&=[C_\eta]_{i,:}(A_\eta+\lambda I)^{\sigma_i}. \label{eq_proof_proposition_K0_3}
\end{align}
Since $sI-A_\eta=(s+\lambda)I-(A_\eta+\lambda I)$, the definition of $S_i(s)$ in \eqref{eq_def_Si} gives
\begin{align}
S_i(s)(sI-A_\eta)
&=[q]_i[C_\eta]_{i,:}(s+\lambda)^{\sigma_i}\notag\\
&\quad-[q]_i[C_\eta]_{i,:}(A_\eta+\lambda I)^{\sigma_i}, \label{eq_proof_proposition_K0_4}
\end{align}
for all $i=1,\ldots,n_y$. Using the expression for $[F_\eta]_{i,:}$ in \eqref{eq_proof_proposition_K0_3}, we obtain
\begin{align}
S_i(s)(sI-A_\eta)&=[q]_i\bigl([C_\eta]_{i,:}(s+\lambda)^{\sigma_i}-[F_\eta]_{i,:}\bigr), \label{eq_proof_proposition_K0_5}
\end{align}
which, together with the definition of $\ell$ in \eqref{eq_def_ell}, implies
\begin{align}
&\ell(sI-A_\eta)
=p(sI-A_\eta)-\sum_{i=1}^{n_y}[q]_i((s+\lambda)^{\sigma_i}[C_\eta]_{i,:}\notag \\
&\hspace{30pt} -[F_\eta]_{i,:})\notag\\
&~ =p(sI-A_\eta)-\sum_{i=1}^{n_y}[q]_i(s+\lambda)^{\sigma_i}[C_\eta]_{i,:}+qF_\eta. \label{eq_proof_proposition_K0_6}
\end{align}
Then, using $q=pB_\eta E_\eta^\dagger$ and $A_0=A_\eta-B_\eta E_\eta^\dagger F_\eta$, we obtain
\begin{align}
&\ell(sI-A_\eta)+\sum_{i=1}^{n_y}[q]_i(s+\lambda)^{\sigma_i}[C_\eta]_{i,:}\notag\\
&=p(sI-A_\eta)+qF_\eta\notag\\
&=p(sI-A_\eta+B_\eta E_\eta^\dagger F_\eta)=p(sI-A_0). \label{eq_proof_proposition_K0_7}
\end{align}
\end{enumerate}

Based on properties \eqref{eq_proof_proposition_K0_2} and \eqref{eq_proof_proposition_K0_7}, we have
\begin{align}
&[\ell,[q]_1(s+\lambda)^{\sigma_1},\ldots,[q]_{n_y}(s+\lambda)^{\sigma_{n_y}}]
[sI-A_\eta,B_\eta;C_\eta,-D_\eta]\notag\\
&=[\ell(sI-A_\eta)+\sum_{i=1}^{n_y}[q]_i(s+\lambda)^{\sigma_i}[C_\eta]_{i,:},~\ell B_\eta-qD_\eta]\notag\\
&=[p(sI-A_0),~pB_\eta P_\eta]=0, \label{eq_proof_ell_equalities}
\end{align}
where the first equality uses \eqref{eq_proof_weighted_Deta_arxiv}, the second equality follows from properties \eqref{eq_proof_proposition_K0_2} and \eqref{eq_proof_proposition_K0_7}, and the last equality holds by \eqref{eq_proof_K0_PBH}.

By the full-row-rank condition \eqref{eq_proof_observer_rank_equivalence_arxiv}, the matrix $[sI-A_\eta,B_\eta;C_\eta,-D_\eta]$ has full row rank. Hence, the row vector $[\ell,[q]_1(s+\lambda)^{\sigma_1},\ldots,[q]_{n_y}(s+\lambda)^{\sigma_{n_y}}]$ in \eqref{eq_proof_ell_equalities} must be zero, giving $\ell=0$ and $[q]_i(s+\lambda)^{\sigma_i}=0$ for all $i=1,\ldots,n_y$. Since $\lambda>0$ and $s\in\bar{\mathbb{C}}_+$, we have $s+\lambda\ne0$, which implies $[q]_i=0$ for every $i$ and thus $q=0$. The definition of $S_i(s)$ in \eqref{eq_def_Si} then gives $S_i(s)=0$ for all $s\in\bar{\mathbb{C}}_+$ and all $i=1,\ldots,n_y$. Substituting these equalities into \eqref{eq_def_ell} gives $p=\ell=0$, contradicting $p\ne0$. Therefore, $(A_0,B_\eta P_\eta)$ is stabilizable.
\fi
% ~~~~放到arxiv版本的部分~~~~

{\it 3) Velocity-regulation property \eqref{eq_observer_tracking_convolution}.}
We establish the property \eqref{eq_observer_tracking_convolution} by deriving upper bounds on $|[\tilde{v}(t)]_i|$ in the two cases $\sigma_i\ge1$ and $\sigma_i=0$ and combining the resulting upper bounds. To this end, let $\tilde{\eta}=\eta-\hat{\eta}$ and, for each $i=1,\ldots,n_y$, define $e_i=[e_{i,1};\ldots;e_{i,\max\{1,\sigma_i\}}]$ with $e_{i,1}=[\tilde{v}]_i$ and $e_{i,q}=[C_\eta]_{i,:}A_\eta^{q-1}\eta$ for $q=2,\ldots,\sigma_i$.

Consider the case $\sigma_i\ge1$. Recall the dynamics \eqref{eq_yeta_system} and the velocity-regulation controller \eqref{eq_regulation_controller}. Taking the time derivatives of $e_i$ and $\tilde{\eta}$ and using the definition of $\sigma_i$ in \eqref{eq_def_ri}, the property $C_\eta A_\eta^kB_\eta=CA^{k+1}B$ for any nonnegative integer $k$, the full-row-rank condition \eqref{eq_fullrowrank}, the binomial formula \cite[Section 3.2.5]{Horn-Johnson-book-2012}, and the definitions in \eqref{eq_KT_selection_condition}, we obtain
\begin{subequations}
\label{eq_proof_joint_system}
\begin{align}
\begin{bmatrix}\dot{e}_i\\ \dot{\tilde{\eta}}\end{bmatrix}
&=\begin{bmatrix}A_i&-J_i[E_\eta K]_{i,:}\\0&A_\eta-LC_\eta\end{bmatrix}
\begin{bmatrix}e_i\\\tilde{\eta}\end{bmatrix}
+\begin{bmatrix}B_i\\0\end{bmatrix}[\dot{v}_*]_i,\\
[\tilde{v}]_i&=C_i e_i
\end{align}
\end{subequations}
for almost every $t\ge0$, where $A_i$ is the companion matrix whose last-row entries are $[A_i]_{\sigma_i,q}=-{\sigma_i!\lambda^{\sigma_i-q+1}}/((q-1)!(\sigma_i-q+1)!)$ for $q=1,\ldots,\sigma_i$, $B_i=[-1,0,\ldots,0]^T$, $C_i=[1,0,\ldots,0]$, and $J_i=[0,\ldots,0,1]^T$. Here, $E_\eta$ is defined in \eqref{eq_def_Eeta}, and $L$ is chosen to satisfy \eqref{eq_L_condition}.

Since $\lambda=\bar{\sigma}/\gamma$, defined below \eqref{eq_def_ri}, is positive, $A_i$ is Hurwitz. This, together with \eqref{eq_L_condition} and the block triangular structure of the state matrix in \eqref{eq_proof_joint_system}, ensures that the zero-input response of the joint system decays exponentially. Since neither $[\dot{v}_*]_i$ nor $e_i$ enters the observation-error dynamics, the zero-state response of $[\tilde{v}]_i$ to $[\dot{v}_*]_i$ is $(C_i e^{A_i(\cdot)}B_i\ast[\dot{v}_*(\cdot)]_i)(t)$. By the repeated-root solution formula \cite[Section 4.2, Eq. (18)]{Boyce-DiPrima-Book-2012}, $\sigma_i\le\bar{\sigma}$, and \eqref{eq_observer_tracking_convolution_h}, we have $|C_i e^{A_i t}B_i|=e^{-{\bar{\sigma}}t/{\gamma}}\Sigma_{q=0}^{\sigma_i-1}{(\bar{\sigma}t/\gamma)^q}/{q!}\le h(t)$. Applying the solution formula \cite[Theorem 5.2]{Hespanha-Book-2018}, we therefore obtain positive constants $a_i,b_i$ such that
\begin{align}
|[\tilde{v}(t)]_i|
&\le a_i e^{-b_i t}(|e_i(0)|+|\tilde{\eta}(0)|)
+(h(\cdot)\ast|[\dot{v}_*(\cdot)]_i|)(t)
\label{eq_proof_observer_channel_bound}
\end{align}
for all $t\ge0$ and all $i\in\{1,\ldots,n_y\}$ with $\sigma_i\ge1$. 

Consider the case $\sigma_i=0$. The definitions of $E_\eta$, $F_\eta$, $K$ and $T$ in \eqref{eq_KT_selection_condition}, under condition \eqref{eq_fullrowrank}, reduce the velocity equation in \eqref{eq_eta_subsystem_new} with the control law \eqref{eq_controller} to $e_i=[\tilde{v}]_i=[C_\eta]_{i,:}\tilde{\eta}$. The exponential stability of the $\tilde{\eta}$-subsystem in \eqref{eq_proof_joint_system} under condition \eqref{eq_L_condition} then yields \eqref{eq_proof_observer_channel_bound} for suitable $a_i,b_i>0$.

Therefore, using the definitions of $e_i$ and $\tilde{\eta}$ to estimate their initial values, stacking \eqref{eq_proof_observer_channel_bound} over all channels, and applying Minkowski's integral inequality \cite[Theorem 201]{Hardy-Littlewood-Polya-Book-1934}, we obtain \eqref{eq_observer_tracking_convolution} for suitable positive constants $c_1,c_2$.

\subsection{Proof of Proposition \ref{proposition_noisy_controller_bound}}\label{appendix_proposition_noisy_controller_bound}

For any $(y,t)$ satisfying $d(y,t)>D$, we show
\begin{align}
&|\rho_s(\mathcal{M}_m(y,t))-\rho_s(\mathcal{M}_a(y,t))|\le\max\{5k_\alpha\bar\delta_m,\notag\\
&\qquad k_\alpha\bar\delta_m+2k_\alpha(D_s-D)\sqrt{\bar\delta_m/D}\}=:\kappa_s(\bar\delta_m). \label{eq_noisy_rhos_bound}
\end{align}
By \eqref{eq_feasibility_point}, \eqref{eq_robust_measured_range}, and \eqref{eq_assumption_delta}, we obtain the upper bound $5k_\alpha\bar{\delta}_m$ for $d(y,t)\ge D_s-2\bar{\delta}_m$. For $d(y,t)<D_s-2\bar{\delta}_m$, additionally using the convexity and separation of the obstacles in Assumption \ref{assumption_obstacle_new} and the projection theorem \cite[Proposition B.11(a)]{Bertsekas-Book-1997}, we obtain the upper bound $k_\alpha\bar{\delta}_m+2k_\alpha(D_s-D)\sqrt{\bar{\delta}_m/D}$.

Moreover, the definition of $r_L$ in \eqref{eq_def_rL}, the measurement model \eqref{eq_robust_measured_range}, and the error bound \eqref{eq_assumption_delta} imply that $|r_L(\mathcal{M}_m(y,t),l)-r_L(\mathcal{M}_a(y,t),l)|\le\bar{\delta}_m$ for any unit vector $l$. Combining this estimate with the definition of $b_L$ in \eqref{eq_def_bL_main}, the unit-row condition \ref{condition_AL_unit_rows} on $A_L$, and the bound \eqref{eq_noisy_rhos_bound} gives
\begin{align}
&|b_L(\mathcal{M}_m(y,t))-b_L(\mathcal{M}_a(y,t))|\le\sqrt{n_L}(k_\alpha\bar{\delta}_m+\kappa_s(\bar{\delta}_m)). \label{eq_noisy_bL_bound}
\end{align}
The Lipschitz continuity of the solution to the QP \eqref{eq_safety_critical_controller} with respect to $\rho_s$ and $b_L$, established in the proof of Proposition \ref{proposition_safety}, together with \eqref{eq_noisy_rhos_bound} and \eqref{eq_noisy_bL_bound}, yields \eqref{eq_noisy_controller_bound}.

\section{A Convolution-Based Small-Gain Lemma}\label{appendix_lemmas}

Consider measurable and locally bounded signals $y_i:[0,T)\to\mathbb{R}_+$, $i=1,2$, with $0<T\le\infty$, satisfying
\begin{align}
y_i(t)\le E_i e^{\lambda_i t}+\sum_{j=1}^{2}(a_{i,j}(\cdot)\ast y_j(\cdot))(t)+c_i \label{eq_lemma_symmetric_convolution_inequalities}
\end{align}
for all $t\in[0,T)$, where $E_i,c_i\ge0$ and $\lambda_i<0$ are constants. With the symbol ``$\ast$'' defined in the Notations section, $a_{i,j}:\mathbb{R}_+\to\mathbb{R}_+$ represents the convolution-based gain from $y_j$ to $y_i$ for $i,j=1,2$. We make the following assumption.

\begin{assumption}\label{assumption_convolution_laplace}
For $i,j=1,2$, the functions $a_{i,j}$ are measurable, and there exists a constant $\bar{\lambda}<0$ such that $\sint_0^\infty e^{-\bar{\lambda}t}a_{i,j}(t)\mathrm{d}t<\infty$.
\end{assumption}

Assumption \ref{assumption_convolution_laplace} ensures that the Laplace transform
\begin{align}
\Theta(s)&=\int_0^\infty e^{-st}\begin{bmatrix}
a_{1,1}(t) & a_{1,2}(t)\\
a_{2,1}(t) & a_{2,2}(t)
\end{bmatrix}\mathrm{d}t \label{eq_lemma_symmetric_weighted_small_gain}
\end{align}
is well defined for all $s\ge\bar{\lambda}$. Note that $\Theta(0)$ is the standard gain matrix defined with the interconnection variables taking supremum norms. The following lemma gives an upper bound estimation on $[y_1(t);y_2(t)]$.

\begin{lemma}\label{lemma_symmetric_convolution_small_gain}
Under Assumption \ref{assumption_convolution_laplace}, if the spectral radius of $\Theta(0)$ is less than one, i.e., the small-gain condition holds as
\begin{subequations} \label{eq_lemma_smallgain_condition}
\begin{align}
[\Theta(0)]_{1,1} <1,\quad [\Theta(0)]_{2,2} &<1, \\
\frac{[\Theta(0)]_{1,2}[\Theta(0)]_{2,1}}{(1-[\Theta(0)]_{1,1})(1-[\Theta(0)]_{2,2})}&<1,
\end{align}
\end{subequations}
then there exists a constant $\mu$ satisfying $\max\{\lambda_1,\lambda_2,\bar{\lambda}\}<\mu<0$ such that
\begin{align}
\begin{bmatrix} y_1(t)\\ y_2(t) \end{bmatrix} &\le 
[I_2-\Theta(\mu)]^{-1}\begin{bmatrix} E_1\\ E_2 \end{bmatrix} e^{\mu t}+[I_2-\Theta(0)]^{-1}\begin{bmatrix} c_1\\ c_2 \end{bmatrix} 
\label{eq_lemma_symmetric_convolution_bounds}
\end{align}
for $t\in[0,T)$.
\end{lemma}

\begin{proof}
From the definition of $\Theta(s)$ in \eqref{eq_lemma_symmetric_weighted_small_gain} and Assumption \ref{assumption_convolution_laplace}, by \cite[Theorem 10.38]{Apostol-book-1974}, we have that $\Theta(s)$ is continuous at $s=0$. Thus, there exists a negative constant $\mu>\max\{\lambda_1,\lambda_2,\bar{\lambda}\}$ such that the spectral radius of $\Theta(\mu)$ is less than one. Consequently, $I_2-\Theta(\mu)$ and $I_2-\Theta(0)$  are nonsingular, with nonnegative inverses.

Let
\begin{subequations}
\begin{align}
W_i(t)&=\sup_{0\le\tau\le t}e^{-\mu\tau}[y_i(\tau)-[\bar c]_i]^+,~i=1,2, \label{eq_lemma_def}\\
\bar{c}&=[I_2-\Theta(0)]^{-1}[c_1;c_2].
\end{align}
\end{subequations}
The local boundedness of $y_i$ ensures that $W_i(t)$ is well-defined for $t<T$. Using \eqref{eq_lemma_symmetric_convolution_inequalities} and the fact $y_j\le[y_j-[\bar c]_j]^++[\bar c]_j$, we obtain
\begin{align}
&[y_i(\tau)-[\bar c]_i]^+\le[E_i e^{\lambda_i\tau}+\sum_{j=1}^{2}(a_{i,j}(\cdot)\ast[y_j(\cdot)-[\bar c]_j]^+)(\tau)\notag\\
&\quad+\sum_{j=1}^{2}[\bar c]_j\int_0^\tau a_{i,j}(s)\mathrm{d}s+c_i-[\bar c]_i]^+\notag\\
&\quad\le E_i e^{\lambda_i\tau}+\sum_{j=1}^{2}(a_{i,j}(\cdot)\ast[y_j(\cdot)-[\bar c]_j]^+)(\tau)
\label{eq_proof_symmetric_shifted}
\end{align}
for $0\le\tau<T$, where the last inequality is obtained by using $\sint_0^\tau a_{i,j}(s)\mathrm{d}s\le[\Theta(0)]_{i,j}$ obtained from the definition of $\Theta(s)$ in \eqref{eq_lemma_symmetric_weighted_small_gain} and $\bar c=\Theta(0)\bar c+[c_1;c_2]$. Multiplying both sides of \eqref{eq_proof_symmetric_shifted} by $e^{-\mu\tau}$, we obtain
\begin{align}
&e^{-\mu\tau}[y_i(\tau)-[\bar c]_i]^+\notag\\
&\le E_i e^{(\lambda_i-\mu)\tau}+\sum_{j=1}^{2}W_j(t)\int_0^\tau e^{-\mu s}a_{i,j}(s)\mathrm{d}s\notag\\
&\le E_i+\sum_{j=1}^{2}[\Theta(\mu)]_{i,j}W_j(t)
\label{eq_proof_symmetric_weighted_inequality}
\end{align}
for $0\le\tau\le t<T$, where we used the definition of $W_i$ in \eqref{eq_lemma_def} for the first inequality, and used $\lambda_i<\mu$ and \eqref{eq_lemma_symmetric_weighted_small_gain} for the last inequality. Taking the supremum of both sides of \eqref{eq_proof_symmetric_weighted_inequality} over $\tau\in[0,t]$ gives
\begin{align}
[W_1(t);W_2(t)]&\le[E_1;E_2]+\Theta(\mu)[W_1(t);W_2(t)]\notag\\
\Rightarrow [W_1(t);W_2(t)]&\le[I_2-\Theta(\mu)]^{-1}[E_1;E_2]. \label{eq_proof_symmetric_composite_gain}
\end{align}
This, together with $y_i(t)\le[\bar c]_i+e^{\mu t}W_i(t)$ and the definition of $\bar c$, gives \eqref{eq_lemma_symmetric_convolution_bounds}.
\end{proof}

The convolution-based small-gain analysis preserves the fading-memory effects of subsystem inputs, and helps evaluate the influence of initial states.
% yields a safety-margin condition independent of initial states; see Remark \ref{remark_smallgain_comparison} for details.

\section*{References}

\bibliographystyle{IEEEtran}
\bibliography{material/bpreference}

\vspace{\biographyvspace}

\begin{IEEEbiography}[{\includegraphics[width=1in, height=1.25in, clip, keepaspectratio]{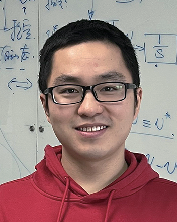}}]{Si Wu} 
\interlinepenalty=-1000
received his Ph.D. from Northeastern University, Shenyang, China, in 2025, where he currently serves as a Postdoctoral Fellow and an Associate Research Fellow. His research interest focuses on nonlinear control theory and applications to vehicles and robotic systems. He was a leading team member winning the IMAV 2019 outdoor competition.
\end{IEEEbiography}

\vspace{\biographyvspace}

\begin{IEEEbiography}[{\includegraphics[width=1in, height=1.25in, clip, keepaspectratio]{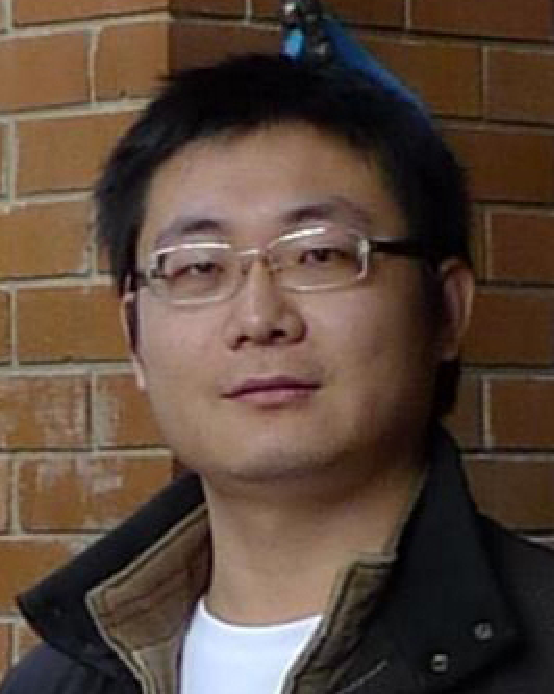}}]{Tengfei Liu} 
\interlinepenalty=-1000
received his Ph.D. from the Australian National University, Canberra, Australia, in 2011. From 2011 to 2013, he was a Postdoctoral Fellow with a faculty fellowship at the Polytechnic Institute of New York University. Since 2014, he has been a faculty member at Northeastern University, China, where he is currently a full professor. His research interests focus on the stability and control of interconnected nonlinear systems. He serves as an Associate Editor for IEEE Transactions on Automatic Control.
\end{IEEEbiography}

\vspace{\biographyvspace}

\begin{IEEEbiography}[{\includegraphics[width=1in, height=1.25in, clip, keepaspectratio]{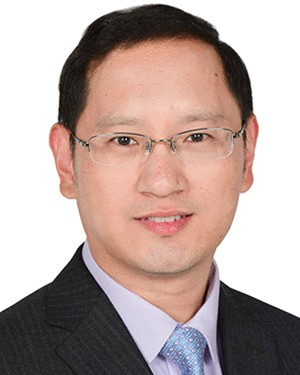}}]{Jinliang Ding}
\interlinepenalty=-1000
received the bachelor's, master's, and Ph.D. degrees in control theory and control engineering from Northeastern University, Shenyang, China, in 2001, 2004, and 2012, respectively. He is currently a professor with the State Key Laboratory of Synthetical Automation for Process Industries, Northeastern University. He has authored or co-authored more than 300 refereed journal papers and refereed papers at international conferences. He has also invented or co-invented more than 50 patents. His research interests include modeling, plant-wide control and optimization for complex industrial systems, machine learning, industrial artificial intelligence, and computational intelligence and applications. 

Dr. Ding was a recipient of the Young Scholars Science and Technology Award of China in 2016, the National Science Fund for Distinguished Young Scholars in 2015, and the National Technological Invention Award in 2013. One of his articles published in Control Engineering Practice was selected for the Best Paper Award from 2011 to 2013. He has served as an Associate Editor of IEEE Transactions on Evolutionary Computation, IEEE Transactions on Emerging Topics in Computational Intelligence, and IEEE Transactions on Circuits and Systems II: Express Briefs.
\end{IEEEbiography}

\vspace{\biographyvspace}

\begin{IEEEbiography}[{\includegraphics[width=1in, height=1.25in, clip, keepaspectratio]{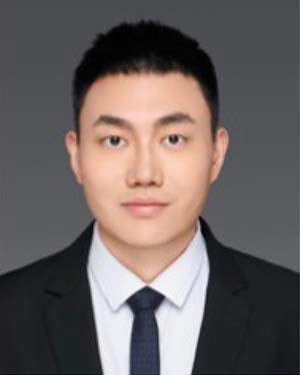}}]{Yuzhe Li} 
\interlinepenalty=-1000
received the B.S. degree in mechanics from Peking University, Beijing, China, in 2011, and the Ph.D. degree in electronic and computer engineering from the Hong Kong University of Science and Technology, Hong Kong, in 2015.

He is currently a Professor with the State Key Laboratory of Synthetical Automation for Process Industries, Northeastern University, Shenyang, China. From June 2013 to August 2013, he was a Visiting Scholar with the University of Newcastle, Callaghan, NSW, Australia. From September 2015 to September 2017, he was a Postdoctoral Fellow with the Department of Electrical and Computer Engineering, University of Alberta, Edmonton, AB, Canada. His research interests include networked control systems, cyber-physical systems security, and state estimation.
\end{IEEEbiography}

\vspace{\biographyvspace}

\begin{IEEEbiography}[{\includegraphics[width=1in,height=1.25in, clip,keepaspectratio]{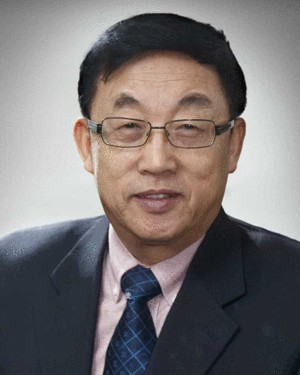}}]{Tianyou Chai} 
\interlinepenalty=-1000
received the Ph.D. degree in control theory and engineering from Northeastern University in 1985, where he became a Professor in 1988. He is the Founder and Director of the Center of Automation, which became a National Engineering and Technology Research Center and a State Key Laboratory. He has served as Director of Department of Information Science of National Natural Science Foundation of China from 2010 to 2018. His current research interests include modeling, control, optimization and integrated automation and intelligence of complex industrial processes.

His paper titled Hybrid intelligent control for optimal operation of shaft furnace roasting process was selected as one of three best papers for the Control Engineering Practice Paper Prize for 2011-2013. For his contributions, he has won 5 prestigious awards of National Natural Science, National Science and Technology Progress and National Technological Innovation, the 2007 Industry Award for Excellence in Transitional Control Research from IEEE Multiple-Conference on Systems and Control, and the 2017 Wook Hyun Kwon Education Award from Asian Control Association. He is a Member of Chinese Academy of Engineering, IFAC Fellow and IEEE Fellow.
\end{IEEEbiography}

\vspace{\biographyvspace}

\begin{IEEEbiography}[{\includegraphics[width=1in,height=1.25in, clip,keepaspectratio]{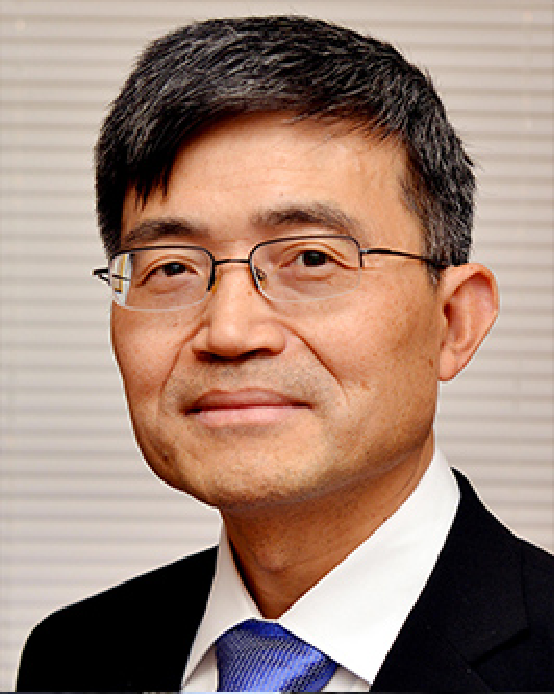}}]{Zhong-Ping Jiang} 
\interlinepenalty=-1000
received the M.Sc. degree in statistics from the University of Paris XI, France, in 1989, and the Ph.D. degree in automatic control and mathematics from the Ecole des Mines de Paris, France, in 1993. Currently, he is an Institute Professor at the Tandon School of Engineering, New York University. His main research interests include stability theory, robust/adaptive/distributed nonlinear control, learning-based control and their applications to information, mechanical, transportation and biological systems.

He has served as Deputy Editor-in-Chief, Senior Editor and Associate Editor for numerous journals. Prof. Jiang is a Fellow of IEEE, IFAC, CAA, AAIA and AAAS and is among the Clarivate Analytics Highly Cited Researchers and World's Top 2\% Highly Cited Scientists by Stanford. He is elected to the Academia Europaea (Academy of Europe) in 2021 and to the European Academy of Sciences and Arts in 2023 and, in 2022, he received the Excellence in Research Award from NYU Tandon School of Engineering.
\end{IEEEbiography}

\end{document}